\documentclass[11pt]{article}
\pdfoutput=1

\usepackage{setspace}
\usepackage[letterpaper,total={15.5cm, 22.9cm}]{geometry}

\usepackage{titlesec}

\titleformat*{\section}{\Large\bfseries}
\titleformat*{\subsection}{\large\bfseries}
\titleformat*{\subsubsection}{\bfseries}

\usepackage{microtype}
\usepackage{upgreek}

\usepackage{empheq}
\usepackage{amsmath,amsthm,amssymb}
\usepackage{booktabs}
\usepackage{cite}
\usepackage{enumitem}

\usepackage{ifpdf}
\usepackage{lmodern}
\usepackage[T1]{fontenc}
\usepackage[ansinew]{inputenc}
\usepackage{relsize}
\usepackage{comment}
\usepackage{graphicx}
\usepackage{color}
\definecolor{darkblue}{cmyk}{0.9,0.9,0,0}
\definecolor{darkgreen}{rgb}{0,0.55,0}
\usepackage[colorlinks=true,linkcolor=darkblue,citecolor=darkblue,urlcolor=darkblue]{hyperref}
\usepackage{epsfig}
\usepackage{epstopdf}
\usepackage{mathtools}

\newtheorem{innercustomres}{Result}
\newenvironment{result}[1]
  {\renewcommand\theinnercustomres{#1}\innercustomres}
  {\endinnercustomres}

\makeatletter
\long\def\@makecaption#1#2{
  \vskip\abovecaptionskip
  \sbox\@tempboxa{{\captionfonts #1: #2}}
  \ifdim \wd\@tempboxa >\hsize
    {\captionfonts #1: #2\par}
  \else
    \hbox to\hsize{\hfil\box\@tempboxa\hfil}
  \fi
  \vskip\belowcaptionskip}
\makeatother

\usepackage{array, booktabs, makecell, multirow}
\setcellgapes{3pt}
\def\ccrit{\mathcal{C}_{\rm crit}}
\def\sfa{\mathsf{a}}
\def\sfb{\mathsf{b}}
\def\sfc{\mathsf{c}}
\def\RMT{{\rm RMT}}
\DeclareMathOperator*{\sumint}{%
\mathchoice%
  {\ooalign{$\displaystyle\sum$\cr\hidewidth$\displaystyle\int$\hidewidth\cr}}
  {\ooalign{\raisebox{.14\height}{\scalebox{.7}{$\textstyle\sum$}}\cr\hidewidth$\textstyle\int$\hidewidth\cr}}
  {\ooalign{\raisebox{.2\height}{\scalebox{.6}{$\scriptstyle\sum$}}\cr$\scriptstyle\int$\cr}}
  {\ooalign{\raisebox{.2\height}{\scalebox{.6}{$\scriptstyle\sum$}}\cr$\scriptstyle\int$\cr}}
}

\usepackage{physics}

\def\sl{SL(2,\Z)}

\def\spec{{\rm spec}}

\def\s{\sigma}

\def\L{\Lambda}

\def\z{\zeta}

\def\cA{{\cal A}}

\def\cC{{\cal C}}

\def\cF{{\cal F}}

\def\cI{{\cal I}}

\def\cM{{\cal M}}
\def\cN{{\cal N}}
\def\cO{{\cal O}}
\def\cP{{\cal P}}

\def\cR{{\cal R}}
\def\cS{{\cal S}}

\def\RR{{\mathbb R}}

\def\HH{{\mathbb H}}

\def\Re{{\rm Re \,}}
\def\Im{{\rm Im \,}}

\def\Tr{{\rm Tr}}
\def\det{{\rm det \,}}

\def\half{{1\over 2}}
\def\p{\partial}

\def\a{\alpha}

\def\b{\beta}
\def\g{\gamma}
\def\G{\Gamma}
\def\l{\lambda}
\def\eps{\epsilon}

\def\cI{\mathcal{I}}
\def\HH{\mathbb{H}}
\def\sl{SL(2,\Z)}
\def\crit{{\half+i\w}}
\def\sn{\,\sum_{n=1}^\i}

\def\ds{\int_{\Re \,s\,=\,\half} ds\,}
\def\CJ{{\rm CJ}}
\def\WH{{\rm WH}}
\def\cP{\mathcal{P}}

\def\spec{{\rm spec}}
\def\diag{{\rm diag}}
\def\Hecke{{\rm Hecke}}
\def\Zh{Z_{\rm Hecke}}
\def\Zd{Z_{\rm diag}}
\def\Zs{Z_{\rm spec}}
\def\Rs{\rho_{\rm spec}}
\def\Rsjt{\rho_{{\rm spec},\,j}(t)}
\def\Zl{\widehat{Z}_L}
\def\cF{\mathcal{F}}
\def\w{\omega}
\def\L{\Lambda}
\def\half{{1\o2}}

\def\R{\mathbb{R}}
\def\qb{\overline{q}}

\def\Re{\text{Re}}
\def\Im{\text{Im}}

\def\x{\times}

\def\eps{\epsilon}

\def\hb{\overline h}
\def\Z{\mathbb{Z}}
\def\M{\mathcal{M}}

\def\1{{\rm 1-loop}}

\def\Tr{{\rm Tr}}

\def\c{\cite}

\def\cA{\mathcal{A}}

\def\cR{\mathcal{R}}
\def\cM{\mathcal{M}}

\def\cN{\mathcal{N}}
\def\cA{\mathcal{A}}

\def\c{\cite}

\def\vs{\vskip .1 in}

\def\G{\Gamma}
\def\p{\partial}
\def\o{\over}
\def\g{\gamma}
\def\D{\Delta}
\def\rar{\rightarrow}

\def\eqr{\eqref}

\def\O{{\cal O}}
\def\ra{\rangle}
\def\la{\langle}
\def\ssec{\subsection}
\def\sssec{\subsubsection}
\def\sec{\section}
\def\i{\infty}
\def\foot{\footnote}
\newcommand{\es}[2] {\begin{equation} \label{#1} \begin{split} #2 \end{split} \end{equation}}
\newcommand{\e}[2] {\begin{equation} \label{#1} #2 \end{equation}}

\newcommand{\beq}{\begin{equation}}
\newcommand{\eeq}{\end{equation}}
\newcommand{\beqy} {\begin{eqnarray}}
\newcommand{\eeqy} {\end{eqnarray}}
\newcommand{\bsmat}{\begin{smallmatrix}}
\newcommand{\esmat}{\end{smallmatrix}}
\newcommand{\bmat}{\begin{matrix}}
\newcommand{\emat}{\end{matrix}}

\def\({\left(}
\def\){\right)}
\def\[{\left[}
\def\]{\right]}

\def\<{\langle}
\def\>{\rangle}

\def\a{\alpha}
\def\b{\beta}
\def\g{\gamma}
\def\G{\Gamma}
\def\d{\delta}
\def\D{\Delta}
\def\z{\zeta}

\def\l{\lambda}

\def\t{\tau}
\def\s{\sigma}
\def\vs{\vskip .1 in}

\usepackage{varioref}
\usepackage{makeidx}
\makeindex

\usepackage[english]{babel}
\usepackage{subfig}

\numberwithin{equation}{section}

\usepackage{tabularx}

\begin{document}

\begin{spacing}{1.15}

\begin{titlepage}

\vspace*{3cm}
\begin{center}
{\LARGE \bfseries AdS$_3$/RMT$_2$ Duality}

\vspace*{1cm}

Gabriele Di Ubaldo, Eric Perlmutter

\vspace*{2mm}

\textit{\small Universit\'e Paris-Saclay, CNRS, CEA, Institut de Physique Th\'eorique, 91191, Gif-sur-Yvette, France}

\vspace{2mm}

{\tt \small gdiubaldo,\,perl@ipht.fr}

\vspace*{1cm}
\end{center}
\begin{abstract}

We introduce a framework for quantifying random matrix behavior of 2d CFTs and AdS$_3$ quantum gravity. We present a 2d CFT trace formula, precisely analogous to the Gutzwiller trace formula for chaotic quantum systems, which originates from the $SL(2,\mathbb{Z})$ spectral decomposition of the Virasoro primary density of states. An analogy to Berry's diagonal approximation allows us to extract spectral statistics of individual 2d CFTs by coarse-graining, and to identify signatures of chaos and random matrix universality. This leads to a necessary and sufficient condition for a 2d CFT to display a linear ramp in its coarse-grained spectral form factor.

Turning to gravity, AdS$_3$ torus wormholes are cleanly interpreted as diagonal projections of squared partition functions of microscopic 2d CFTs. The projection makes use of Hecke operators. The Cotler-Jensen wormhole of AdS$_3$ pure gravity is shown to be extremal among wormhole amplitudes: it is the minimal completion of the random matrix theory correlator compatible with Virasoro symmetry and $SL(2,\mathbb{Z})$-invariance. We call this {\sf MaxRMT}: the maximal realization of random matrix universality consistent with the necessary symmetries. Completeness of the $SL(2,\mathbb{Z})$ spectral decomposition as a trace formula allows us to factorize the Cotler-Jensen wormhole, extracting the microscopic object $Z_{\rm RMT}(\tau)$ from the coarse-grained product. This captures details of the spectrum of BTZ black hole microstates. $Z_{\rm RMT}(\tau)$ may be interpreted as an AdS$_3$ half-wormhole. We discuss its implications for the dual CFT and modular bootstrap at large central charge.

\end{abstract}

\end{titlepage}
\end{spacing}

\pagenumbering{roman}
\begin{spacing}{0.8}

\setcounter{tocdepth}{2}
\tableofcontents
\end{spacing}

\newpage

\pagenumbering{arabic}
\setcounter{page}{1}

\begin{spacing}{1.12}

\sec{Introduction}

To establish whether AdS$_3$ pure gravity exists, one must understand the random matrix behavior of its black hole microstates.

Such is the view suggested by recent work on holographic duality in low dimensions, both for the AdS$_3$ quantum theory and its semiclassical limit. Perhaps the main justification comes from the celebrated JT/RMT duality in two bulk dimensions \cite{Saad:2019lba}, in which the boundary theory is an ensemble of (double-scaled) random matrices. This work (and its dilaton gravity generalizations, e.g. \cite{Stanford:2019vob,Witten:2020wvy,Turiaci:2020fjj,Maxfield:2020ale,Mertens:2020hbs}) was a combined evolution of earlier works drawing direct connections between black hole dynamics and random matrix statistics in AdS/CFT \cite{Cotler:2016fpe,Saad:2018bqo} and the emergence of the SYK model as a tractable yet strongly-coupled quantum system \cite{Sachdev_1993,KitaevTalk,Maldacena:2016hyu,You_2017,GarcaGarca2016}.  In higher-dimensional holography, the boundary theory is a continuum conformal field theory (CFT), endowed with extra structure. For 2d CFTs, this structure includes locality, Virasoro symmetry and modular invariance of the torus partition function, but more generally is comprised of some set of fundamental bootstrap axioms. How is random matrix theory (RMT) ``allowed'' to manifest itself in the observables of an individual 2d CFT while respecting the necessary constraints? 

We focus our attention on AdS$_3$/CFT$_2$ henceforth, and the ongoing quest for AdS$_3$ pure gravity.\foot{We will not spell out the full history of this subject, whose modern incarnation started in \cite{Witten:2007kt}; a recent account was given in \cite{Benjamin:2020mfz}.} As is well-known, the natural idea \cite{Maloney:2007ud} for computing the semiclassical bulk path integral with a single torus boundary (sum over all smooth bulk saddles $\cM$ with $\p\cM = T^2$) fails to produce a unitary result, instead carrying exponentially large negative degeneracies \cite{Benjamin:2019stq}: if a consistent partition function exists, something more must contribute. But what? Reckoning with path integral contours is not a simple endeavor in quantum field theory, much less in gravity. In the proposal of  \cite{Maxfield:2020ale} -- still fairly implicit, but currently the only one which preserves a spectral gap to the black hole threshold -- what ostensibly fixes the problem is a specific infinite family of off-shell geometries (Seifert manifolds), whose circle reductions are identified with JT gravity backgrounds in the presence of defects. This gives an elegant hint of random matrices in AdS$_3$ pure gravity in the near-extremal regime.

Stronger hints come from wormholes. Two-boundary path integrals have been computed in JT/RMT duality: the double-trumpet wormhole, together with the all-orders genus sum over higher topologies, exhibits the famous RMT level repulsion in the ensemble-averaged density-density correlator \cite{Saad:2019lba}. After analytic continuation to complex temperature, the double-trumpet leads to a linear ramp in the spectral form factor (SFF); the wormholes with higher topology, exponentially suppressed in entropy, collectively initiate the transition from ramp to plateau \cite{Altland:2022xqx,SSSY,Blommaert:2022lbh,Weber:2022sov}. In seeking an AdS$_3$/CFT$_2$ lesson from (or version of) the JT/RMT ensemble duality, it is an AdS$_3$ analog of the double-trumpet geometry that one should understand. This was the motivation of Cotler and Jensen (CJ) \cite{Cotler:2020ugk}, who computed the contribution of such a geometry -- an off-shell, connected, two-boundary torus wormhole -- to the AdS$_3$ gravity path integral. Let us call this $Z_\CJ(\t_1,\t_2)$. Being off-shell, the computation is non-standard, requiring the technique of constrained instantons instead of familiar-but-unavailable saddle point techniques. 

The CJ result is at once highly mysterious, remarkably simple, and deceptively rich. It contains unmistakable signs of random matrices or 2d CFT avatars thereof: at leading order in the low-temperature limit, $Z_\CJ(\t_1,\t_2)$ reproduces the universal result of double-scaled matrix integrals. It also contains infinite series of corrections that are apparently tied to modular invariance, and generalizes the RMT result to include spacetime spin. This indicates the presence of some underlying Virasoro generalization of RMT. Less clear is the sense in which an ensemble interpretation of the result is necessary, and if so, how this squares with reasonable expectations of the space of irrational 2d CFTs as a sporadic, generically discrete set of points. This constellation of ideas was playfully labeled ``random CFT'' in \cite{Cotler:2020ugk}. Despite the absence of a proper definition, this much seems certain: whatever ``random CFT'' means, it ought to be relevant for holography. 

In a chaotic 2d CFT in general, how do we extract random matrix behavior hiding within?

In a chaotic 2d CFT dual to AdS$_3$ pure gravity in particular, what does $Z_\CJ(\t_1,\t_2)$ mean? 

Perhaps in spite of appearances, the CJ wormhole does not imply that the boundary dual of semiclassical AdS$_3$ pure gravity is an ensemble of large central charge CFTs. Given a microscopic large $c$ CFT, by which we mean a $c \rar\i$ limit of an unbounded sequence of irrational Virasoro CFTs $\{ \mathcal{T}_c\}$ with a suitable spectral gap, there may be a coarse-graining procedure or kinematic averaging (e.g. with respect to energy or time), as in quantum systems, which is compatible with the bulk wormhole computation. In general, bulk calculations that imply non-factorizing correlations between disconnected boundaries are agnostic about what kind of boundary averaging gives rise to this correlation \cite{Belin:2020hea,Cotler:2020ugk,Pollack:2020gfa}; we are not aware of an effective field theory calculation in AdS$_{D\geq 3}$ gravity that singles out a boundary ensemble interpretation. The more robust concept, as emphasized in \cite{Schlenker:2022dyo}, is not ensemble averaging {\it per se}, but {\it apparent} averaging, which arises essentially because of the chaotic nature of the high-energy spectrum in the large $c$ limit. A nice discussion of this set of ideas is given in the Introduction of \cite{Cotler:2022rud}.  Also relevant for our work are the comments on the role of wormholes in non-averaged theories in \cite{PSSY,Saad:2019lba}.

Condensing the above into a challenge for semiclassical AdS$_3$ holography, the goal is to show how the bulk theory can be dual to a microscopic large $c$ CFT in a manner consistent with nonvanishing bulk wormhole amplitudes (or, perhaps, to show that it cannot). We emphasize that this is a challenge particularly posed by off-shell wormholes, as on-shell wormholes are instead fixed by suitable gluing of universal asymptotic CFT data (spectrum and OPE coefficents), which are themselves fixed by crossing symmetry in terms of low-energy inputs -- insensitive to level statistics of black hole microstates.\foot{Recent work establishes an impressive match between partition functions of individual saddle points of AdS$_3$ gravity (possibly coupled to point particles) of some fixed topology, and certain boundary computations. The latter recast the partition function either as a moment problem of a (near-)Gaussian ``large $c$ ensemble'' of CFT data \cite{Chandra:2022bqq} or using a novel topological quantum field theory \cite{Collier:2023fwi}. (See e.g. \cite{Chandra:2023rhx,Chandra:2022fwi,jafferisetall} for further developments.) This extends earlier ideas about AdS$_3$ gravity as an effective field theory, by incorporating a version of ETH for 2d CFTs \cite{Chandra:2022bqq,Belin:2020hea} and allowing multiple disconnected boundaries. Because these works establish a match at the level of individual saddles, irrespective of the full sum over topologies and of level statistics, the questions of whether pure gravity exists and what its boundary dual is (e.g. ensemble or not) are of course not addressed.}

Independent of applications to AdS$_3$ wormholes, we would like to develop a quantitative toolkit to derive emergent RMT physics from microscopic 2d CFT data. It may be useful to phrase this yet another way, using the vocabulary of the bootstrap approach to CFTs \cite{Hartman:2022zik}. The modular and conformal bootstrap have focused so far on constraining single-copy observables. This seems to obscure chaotic microstructure of the spectrum which is revealed in ``two-copy observables'' like the SFF. One would like to know whether the dip-ramp-plateau structure of the coarse-grained SFF of a chaotic CFT can be {\it bootstrapped} from a minimal set of CFT data. Lowering our sights by focusing on the linear ramp region, the CJ wormhole suggests a more specific program toward addressing this question vis-\`a-vis gravity: first, use the constraints of Virasoro symmetry and modular invariance to carve out the space of possible wormhole amplitudes; then, upon imposing the features of a 2d CFT dual to {\it pure} gravity in particular, determine where the CJ wormhole sits in this space. An emergence of RMT-like physics from this analysis would give one answer to what ``random CFT'' could mean in two dimensions, compatible with the necessary symmetry constraints. 

\ssec{Summary of results}

This work makes some headway on these questions for generic $c>1$ Virasoro CFTs and their AdS$_3$ gravity duals, guided in part by the theory of trace formulas for chaotic quantum systems, and by new perspectives on modular invariance.

We begin in {\bfseries Section \ref{sec:s2}} by recalling the $\sl$ spectral decomposition of torus partition functions of parity-invariant Virasoro CFTs (with no extra conserved currents) \cite{Benjamin:2021ygh}. A key player in our framework is $\Zs(\t)$, a certain subtracted partition function, defined by removing ``light'' primary operators from the partition function in a modular-invariant way. $\Zs(\t)$ admits a decomposition into a complete $\sl$-invariant eigenbasis.\foot{Leaving details to the main text, the eigenbasis is comprised of real-analytic Eisenstein series $E_s(\t)$ with $s=\half+i\w$ and $\w\in\R$, and Maass cusp forms $\phi_n(\t)$ with $n\in\Z_{\geq 0}$.} We review some suggestive hints from AdS$_3$ gravity and Narain CFT about the physical meaning of $\Zs(\t)$; these lead us to view $\Zs(\t)$ as the ``chaotic part'' of the partition function that suitably incorporates the symmetries. 

\begin{figure}[t]
\centering
{
\subfloat{\includegraphics[scale=0.17]{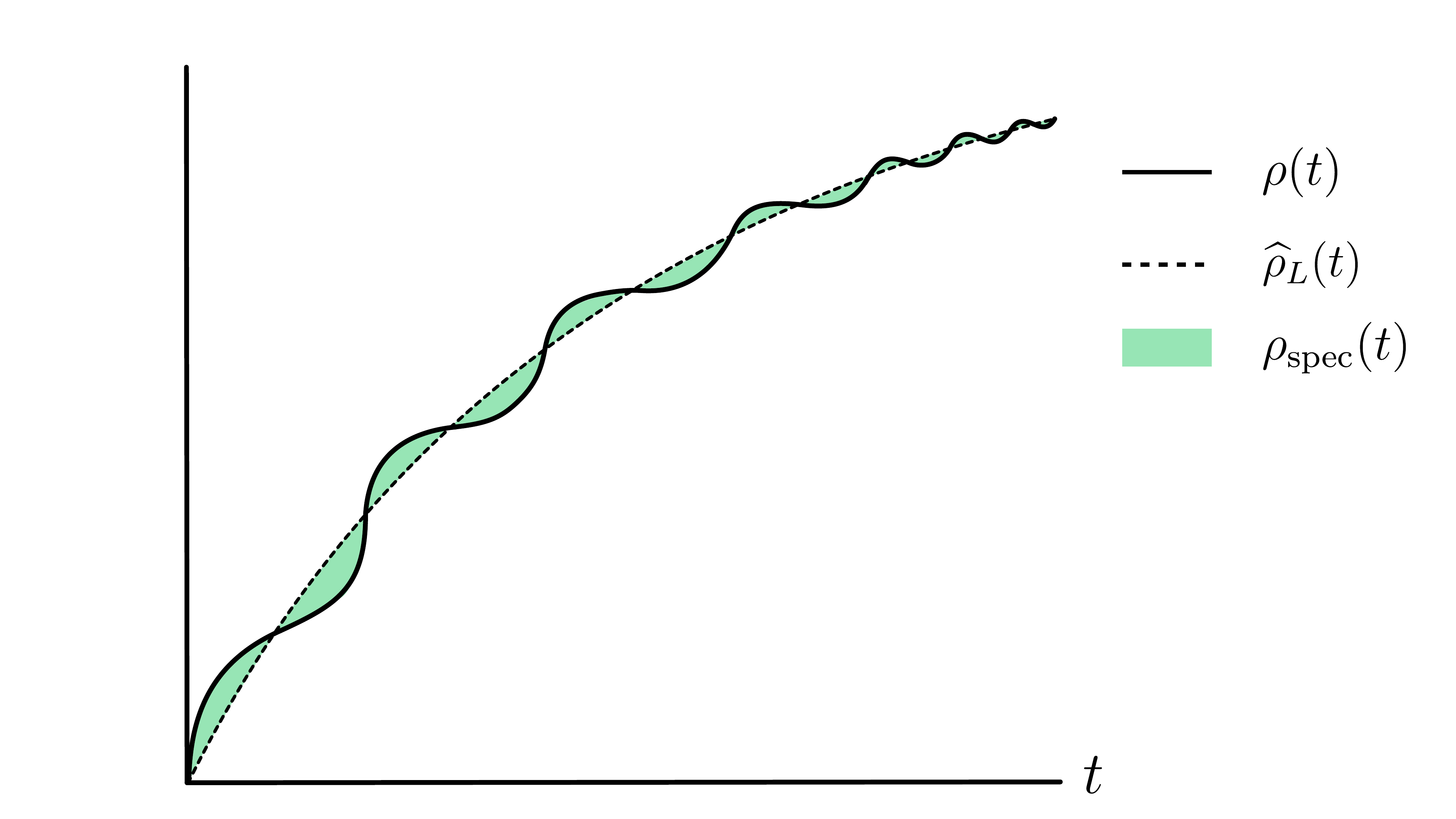}}
}
\caption{The density of heavy states $\rho(t)$ of a compact CFT, here approximated by the smooth black curve  (as in a large $c$ limit) for illustrative purposes, is highly oscillatory on wavelengths of order $e^{-S_{\rm Cardy}(t)}$, the mean level spacing, while the modular completion of light states, $\widehat{\rho}_{L}(t)$, contains the smooth, universal Cardy-like growth (dashed line). Their difference, $\rho_{\rm spec}(t)$, captures the oscillations, encoding chaotic statistics. We have suppressed the spin index $j$.}
\vspace{.1in}
\label{densityfig}
\end{figure}

In {\bfseries Section \ref{sec:s3}} we substantiate this point of view by presenting a 2d CFT trace formula. It mimics the Gutzwiller trace formula for chaotic quantum systems, which we first review. To make the connection, we proceed to transform $\Zs(\t)$ to a microcanonical density of states. The total density of spin-$j$ Virasoro primaries splits into two terms:
\e{}{\rho_j(t) = \widehat \rho_{L,j}(t) + \rho_{\spec,\,j}(t)\,, \quad\quad t:=\text{min}(h,\hb) - {c-1\o 24}}
One should think of the spectral density $\rho_{\spec,\,j}(t)$ as having removed all self-averaging contributions $\widehat \rho_{L,j}(t)$ from the total density $\rho_j(t)$: it is supported only on the chaotic, ``heavy'' spectrum $t\geq 0$, computing the difference between the exact density and the smooth asymptotic approximation to it. See Figure \ref{densityfig}.

In seeking a possible analogy to quantum systems, the first term $\widehat \rho_{L,j}(t)$ would map to a mean density $\overline\rho(E)$, while the second term $\rho_{\spec,\,j}(t)$ would map to $\rho_{\rm osc}(E)$, an oscillatory part that can be expanded over periodic orbits. Indeed, such a relation can be made sharp: the $\sl$ spectral decomposition of $\rho_{\spec,\,j}(t)$ is shown to take exactly the form of a Gutzwiller trace formula, for every fixed spin $j$ (see (\ref{eq:2dtrace})). Periodic orbits correspond to elements of the $\sl$ eigenbasis, labeled by $\sl$ spectral frequencies $\w$, with a clean identification of the orbit actions, periods and one-loop determinants for each element. An important aspect of this trace formula is that the $\sl$ eigenbasis, and hence the set of orbits, is complete. 

With this in hand, we analyze correlations and define a coarse-graining procedure in analogy to Berry's diagonal approximation. Mimicking the local energy averaging of quantum systems, coarse-graining a product of spectral densities over mean twist correlates the two copies by pairing the $\sl$ eigenvalues -- a 2d CFT analog of restricting the double sum over orbits to those of equal action. Inspired by this we define a diagonal partition function in the canonical ensemble. First we define $\Zd(\t_1,\t_2)$, by projecting the factorized product $\Zs(\t_1)\Zs(\t_2)$ onto the kernel of a difference of Laplacians, thus correlating the eigenvalues of basis elements. We then introduce an enhanced diagonal projection of $\Zs(\t_1)\Zs(\t_2)$ which pairs $\sl$ {\it eigenfunctions}, not just eigenvalues: there are degeneracies between Eisenstein series and Maass cusp forms. This is defined by projecting onto the kernel of a difference of $\sl$ Hecke operators, $T_j^{(\t_1)} - T_j^{(\t_2)}$, for every spin $j\in\Z_+$: we call this Hecke projection, and the corresponding partition function $\Zh(\t_1,\t_2)$. These are given in \eqr{heckeproj} and \eqr{ZHecke}. From the point of view of the trace formula and periodic orbit theory, Hecke projection does the job of properly pairing {\it identical orbits}. One may view $\Zh(\t_1,\t_2)$ as an enhanced form of coarse-graining in 2d CFTs, carrying extra symmetry and arithmeticity, annihilated as it is by an infinite set of commuting Hecke operators.

In {\bfseries Section \ref{sec:zdiag}}, we analyze $\Zh(\t_1,\t_2)$ for general chaotic 2d CFTs, culminating in a necessary and sufficient condition for the coarse-grained spectral form factor (SFF) to exhibit a linear ramp. In chaotic quantum systems, the SFF, call it $K_\b(T)$, famously exhibits a linear ramp at times $T \gg \b$, with coefficient controlled by the particular RMT ensemble governing the late-time dynamics. The diagonal approximation to the SFF is designed to capture this ramp behavior. Having constructed a diagonal partition function for 2d CFTs, with self-averaging terms judiciously subtracted, we are in position to show the same. 

Focusing on the scalar Fourier mode of $\Zh(\t_1,\t_2)$, it is fully determined by a function we call $\cR(z)$, defined as the inverse Mellin transform of $|(\Zs,E_s)|^2$, the squared spectral overlap of $\Zs(\t)$ with the Eisenstein series $E_s(\t)$. The variable $z:=y_1/y_2$ is the ratio of inverse temperatures $y_i := \Im(\t_i)$. Passing to SFF kinematics via $y_1 = \b+iT$ and $y_2 = \b-iT$, we show that the coarse-grained SFF exhibits a linear ramp at times $T \gg \b$ if and only if $\cR(z)$ has a simple pole at $z=-1$, with the correct RMT residue:
\es{ramppoleintro}{\mathcal{R}(z \rar -1) \sim {\mathsf{C}_{\rm RMT}\o 2\pi}{1\o 1+z}}
The constant $\mathsf{C}_{\rm RMT}$ sets the RMT ensemble (for example, $\mathsf{C}_{\rm GOE}=2$). This simple pole may be recast as a straightforward falloff condition on the partition function in $\sl$ spectral space:
\es{lineardivfinalintro}{|(\Zs,E_{\crit})|^2 \sim e^{-{\pi}\w}f(\w)\qquad (\w\rar\i)}
where $f(\w)$ approaches $\mathsf{C}_{\rm RMT}$ asymptotically. Moreover, Virasoro symmetry and $\sl$-invariance imply a specific set of terms \eqr{Sresult} in $\Zh(\t_1,\t_2)$ that necessarily accompany \eqr{ramppoleintro}; in SFF kinematics, these are power-law corrections at late times.

\begin{figure}[t]
\centering
{
\subfloat{\includegraphics[scale=0.17]{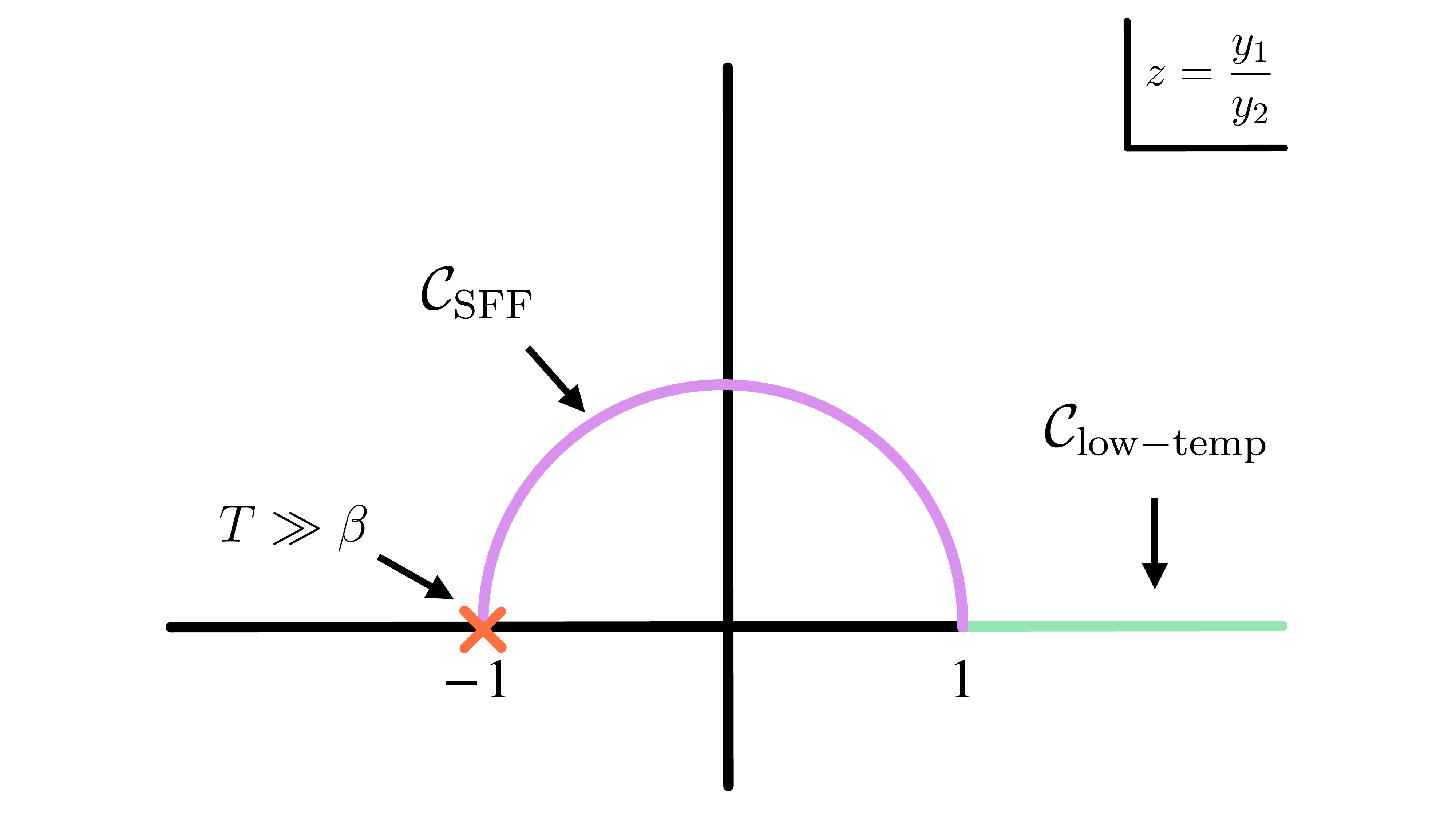}}
}
\caption{Off-shell wormhole amplitudes $Z_\WH(\t_1,\t_2)$ in semiclassical AdS$_3$ gravity are fixed by a single function $\cR(z)$, where $y_i = \Im(\t_i)$ are inverse temperatures. The colored contours show Euclidean and Lorentzian kinematics in the low-temperature limit. The spectral form factor (SFF), for which $z$ lies on the unit circle, has a linear ramp at times $T \gg \b$ if and only if $\cR(z)$ has a simple pole at $z=-1$: see \eqr{ramppoleintro}.}
\vspace{.1in}
\label{fig:contourz}
\end{figure}

In {\bfseries Section \ref{s4}} we prepare for our descent into the wormhole. 

There is a satisfying synergy of the diagonal partition function $\Zh(\t_1,\t_2)$ with AdS$_3$ torus wormholes. In particular, starting from a geometric definition, we demonstrate that torus wormhole amplitudes $Z_\WH(\t_1,\t_2)$ are Hecke symmetric: that is, they exhibit precisely the functional form of $\Zh(\t_1,\t_2)$.  This signals that wormholes may be viewed microscopically, understood as coarse-grained two-copy partition functions of underlying chaotic CFTs. This forms the basis of our ``wormhole Farey tail'': that is, the interpretation of bulk wormhole amplitudes $Z_\WH(\t_1,\t_2)$, constructed as $\sl$ image sums over large diffeomorphisms, as gravitational duals of $Z_\Hecke(\t_1,\t_2)$ in large $c$ CFTs. This is an AdS$_3$ realization of the idea of \cite{PSSY,Saad:2019lba} that bulk spacetime wormholes geometrize the diagonal approximation.

Let us expand on this slightly. In a diffeomorphism-invariant theory of semiclassical gravity, a wormhole amplitude $Z_\WH(\t_1,\t_2)$ that is independently modular-invariant with respect to both $\t_1$ and $\t_2$ can be constructed as an $\sl$ Poincar\'e sum of a suitable seed function: the $\sl$ transformations at the boundary implement the action of large bulk diffeomorphisms. This is a multi-boundary generalization of the familiar black hole Farey tail for thermal partition functions \cite{Dijkgraaf:2000fq,Maloney:2007ud,Keller:2014xba}. Taking this as one definition (given more precisely in Subsection \ref{farey}) of an off-shell torus wormhole, we prove that Poincar\'e sums of this form enjoy a few remarkable properties. Among others, $Z_\WH(\t_1,\t_2)$ is even more constrained than a generic Hecke projection: its Eisenstein and cusp form spectral overlaps are functionally equal, leading to a highly-constrained functional form in spectral space,
\e{zwhintro}{{Z_\WH(\t_1,\t_2) =  \int_{\ccrit} f_\WH(s) E_{1-s}(\tau_1)E_s(\tau_2) +\sum_{n=1}^\i f_\WH(s_n)\phi_n(\tau_1)\phi_n(\tau_2)}}
where $\ccrit$ defined in (\ref{eq:Ccrit}) denotes the straight contour $s=\half+i\R$. Given a $\Zs(\t)$ of an underlying CFT, the identification with the wormhole is simply 
\e{}{f_\WH(s) = |(\Zs,E_s)|^2\,.}
The form of \eqr{zwhintro} makes manifest that wormhole correlations are diagonalized by the $\sl$ spectral basis. Since diagonality is basis-dependent, this affirms the $\sl$ spectral decomposition as a proper trace formula for 2d CFT. 

\begin{figure}[t]
\centering
\vspace{.6in}
{
\subfloat{\includegraphics[scale=0.22]{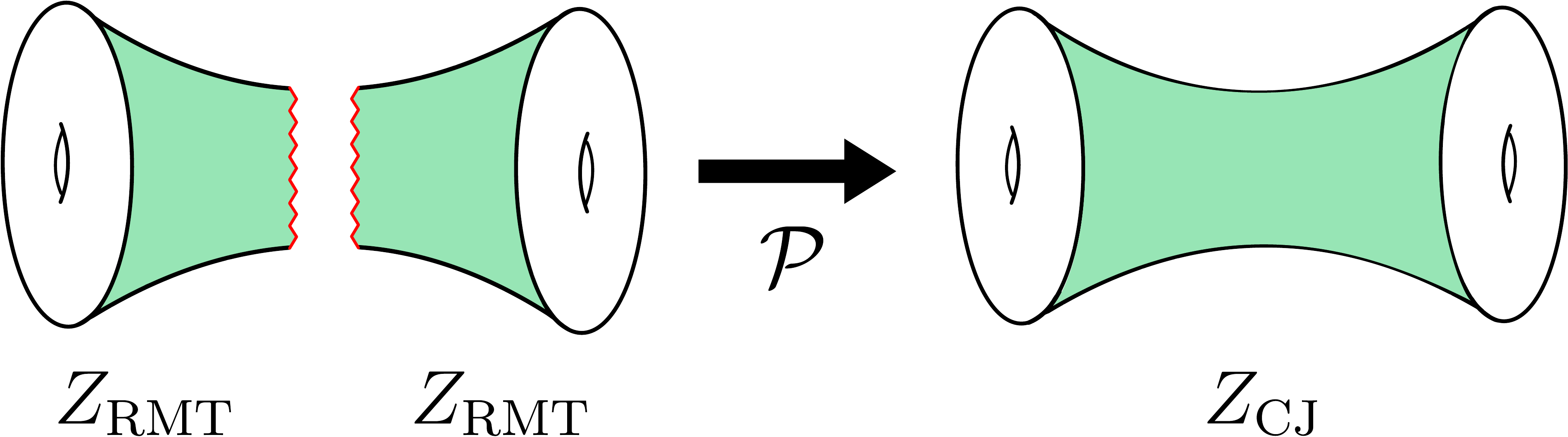}}
}
\vspace{.4in}
\caption{The Cotler-Jensen wormhole $Z_\CJ(\t_1,\t_2)$ is generated by gluing two single boundary partition functions $Z_\RMT(\t_1)$ and $Z_\RMT(\t_2)$. Each factor may be interpreted as a half-wormhole in AdS$_3$ pure gravity. The gluing is performed by coarse-graining over the spectrum of the dual CFT: the (Hecke) projection $\cP$ projects the factorized product onto the diagonal terms with respect to the $\sl$ spectral basis, analogously to trace formulas for non-disordered chaotic systems. }
\vspace{.08in}
\label{fig:halfWH}
\end{figure}

In {\bfseries Section \ref{sec:CJ}} we turn to AdS$_3$ pure gravity and the CJ wormhole; see Figure \ref{fig:halfWH}. This wormhole amplitude was derived in \cite{Cotler:2020ugk} as a Poincar\'e sum of the above type. Its spectral overlap is very simple: 
\e{}{f_{\rm CJ}(s) = {1\o\pi} \G(s)\G(1-s)\,.}
In terms of the function $\cR(z)$ defined earlier, $\cR_\CJ(z)$ not only contains the pole \eqr{ramppoleintro} that generates the linear ramp -- it is exactly {\it equal} to it. Moreover, the corrections prescribed by Virasoro and $\sl$ are exactly those found in \cite{Cotler:2020ugk}. 

This analysis tells us that the CJ wormhole is {\it extremal} within the space of admissible wormhole amplitudes, in the following quantitative sense. Having incorporated the requisite Virasoro symmetry and modular invariance -- that is, upon ``quotienting'' by the symmetries of 2d CFTs and wormholes -- the amplitude is determined solely by the function $\cR(z)$. The CJ wormhole of AdS$_3$ pure gravity then sets $\cR(z)$ exactly equal to the double-scaled RMT result. This signature of pure gravity is what we call {\sf {MaxRMT}:} {\it the maximal realization of random matrix universality consistent with Virasoro symmetry and modular invariance.} The fact that pure gravity exhibits MaxRMT statistics may be viewed as extending the hallmark maximal chaos of pure gravity in the semiclassical, early-time regime of Lyapunov chaos \cite{MSS} to the quantum, late-time regime as defined by RMT. We expand on this and make related comments in Subsection \ref{maxRMT}. 

That our formalism fits the CJ wormhole like a glove strongly indicates that the wormhole may be interpreted in terms of a {\it microscopic} 2d CFT dual, compatible with a traditional holographic interpretation for semiclassical AdS$_3$ pure gravity. The CJ wormhole is generated dynamically from an underlying CFT upon coarse-graining the chaotic spectral correlations as prescribed above. This gives a concrete actualization of the apparent averaging phenomenon of \cite{Schlenker:2022dyo}, which was argued on general grounds to emerge from chaos of the semiclassical black hole spectrum.

\begin{figure}[t]
\centering
\vspace{.6in}
{
\subfloat{\includegraphics[scale=0.22]{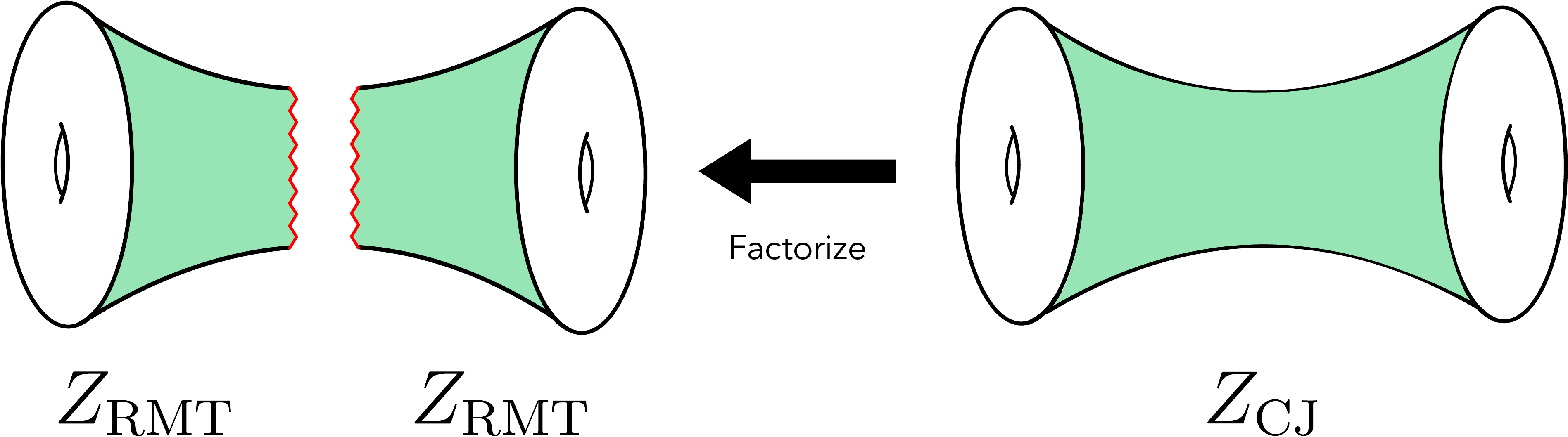}}
}
\vspace{.4in}
\caption{Completeness of the $SL(2,\Z)$ spectral basis permits factorization of the Cotler-Jensen wormhole. $Z_\RMT(\t)$ encodes quantum substructure of the pure gravity black hole spectrum.}
\vspace{.08in}
\label{fig:fact}
\end{figure}

With this understanding, in Subsection \ref{sec:BHmicro} we take a step further by leveraging the completeness of the $\sl$ spectral eigenbasis to factorize the CJ wormhole into its constituent components: see Figure \ref{fig:fact}. We call the resulting microscopic partition function $Z_\RMT(\t)$. The result is unique up to signs, and is given explicitly in \eqr{zrmteis} and \eqr{zrmt}. $Z_\RMT(\t)$ captures exponentially suppressed fine structure of the black hole spectrum of AdS$_3$ pure gravity. As is perhaps clear from these results, $Z_\RMT(\t)$ may be meaningfully viewed as a half-wormhole of  pure gravity. We substantiate this with comparison to 2D gravity half-wormholes, broken cylinders and branes. One intriguing aspect is that $Z_\RMT(\t)$ carries a conspicuous erratic phase: its spectral overlap is dressed by a Riemann zeta phase $\phi(\w) = \text{arg}(\z(1+2i\w))$. This is a property of $\Zs(\t)$ in general. The Riemann zeta function is a famously quantum chaotic object \cite{Montgomery,Odlyzko1987,Berry1999,Keating2000,Bourgade2013}, and $\phi(\w)$ varies wildly along the line $\w\in\R$ (see Figure \ref{zetaphasefig}); upon gluing to form a wormhole, the phases cancel. This nicely exhibits the erratic nature expected of a higher-dimensional generalization of half-wormholes. 

Summarizing the above, the identification of $Z_\RMT(\t)$ gives a new, quantum piece of the torus partition function $Z_{\rm grav}(\t)$ of the putative 2d CFT dual to semiclassical AdS$_3$ pure gravity, which augments the sum over smooth bulk saddles: 
\es{Zgravrmtintro1}{Z_{\rm grav}(\t) \approx Z_{\rm MWK}(\t) + Z_\RMT(\t) \,.}
$Z_{\rm MWK}(\t)$ is the sum over saddles with $\partial \cM = T^2$ \c{Maloney:2007ud,Keller:2014xba}. There are small, as-yet-undetermined corrections to $Z_\RMT(\t)$, call them $\delta Z_\RMT(\t)$, expected to come from other off-shell configurations in the bulk; on the CFT side, these would give corrections to the spectral statistics encoded in $Z_\RMT(\t)$. We explain why $\delta Z_\RMT(\t)$ must be nonzero due to CFT unitarity, and sketch its origins from the point of view of wormholes and string theory. This further implies that the BTZ black hole threshold lies strictly {\it below} the naive semiclassical threshold $t=0$. The structure of the resulting partition function also jibes nicely with the proposal of Maxfield and Turiaci \cite{Maxfield:2020ale}. The inclusion of level statistics of heavy operators poses an interesting challenge for the large $c$ modular bootstrap of 2d CFTs.

We end the paper with a discussion of some future directions, a glossary of partition functions defined in this work, and appendices with details complementing the main text. 

\section{Groundwork}\label{sec:s2}
We begin by briefly recalling some notions in the spectral theory of the Laplacian on the fundamental domain $\cF= \HH/\sl$ and their application to 2d CFTs.  For more detailed treatments and background, see \cite{Terras_2013,Benjamin:2021ygh,Collier:2022emf}; for follow-up work, see \cite{Benjamin:2022pnx,Haehl:2023tkr}. We mostly use conventions of \cite{Collier:2022emf}.  

\subsection{Lightning review of $SL(2,\mathbb{Z})$ spectral theory in 2d CFT}\label{ssec:2.1}

A square-integrable, $\sl$-invariant function admits a spectral decomposition in a complete $\sl$-invariant eigenbasis with three branches:
\es{}{ 
   \begin{cases}
      \Delta_\tau E^*_{s}(\tau)=s(1-s)E^*_s(\tau) & \quad s=\frac{1}{2}+i\w, \qquad \w\in\RR \\
      \Delta_\tau \phi_n(\tau)=s_n(1-s_n)\phi_n(\tau)  & \quad s_n=\half +i \w_n,\quad \w_n\in\RR\\
      \Delta_\t \phi_0 =0
    \end{cases}
}
$E^*_s(\tau)=E^*_{1-s}(\tau)$ is the completed Eisenstein series evaluated on the critical line, $\phi_n(\t)$ is an infinite discrete set of Maass cusp forms labeled by $n\in\Z_+$, and $\phi_0$ is the constant function. We define square integrability with respect to the Petersson inner product with hyperbolic measure ${dx \,dy}/y^{2}$ where $\t := x + iy$. The cusp forms are unit-normalized. The respective Fourier decompositions are\foot{With respect to notation in \cite{Benjamin:2021ygh,Collier:2022emf}, $2 \mathsf{b}_j^{(n)}|_{\rm here} = a_j^{(n)}|_{\rm there}$.}
\es{}{ 
E^*_s(\tau)&= \sum_{j= 0}^\i(2-\delta_{j,0})\sfa_j^{(s)} \cos(2\pi j x)  \sqrt{y} K_{s-\half}(2\pi j y),\\ 
\phi_n(\tau)&= \sum_{j= 1}^\i 2\sfb_j^{(n)} \cos(2\pi j x) \sqrt{y} K_{s_n-\half}(2\pi j y),
}
There are also Maass cusp forms odd under $x \rar -x$, but they are not relevant for the applications ahead. The Eisenstein series has Fourier coefficients
\e{}{ 
\sfa_j^{(s)}=\frac{2\sigma_{2s-1}(j)}{j^{s-\half}}
}
Its scalar mode can be formally obtained as a smooth $j\rightarrow 0$ limit of the spinning modes (see Appendix \ref{app:jzero}),
\es{}{
E^*_{s,0}(y) &= \L(s)y^s+\L(1-s)y^{1-s}\,,\\
&= \lim_{j\rightarrow 0} \sfa_j^{(s)} \sqrt{y}K_{s-\half}(2\pi j y)
} 
where $\L(s)= \pi^{-s} \G(s)\z(2s) = \L\qty(\half-s)$ is the completed Riemann zeta function. The cuspidal spectral parameters $\omega_n$ and Fourier coefficients $\sfb_j^{(n)} $ are sporadic real numbers with no known analytic expression. The multiplicity of Maass cusp forms at a given eigenvalue is conjecturally bounded above by one, i.e. the spectrum is ``simple'' (e.g. \cite{Sarnak2003}). 

Given an $f(\tau)\in L^2(\cF)$, the spectral decomposition is
\es{eq:specdec}{ 
f(\tau)= \la f\ra+ \int_{\cC_{\rm crit}} \{f,E_s\}E^*_s(\tau) +\sum_{n=1}^\i (f,\phi_n)\phi_n(\tau),
}
where we integrate over the critical line,
\e{eq:Ccrit}{\int_{\cC_{\rm crit}}:=\frac{1}{4\pi i}\int_{\Re(s)=\half} ds=\frac{1}{4\pi}\int_{-\infty}^{\infty}d\w}
The (completed) Eisenstein overlaps are
\es{}{
\{f,E_s\}:= \frac{(f,E_s)}{\Lambda(s)}
} 
where $(f,E_s)$ is the Petersson inner product. The constant term, $\< f\>$, is the ``modular average'' of $f(\t)$ over $\cF$. In the remainder of this work, we focus on parity-invariant CFT observables, so the sum over cusp forms is taken over even cusp forms only, with $n\in\Z_+$. We will occasionally find it useful to write the spectral decomposition in a unified notation as 
\e{specunified}{f(\t) = \sumint_\w \tilde f_\w \,\psi_\w(\t)\,,\quad \text{where}~~\psi_\w(\t) := \{E_\crit(\t), \phi_n(\t), \phi_0\}}
are the eigenfunctions and $\tilde f_\w := (f(\t),\psi_\w(\t))$ is a shorthand for the spectral overlap.

Consider now the torus partition function $Z(\tau)$ of a general (non-holomorphic) parity-invariant 2d CFT, which we assume to have only Virasoro symmetry with $c>1$. The primary partition function, which strips off Virasoro descendants\foot{The level-one null descendants of the Virasoro vacuum module are included, but all other descendants are removed.}  while preserving modular invariance, is defined by 
\e{}{Z_p(\tau)= \sqrt{y} |\eta(\tau)|^2 Z(\tau).}
In \cite{Benjamin:2021ygh}, it was shown that $Z_p(\tau)$ can be written in a manifestly modular-invariant way: 
\e{Zpsplit}{ 
Z_p(\tau)= \Zl(\tau) +\Zs(\tau), 
}
$\Zl(\tau)$ is the ``modular completion'' of light states, defined as follows. First, one constructs the partition function of light primaries
\es{ZLdef}{ 
Z_L(\t) := \sqrt{y}\sum_{\min(h,\hb)\leq \xi} q^{h-\xi}\qb^{\hb-\xi},
}
where in terms of conformal weight $\D$ and spin $j$,
\e{}{\D=h+\hb\,,\quad j = |h-\hb|\,,\quad \xi = {c-1\o 24}}
and $q = e^{2\pi i \t}$. One subsequently ``completes'' this into a modular-invariant function $\Zl(\tau)$ by suitably adding heavy states. A convenient, and physically distinguished, mode of modular completion is to perform a Poincar\'e sum over $\sl$ images \cite{Maldacena:1998bw, Dijkgraaf:2000fq,Maloney:2007ud,Keller:2014xba}:
\es{ZHLdef}{ 
\Zl(\tau)=\sum_{\gamma \in \sl/\Gamma_{\infty}} Z_L(\g\t).
}
We have modded out by $\Gamma_{\infty}$, the set of $T$-transformations, under which the summand is invariant (thanks to spin-quantization). Poincar\'e summation is physically distinguished because it is the {\it minimal} modular completion: it adds only the modular images of the light states, and nothing more. For this reason we define $\Zl(\tau)$ using Poincar\'e summation throughout the paper. Having explicitly constructed $\Zl(\t)$, we can define $\Zs(\t) $ by subtraction, which is square-integrable and thus admits an $\sl$ spectral decomposition: 
\es{}{ 
\Zs(\tau)= \la \Zs\ra+ \int_{\cC_{\rm crit}} \{\Zs,E_s \}E^*_s(\tau) +\sum_{n=1}^\i (\Zs,\phi_n)\phi_n(\tau).
}
Whereas $\Zl(\t)$ contains the leading-order Cardy asymptotics of $Z_p(\t)$ -- namely (but not only)  the identity block and its modular images -- $\Zs(\t)$ instead probes the chaotic, high-energy spectrum. We will make this sharper below. Note that $\Zs(\t)$ has support only on heavy states, but it indirectly ``knows about'' the light states of the CFT because of the modular completion used to define it. 

We point out that the analog of $\Zs(\t)$ would be trivial in a holomorphic (or holomorphically factorized) CFT, where the spectrum of states with $h > c/24$ is determined by its complement; this inherent non-holomorphicity is useful for focusing on generic irrational 2d CFTs and their gravity duals (as opposed to toy models such as chiral gravity). 

\sssec*{Two technical remarks}

In our expressions for $\Zs(\t)$, we henceforth set the modular averages $\< \Zs\>$ to zero. For Virasoro CFT partition functions \eqr{Zpsplit}, this is just a choice of convention to ``put'' the constant in $\Zl(\t)$ rather than $\Zs(\t)$: a c-number constant may be freely shuffled between the two terms while keeping $Z_p(\t)$ fixed.\foot{This may be viewed as a kind of physical implementation of Zagier's regularization \cite{zbMATH03796039}, who showed that a constant term in a square-integrable function on $\cF$ may be formally renormalized away by introducing a renormalized Rankin-Selberg transform, in which one cuts off the integral near the cusp at $y=L$ and takes the $L\rar\i$ limit. This amounts to defining a new $\Zs'(\t) = \Zs(\t) - \<\Zs\>$, which leaves $Z_p(\t)$ invariant if we shift $\Zl(\t)$ oppositely.} This choice is motivated by the fact that a constant's contribution to a microcanonical density of states is self-averaging under coarse-graining in energy -- a concept that will show up later. Moreover, we note that in many cases of interest, such as the Narain case to be reviewed in Subsection \ref{sec:hints}, $\<\Zs\>=0$. 

Note that in \eqr{ZLdef} we have defined ``light'' operators with respect to the twist threshold, min$(h,\hb) \leq \xi$, rather than the dimension threshold, $\D \leq 2\xi$. The latter is more commonplace, and was used in the original definition of \cite{Benjamin:2021ygh}; moreover, from the mathematical point of view, one need only subtract the operators with $\D \leq 2\xi$ from $Z_p(\t)$ in order for $\Zs(\t)$ to be square-integrable. On the other hand, operators with min$(h,\hb) \leq \xi$ -- dubbed ``censored'' operators in \cite{Keller:2014xba} -- are not part of the black hole spectrum of AdS$_3$ pure gravity, and more generally, the chaotic spectrum of a 2d CFT (as it is currently understood). Though mathematically unnecessary, it is physically well-motivated to define $\widehat Z_L(\t)$ and hence $\Zs(\t)$ with respect to the twist threshold in the manner above, and we will do so in view of the applications of present interest.\foot{Due to cosmic censorship, states with $h<\xi$ but $\hb>\xi$ are dual to neither black holes nor conical defects in gravity. The most familiar such states are multi-twist composites of light primaries with min$(h,\hb)<{3\o4} \xi$: these composites live on Regge trajectories which asymptote, at large spin, to the Virasoro Mean Field Theory spectrum \cite{Collier:2018exn}. This demonstrates that a 2d CFT may have an infinite number of operators with min$(h,\hb) \leq \xi$, whereas the number of operators with $\D \leq 2\xi$ is strictly finite (at finite $\xi$). The sum \eqr{ZLdef} may thus require regularization. The same is in fact true for the Poincar\'e modular completion of even a single state \cite{Keller:2014xba,Maloney:2007ud}.} See \cite{Haehl:2023tkr} for similar comments.

\ssec{Hints}\label{sec:hints}
 To  better understand the physical significance of the two pieces $\Zl(\tau)$ and $\Zs(\t)$,  the authors of \cite{Benjamin:2021ygh} studied the partition function of Narain CFT in this formalism.  Let us review a few of the lessons gained. A Narain CFT of $c$ free bosons has a local $U(1)^c\times U(1)^c$ symmetry, and moduli which we collectively denote as $m$.  The $U(1)^c\times U(1)^c$ primary partition function $Z_p^{(c)}(\t|m)$ can be decomposed as
 \e{}{ Z_p^{(c)}(\t|m) = \la Z_p^{(c)}(\t|m)\ra_m+Z_\spec^{(c)}(\t|m)}
The first term is the partition function averaged over moduli, and the second term is the spectral partition function. This expression displays two interesting features:

\begin{itemize}
\item The modular completion of light states is equal to the average over Narain moduli space: 
\e{}{\Zl^{(c)}(\tau)= \la Z_p^{(c)}(\t|m)\ra_m = E_{c\o 2}(\t).}
The only primary in Narain CFT that is light with respect to the $U(1)^c$ primary threshold, with $\text{min}(h,\hb)\leq (c-c_{\rm currents})/24=0$, is the vacuum itself: the Eisenstein series is simply the Poincar\'e modular completion of the vacuum state\cite{Afkhami-Jeddi:2020ezh,Maloney:2020nni}.

\item The spectral decomposition averages to zero  under both modular and ensemble averages: 
\es{}{ \la Z_\spec^{(c)}(\t|m) \ra_{\tau}=\la
Z_\spec^{(c)}(\t|m)\ra_{m}=0.}
Thus $Z_\spec^{(c)}(\t|m)$ captures deviations from the average over moduli space. These deviations encode the higher statistical moments of the Narain ensemble, e.g. the variance
\es{narain2m}{ 
\la Z_p^{(c)}(\t_1|m) Z_p^{(c)}(\t_2|m)  \ra^{\rm (conn)}_m= \la Z_\spec^{(c)}(\t_1|m)Z_\spec^{(c)}(\t_2|m)\ra_m.
}
A study of these correlations was undertaken in \cite{Collier:2021rsn,Cotler:2020hgz}.
\end{itemize}

Extrapolating these properties to arbitrary 2d CFTs suggests \cite{Benjamin:2021ygh} that $\Zl(\t)$ is an average partition function of some kind, capturing universal contributions of light states to the heavy spectrum implied by crossing. The same picture is suggested by the MWK partition function, the sum over smooth semiclassical saddles of AdS$_3$ gravity and fluctuations around them, which is a (regularized) modular completion of the vacuum and its $SL(2,\R)$ null descendants: in particular, this sum generates the smooth, Cardy asymptotics (and non-perturbative corrections thereto) present in any large $c$ CFT density of states. Somewhat similar ideas were subsequently suggested in \cite{Schlenker:2022dyo}, motivated by chaos. In such a picture, these universal contributions would encode leading-order coarse-grained data of the original CFT. The remainder, $\Zs(\t)$, would capture fine-grained fluctuations. For example, one might average the product $\Zs(\t_1)\Zs(\t_2)$, \`a la \eqr{narain2m}, to extract statistical correlations among microstates. 

The challenge in making this precise is to extend it to the generic Virasoro setting, sans moduli. In the next section we will realize a version of this idea for generic 2d CFTs. This will show that $\Zs(\t)$ can indeed be viewed as the chaotic part of the partition function in a manner consistent with the symmetries. 
         
\section{Coarse-graining and Trace Formulas}\label{sec:s3}
We start by briefly introducing the Gutzwiller trace formula \cite{Gutzwiller} for chaotic quantum systems which provides useful physical intuition about the density of states  of individual non-disordered chaotic systems. We show that the $\sl$ spectral decomposition of the density of states is directly analogous to a trace formula. This leads us, via a coarse-graining procedure, to an analog of Berry's diagonal approximation for 2d CFT \cite{Berry}.  

\subsection{Trace formulas for chaotic systems}
A well-understood example of quantum chaos is the semiclassical dynamics of few-body quantum systems which are classically chaotic; for a review of trace formulas, see \cite{bogomolny,PerOrb2,Richter:2022sik,haake}.\foot{Since $\hbar=1$, the semiclassical limit corresponds to $E\gg1$.} 
The Gutzwiller trace formula expresses the density of states  in the semiclassical limit as
\e{rhoqs}{\rho(E)= \overline{\rho}(E) + \rho_{\rm osc}(E)}
$\overline{\rho}(E)$ is a mean density. $\rho_{\rm osc}(E)$, the oscillatory part, is given by a sum over semiclassical  periodic orbits $\gamma$, 
\e{rhoosc}{\rho_{\rm osc}(E)=\frac{1}{\pi}\Re\sum_{\gamma} \cA_{\gamma}e^{i S_{\gamma}(E)}}
$S_{\gamma}(E)$ is the orbit action and $\cA_{\gamma}$ is called the stability amplitude. 
The statistical correlations among energy levels are encoded in the highly oscillatory behavior of  $\rho_{\rm osc}(E)$, and can be extracted by microcanonical coarse-graining over an energy window $\delta E$. 

One notable application of the trace formula is the computation of the coarse-grained microcanonical spectral form factor, 
\e{}{K_{E}(T)= \int_{-\infty}^{\infty} d\eps \,e^{i\eps T} \overline{\rho_{\rm osc }\qty(E+\frac{\eps}{2})\rho_{\rm osc }\qty(E-\frac{\eps}{2})}}
where $T$ is real Lorentzian time. Inserting the trace formula,
\es{}{ 
K_{E}(T)=\frac{1}{4\pi^2}\int_{-\infty}^{\infty}  d\eps \,e^{i\eps T} \sum_{\gamma_1,\gamma_2}  \cA_{\gamma_1}\cA^*_{\gamma_2}\overline{e^{i (S_{\gamma_1}(E+\frac{\eps}{2})-S_{\gamma_2}(E-\frac{\eps}{2}))}} + \text{c.c.}
}
In the theory of chaotic quantum systems, the difference between periodic orbit actions $\Delta S=S_{\gamma_1}-S_{\gamma_2}$ is the central quantity which allows one to systematically organize the sum over orbits.  Periodic orbits with $\Delta S=0$ give the leading diagonal contribution to level statistics.\footnote{In a chaotic system we expect that $S_{\gamma}=S_{\gamma'}$ only if the orbits are identical $\gamma=\gamma'$ or related by a symmetry, such as time-reversal in the GOE class.} Berry \cite{Berry,hannay1984periodic} showed that the diagonal approximation, which restricts the sum to $\gamma_1=\gamma_2$, reproduces the linear ramp as $T\rar\i$: 
\es{}{
K_{E}(T)\Big|_{\rm diag}&=\frac{1}{4\pi^2}\int_{-\infty}^{\infty}  d\eps \,e^{i\eps T}\sum_{\gamma}  |\cA_{\gamma}|^2 \overline{e^{i (S_{\gamma}(E+\frac{\eps}{2})-S_{\gamma}(E-\frac{\eps}{2}))}} + \text{c.c.} \\&\approx  \frac{T}{2\pi}\mathsf{C}_\RMT \,.
}
The factor $\mathsf{C}_\RMT$ controls the universality class of random matrix behavior (e.g. $\mathsf{C}_{\rm GUE}=1$ and $\mathsf{C}_{\rm GOE}=2$).

Berry's analysis extracts the random matrix behavior of an individual chaotic quantum system, with no ensemble averaging, from the oscillatory behavior of the density of states. While the original double sum over orbits is manifestly factorized, the diagonal approximation exhibits an emergent non-factorization, by discarding microscopic details of the spectrum.\foot{It is possible to go beyond the diagonal approximation by systematically including subleading contributions from pairs of orbits with $\Delta S\ll 1$. This is the role of encounter theory\cite{Sieber2001,Heusler_2004,PerOrb1,PerOrb2}.}

\ssec{A trace formula for 2d CFT}

Is there a decomposition of a 2d CFT density of states of the form \eqr{rhoqs}, with the structure \eqr{rhoosc}? As it happens, the $\sl$ spectral decomposition does the trick. 

\sssec{Density of states}

The density of primary spin-$j$ states $\rho_j(\D)$ in a parity-invariant CFT is defined as
\es{}{ 
Z_p(\tau)=\sqrt{y}\sum_{j= 0}^\i (2-\delta_{j,0})\cos(2\pi j x)\int_{j}^\i d\Delta \,e^{-2\pi y (\Delta-2\xi)}\rho_{j}(\Delta)\,.
}
The lower bound is set by unitarity. It will be convenient to introduce the ``reduced twist'' $t$, defined as \cite{Benjamin:2020mfz}
\e{}{t := {\D-j\o 2}-\xi=\min(h,\hb)-\xi\,.}
As a first step toward a trace formula, let us transform the spectral decomposition \eqr{eq:specdec} into the microcanonical ensemble, decomposing the density as 
\e{rhodecomp}{\rho_j(t)= \widehat{\rho}_{L,j}(t)+\rho_{\spec,\,j}(t)\,.}
By inverse Laplace transform, we obtain a manifestly modular-invariant decomposition
\es{eq:rhosp}{ 
\Rs{}_{,j}(t)&= \int_{\cC_{\rm crit}} \{\Zs,E_\crit\} \rho_{\crit,j}^*(t) +\sum_{n=1}^\i (\Zs,\phi_n)\rho_{n,j}(t)
}
The basis elements are given for $j\neq 0$ by
\es{eq:rho_s}{ 
\rho_{\crit,j}^*(t) &=
\sfa_j^{(s) }\theta(t)\frac{\cos(\omega \cosh^{-1}\qty(\frac{2t}{j}+1))}{\sqrt{t(t+j)}} \\
\rho_{n,j}(t) &=
\sfb_j^{(n)}\theta(t)\frac{\cos(\omega_n \cosh^{-1}\qty(\frac{2t}{j}+1))}{\sqrt{t(t+j)}} 
}
and for $j=0$ by
\e{eq:rho_0}{\rho_{\crit,0}^*(t)=\frac{\zeta(2i\omega)}{t}(4t)^{i\omega}+(\omega\rightarrow -\omega)}
Note that $\rho_{\crit,0}^*(t)$ may be obtained as the $j\rar 0$ limit of \eqr{eq:rho_s} (see Appendix \ref{app:jzero}).\foot{In a more symmetric notation, if we define $\bar t = \text{max}(h,\hb)-\xi$ then 
\e{eq:rho_s_foot}{ \rho_{\crit,j}^*(t) =\sfa_j^{(s) }\theta(t)\frac{\cos(\omega \cosh^{-1}\qty(\frac{t+\bar t}{j}))}{\sqrt{t \bar t}}. }
and likewise for $\rho_{n,j}(t)$.}

Let us make a few comments about the physical features of these formulas. While $\rho_j(t)$ is a sum of delta functions in a compact CFT, both $\widehat{\rho}_{L,j}(t)$ and $\rho_{\spec,\,j}(t)$ are continuous functions of $t$. The latter is highly oscillatory, with the spectral parameters $\{\w,\w_n\}$ appearing as frequencies in the basis elements \eqr{eq:rho_s} (hence the notation). We can understand this oscillatory behavior as follows. In the asymptotic spectrum (or for $t \gtrsim \xi$ in sparse large $c$ CFTs \cite{Hartman:2014oaa}), the mean level spacing of $\rho_j(t)$ is approximately $e^{-S_{{\rm Cardy},j}(t)}$, the inverse of the spin-$j$ Cardy degeneracy, where
\e{}{S_{{\rm Cardy},\,j}(t) = 4\pi \sqrt{\xi t} + 4\pi\sqrt{\xi(t+j)}}
Because $\widehat{\rho}_{L,j}(t)$ is continuous and has exponential Cardy growth at $t\rar\i$, $\rho_{\spec,\,j}(t)$ must oscillate over extremely small wavelengths, of the order of the mean level spacing, to encode the microscopic information about the discrete spectrum.\foot{A nonzero constant term $\la \Zs \ra$ in the spectral decomposition would contribute microcanonically as $\delta_{0,j} \la \Zs\ra \sqrt{2/t}$. This is not oscillatory, so we associate it with $\widehat{\rho}_{L,0}(t)$, cf. Subsection \ref{ssec:2.1}.} See Figure \ref{densityfig} (note that the exact spectrum $\rho_j(t)$ is drawn there as a smooth curve, both for illustrative purposes and to evoke the large $c$ limit). 

\sssec{Trace formula}

As explained above, $\widehat{\rho}_{L,j}(t)$ has smooth exponential growth in $t$, whereas $\Rsjt$ is a sum/integral over infinitely many oscillatory terms. This dichotomy suggests a direct analogy with trace formulas, with an identification along the following lines:
\es{}{ 
\overline{\rho}(E)\qquad  &\longleftrightarrow \qquad\widehat{\rho}_{L,j}(t)\\
\rho_{\rm osc}(E)\qquad &\longleftrightarrow\qquad \Rsjt
} 
This is to be understood as holding at every fixed spin $j$. 

In fact, this can be made quite precise. Let us return to the Gutzwiller trace formula in \eqr{rhoosc}. The stability amplitude is canonically decomposed into two pieces: 
\e{}{\cA_{\gamma}=A_{\gamma} T_{\gamma}(E)}
 $T_{\gamma}(E)$ is the period of the orbit,
\e{}{T_{\gamma}(E)= \frac{\p S_{\gamma}(E)}{\p E}\,,}
while $A_{\gamma}$ is a one-loop determinant. We notice that the basis densities \eqr{eq:rho_s} can be written precisely as contributions of individual periodic orbits. Using $\rho_{\crit,j}(t)$ to denote either $\rho_{\crit,j}^*(t)$ or $\rho_{n,j}(t)$, we have
\es{rhoorbit}{
\rho_{\crit,j}(t)= \frac{1}{\pi} \Re\qty(A_{\omega,j} T_{\omega,j}(t)e^{iS_{\omega,j}(t)})\,,
}
with the following identifications:
\es{traceid}{S_{\omega,j}(t)&=\omega \cosh^{-1}\qty(\frac{2t}{j}+1)+\omega\log j\\
T_{\omega,j}(t)&= \frac{\omega}{\sqrt{t(t+j)}}=\frac{\p S_{\omega,j}}{\p t}\\
A_{\omega,j}&= \frac{\pi}{\omega} \sfc_j^{(s)} j^{-i\omega}\,,
}
where $\sfc_j^{(s)}$ stands for either the Eisenstein ($\sfa_j^{(s)}$) or cusp form ($\sfb_j^{(n)}$) Fourier coefficient. The above identifications of the different terms in \eqr{rhoorbit} follow simply from demanding that $S_{\w,j}(t)\in\R$ for $\w\in\R$, and that the $j\rar 0$ limit is smooth for each individual piece. In particular, this instructs us to include $\omega \log j$ in the action: doing so generates $j=0$ orbit data that correctly reproduces \eqr{eq:rho_0}, namely,
\es{traceid2}{S_{\omega,0}(t)&=\omega\log(4t)\\
T_{\omega,0}(t)&= \frac{\omega}{t}\\
A^*_{\omega,0}&= \frac{\pi}{\omega} \z(2i\w)\,.
}
Denoting $A^*_{\w,j}$ and $A_{n,j}$ as the specialization of $A_{\w,j}$ to Eisensteins and cusp forms, respectively, we land on the rewriting of the $\sl$ spectral decomposition as a trace formula for 2d CFTs:
\es{eq:2dtrace}{\boxed{
\Rsjt=\frac{1}{\pi} \,\Re\qty( \int_{\cC_{\rm crit}} \{Z_{\spec},E_{\half+i\omega}\} A^*_{\omega,j} T_{\omega,j}e^{iS_{\omega,j}(t)} +\sum_{n=1}^{\infty} (Z_{\spec},\phi_n) A_{n,j} T_{\omega_n,j}e^{iS_{\omega_n,j}(t)})}}
This holds for any fixed $j$. 

A key fact about this formula, to be used later, is that the set of orbits forms a complete and explicit ($\sl$-invariant) eigenbasis: given a determinant factor $A_{\w,j}$, one can uniquely reconstruct the corresponding orbit action. 

A small but notable point is that $A_{\omega,j}$ is complex by a pure phase. This is reminiscent of something in periodic orbit theory: there exists a phase, the so-called ``Maslov phase'' $e^{i{\pi\o2}\mu_\g}$, that one can choose to absorb into $A_\g$ rather than $S_\g(E)$ such that $S_\g(E)\in\R$.\foot{The Maslov index for periodic orbits is a generalized topological invariant which counts windings around a certain submanifold of phase space \cite{haake}.} This suggests that we identify a CFT analog of the Maslov index, call it $\mu_{\omega,j}$, defined so that $ A_{\omega,j}e^{i\frac{\pi}{2}\mu_{\w,j}}\in \RR$:
\e{maslov}{\frac{\pi}{2}\mu_{\omega,j}:=\omega \log j\,.
}
While \eqr{maslov} is so far intended to be a purely functional observation, it raises the obvious question of the physical meaning of this CFT Maslov index. In any case, we see that the identification \eqr{traceid} admits a non-trivial compatibility with the analytic structure of periodic orbits of quantum systems, revealed by imposing smoothness of the $j \rar 0$ limit. 

Before making use of this 2d CFT trace formula, we pause to comment on the relation to trace formulas more broadly. The Gutzwiller trace formula can be applied only for systems which admit a semiclassical limit. On the other hand, the $\sl$ spectral decomposition is valid for any 2d CFT. Moreover, the Gutzwiller trace formula, when applicable, describes only the high energy ($E\gg 1$) part of the spectrum, while the spectral decomposition is valid for all energies above threshold, thanks to the inherent distinction between light and heavy states due to modular invariance. In these respects, the spectral decomposition is more similar to the Selberg trace formula, an exact relation between the spectrum of the automorphic Laplacian on hyperbolic quotient manifolds and their classical periodic orbits. Nevertheless, we frame the analogy with respect to Gutzwiller, which is more generally applicable, in order to emphasize the universal structure of the $\sl$ spectral decomposition. 

\subsection{Berry's diagonal approximation for 2d CFT }\label{diagsec}
Inspired by the treatment of coarse-grained spectral correlations in chaotic systems, we now proceed analogously, in order to motivate a precise and mathematically well-defined version of Berry's diagonal approximation for 2d CFT. 

A standard assumption in the theory of periodic orbits is that coarse-graining over an energy window $\delta E$ smooths out the oscillatory part of the density of states,
\e{orbitav}{\overline{\rho_{\rm osc}(E)}=0\,.}
The energy window is taken to be a mesoscopic scale, much larger than the mean level spacing $\bar{\rho}(E)^{-1}$ but much smaller than the scale over which the mean density varies noticeably.\footnote{See for example Chapter 10.6 of \cite{haake} and Section 2 of \cite{bogomolny}.} This assumption of phase incoherence means that any nonzero phase factor appearing in a density correlator is taken to approximately vanish upon coarse-graining. 

Naturally, we employ the same approach to the CFT density $\rho_j(t)$. By analogy to \eqr{orbitav}, the density of states $\rho_{\spec,\,j}(t)$ is taken to average to zero upon microcanonical coarse-graining over a mesoscopic window $\delta t$,
\e{}{\overline{\rho_{\spec,\,j}(t)}=0\,.}
We expect that this holds for $\delta t$ larger than the mean level spacing, yet smaller than the characteristic variation of the mean density, both controlled by $\widehat{\rho}_{L,j}(t)$. One can, for instance, perform microcanonical coarse-graining by convolution against a window function $W(t-t')$ with characteristic width $\delta t$:\foot{Coarse-graining the individual basis elements of the $\sl$ decomposition gives
\e{}{ 
\overline{\rho_{\half +i\omega,j}(t)}= \frac{1}{\pi \delta t}\Re \[A_{\omega,j}\int_{t-\delta t}^{t+\delta t} T_{\omega,j}(t') e^{iS_{\omega,j}(t')} dt'\]= \frac{1}{\pi \delta t}\Re \[A_{\omega,j}\int_{S_{\omega,j}(t-\delta t)}^{S_{\omega,j}(t+\delta t)}  e^{iS_{\omega,j}(t)}dS_{\omega,j} \]\,.
}
This is a simple oscillatory integral which vanishes for a suitably chosen scale $\delta t$, derivable from the action \eqr{traceid}. This scale depends on the frequency $\w$ and spin $j$. The assumption from periodic orbit theory is that there exists a finite scale $\delta t$ with respect to which the average of the entire sum vanishes.}
\e{}{ 
\overline{f(t)}:= \int_0^\infty W(t-t')f(t')dt', \quad  \text{with}\quad \int_{0}^{\infty} W(t)dt =1.
} 
The density of light states $\widehat{\rho}_{L,j}(t)$ is instead approximately self-averaging, as its behavior in its argument  is exponential rather than oscillatory. Hence we have that 
\e{}{\overline{\rho_j(t)}\approx\widehat{\rho}_{L,j}(t)\,.}

We now turn to two-point correlations. Again following a standard approach in quantum chaos, we define the microcanonical coarse-graining for the product of two densities by integrating over the mean twist $t:=\frac{t_1+t_2}{2}$ while keeping the difference $\eps:={t_1-t_2\over 2}$ fixed:
\e{}{ 
\overline{f(t_1)f(t_2)}:= \int_0^\i dt' f(t'+\eps)f(t'-\eps)W(t-t')
}
Coarse-graining the product of densities $\overline{\rho_{\spec,\,j_1}(t+\eps)\rho_{\spec,\,j_2}(t-\eps)}$ produces a sum/integral of terms of the following form:
\es{}{
\overline{\rho_{\half+i \omega_1,j_1}(t+\eps)\rho_{\half+i \omega_2,j_2}(t-\eps)}=\frac{A_{\omega_1,j_1} \overline{A}_{\omega_2,j_2}}{4\pi^2} T_{\omega_1,j_1}T_{\omega_2,j_2} \overline{e^{i(S_{\omega_1,j_1}(t+\eps)-S_{\omega_2,j_2}(t-\eps))}} + \text{c.c.}
}
By way of the earlier assumption, terms involving the sum of actions are smoothed out by the averaging,
\e{eq329}{ \overline{e^{i(S_{\omega_1,j_1}(t+\eps)+S_{\omega_2,j_2}(t-\eps))}} =0\,.}
Instead, terms involving the difference in actions $\Delta S= S_{\w_1,j_1}(t)-S_{\w_2,j_2}(t)$ can give large contributions if the actions cancel. Concentrating on the leading terms, with $\Delta S=0$, is the analog of Berry's diagonal approximation. 

To see what this approximation means in the case of 2d CFT, we choose to consider the Fourier mode of equal spins, $j_1=j_2=j$. The above microcanonical coarse-graining then selects the terms with equal actions $S_{\w_1,j}(t)=S_{\w_2,j}(t)$, which sets $\omega_1=\omega_2$:
\es{}{ 
\overline{\rho_{\half+i\w_1,j}(t+\epsilon) \rho_{\half+i \w_2,j}(t-\epsilon)}&\propto  \overline{e^{i(S_{\omega_1,j_1}(t+\eps)-S_{\omega_2,j_2}(t-\eps))}} + \text{c.c.} \\
&\propto  \delta_{\omega_1,\,\omega_2}
}
where we used \eqr{eq329} and took $\eps \rightarrow 0$ to focus on nearby energy levels. We call the resulting diagonal density $\rho_\diag(t_1,t_2)$, the projection of the product of $\sl$ spectral decompositions onto terms with equal eigenvalues. Subsequently, all spin sectors $(j_1,j_2)$ are fixed by $\sl$-invariance of the full density, in terms of the respective basis densities of the Eisensteins and cusp forms:
\es{rhodiag}{
\rho_\diag^{(j_1,j_2)}(t_1,t_2)&=\int_{\ccrit} \{Z_{\spec},E_{\half+i\omega}\}^2 \rho_{\half+i\omega,j_1}^*(t_1) \rho_{\half+i\omega,j_2}^*(t_2)+\sum_{n=1}^\i (Z_{\spec},\phi_n)^2 \rho_{n,j_1}(t_1)\rho_{n,j_2}(t_2)\\ &+ \sum_{n=1}^\i \{Z_{\spec},E_{\half+i\omega_n}\}(Z_{\spec},\phi_n) \Big(\rho^*_{\half+i\omega_n,j_1}(t_1)\rho_{n,j_2}(t_2)+\rho_{n,j_1}(t_1)\rho^*_{\half+i\omega_n,j_2}(t_2)\Big)
}
We point out that the density is not diagonal in spin: $\rho_\diag^{(j_1,j_2)}(t) \neq 0$ for $|j_1| \neq |j_2|$. This is a manifestation of the general fact that modular invariance correlates different spin sectors of 2d CFT data. In Appendix \ref{app:pindelta} we give a quick proof that diagonality in spin is incompatible with modular invariance. In contrast, one expects different spin sectors to be statistically independent in generic chaotic systems. 

Note that there are cross-terms on the second line of \eqr{rhodiag} because of the spectral degeneracy between cusp forms and Eisenstein series at $\w=\w_n$. These cross-terms pair {\it distinct} orbits, with the same action $S_{\w,j}(t)$ but different one-loop determinants $A_{\w,j}$. This suggests that one should really seek a diagonal projection which pairs eigenfunctions rather than eigenvalues. 

This concludes our motivation from microcanonical coarse-graining and the diagonal approximation in the periodic orbit approach to quantum systems. We now proceed to rigorously define the 2d CFT diagonal approximation in the canonical ensemble, by constructing diagonal products of partition functions. We first define what we call $\Zd(\t_1,\t_2)$, the canonical counterpart of $\rho_{\rm diag}(t_1,t_2)$. We then pass to an object that we call $Z_\Hecke(\t_1,\t_2)$, a diagonal partition function that properly pairs identical orbits.

\sssec{Diagonal projection I}
The density $\rho_{\rm diag}(t_1,t_2)$ corresponds, in the canonical ensemble, to the following {\it diagonal partition function}:
\es{}{ Z_{\diag}(\tau_1,\tau_2)&= \int_{\cC_{\rm crit}}  \{Z_{\spec},E_{s}\}^2 E^*_s(\tau_1)E^*_s(\tau_2)+\sum_{n=1}^\i (Z_{\spec},\phi_n)^2 \phi_n(\tau_1)\phi_n(\tau_2)\\&+ \sum_{n=1}^\i \{Z_{\spec},E_{s_n}\}(Z_{\spec},\phi_n) \big(\phi_n(\tau_1)E^*_{s_n}(\tau_2)+\phi_n(\tau_2)E^*_{s_n}(\tau_1)\big),
}
The Laplace eigenvalues of the $\sl$ eigenbasis elements are paired. This is the \textit{diagonal projection}, $Z_{\rm diag}(\tau_1,\tau_2)$, of the factorized product $\Zs(\t_1)\Zs(\t_2)$. To formalize this, consider the operator 
\e{}{\D_{12} := \D_{\t_1} - \D_{\t_2}}
acting on functions $f(\t_1,\t_2)\in L^2(\cF\times \cF)$, the space of square-integrable functions of two moduli $\tau_1,\tau_2$ valued in $\cF$. Such functions admit a double spectral decomposition in a joint basis of eigenfunctions.\foot{A few remarks on the $L^2(\cF\times \cF) $ spectral decomposition, and some technical remarks on scheme-dependence of the projection $\cP_\diag$, are given in Appendix \ref{app:l2space}.} Since $f(\t_1,\t_2)$ has paired eigenvalues if and only if $\D_{12}f(\t_1,\t_2)=0$ for all $\t_1,\t_2$, we can define $Z_{\rm diag}(\t_1,\t_2)$ as the diagonal projection of $\Zs(\t_1)\Zs(\t_2)$ onto $\text{ker}(\D_{12})$, the kernel of $\D_{12}$:
\es{diagproj}{\boxed{Z_{\rm diag}(\t_1,\t_2) :=\cP_{\diag} \[Z_{\spec}(\t_1)Z_{\spec}(\t_2)\]\,, \quad \text{where} \quad \cP_{\diag}:= \cP_{\text{ker}(\D_{12})}}}
$Z_{\rm diag}(\t_1,\t_2)$ is a manifestly modular-invariant analog of Berry's diagonal approximation for 2d CFT. $Z_{\rm diag}(\t_1,\t_2)$ does not factorize, unlike $Z_{\spec}(\t_1)Z_{\spec}(\t_2)$. Factorization can be explicitly restored by including the off-diagonal terms. In Section \ref{sec:BHmicro} we will develop an analogy to the approach to factorization in 2D gravity and random matrix theory.

\sssec{Diagonal projection II: Hecke projection}
In order to eliminate the cross terms in $\Zd(\t_1,\t_2)$ which do not pair identical orbits, we introduce a slightly souped-up diagonal partition function which we call $Z_\Hecke(\t_1,\t_2)$. Stated succinctly, whereas $\Zd(\t_1,\t_2)$ pairs {\it eigenvalues}, $\Zh(\t_1,\t_2)$ pairs {\it eigenfunctions}. 

The new projection may be defined using $SL(2,\Z)$ Hecke operators. Let us recall their definition. Hecke operators $T_j$ exist for every spin $j\in\Z_+$. Their action on  $SL(2,\Z)$-invariant functions $f(\t)$ is\foot{Hecke operators for non-prime spins $j$ are fixed in terms of the prime spins via the Hecke multiplication rule. For a recent review of Hecke operators in a physics context see \cite{DHoker:2022dxx}.}
\e{heckedef}{T_j f(\t) =  {1\o \sqrt{j}} \sum_{ad = j,\, d>0}\sum_{b=0}^{d-1} f\({a\t+b\o d}\)}
In this normalization, the Hecke action on the $SL(2,\Z)$ eigenbasis elements is 
\es{Heckeeigen}{T_j E^*_s(\t) &= \frac{\mathsf{a}^{(s)}_{j}}{2} E^*_s(\t)\\
T_j \phi_n(\t) &= \tilde \sfb^{(n)}_{j} \phi_n(\t)}
where $\tilde \sfb_j^{(n)} :=  \sfb_j^{(n)}/ \sfb_1^{(n)}$. An equivalent definition, which will be useful momentarily, is  
\e{}{T_j f(\t) = {1\o \sqrt{j}}\sum_{A\in SL(2,\Z)\backslash M_j} f(A\t)}
where $M_j$ denotes the following set of matrices,
\es{}{ 
M_j=\bigg\{ \begin{pmatrix}
a & b \\
c & d 
\end{pmatrix} \bigg| ~ a,b,c,d \in \Z; ~  ad-bc=j\bigg\}
}
and $A\t$ denotes the fractional linear transformation, 
\e{}{A\t = {a\t+b \o c\t+d}\,,\quad A\bar \t = {a\bar\t+b \o c\bar \t+d}\,.}
The quotient $SL(2,\Z)\backslash M_j$ is taken by identifying elements under the equivalence relation $A\sim B $ if there exists a $\gamma\in SL(2,\Z)$ such that  $B=\gamma A$. We emphasize that $j\in\Z_+$.

We now introduce $\Zh(\t_1,\t_2)$, the projection of $\Zs(\t_1)\Zs(\t_2)$ onto the kernel of the {\it difference} of Hecke operators at every spin. Defining
\es{}{ T_j^{(12)} := T_j^{(\t_1)} - T_j^{(\t_2)},
}
we define the {\it Hecke projection} onto $\text{ker}\big(T_j^{(12)}\big)$,
\e{heckeproj}{\boxed{\Zh(\t_1,\t_2) :=\cP_\Hecke \big[\Zs(\t_1)\Zs(\t_2)\big]\,,\quad \text{where}\quad \cP_{\text{Hecke}}:=\cP_{\text{ker}\big(T_j^{(12)}\big)}~\forall~j\in\Z_+}}
The essential feature of $\Zh(\t_1,\t_2)$ is seen upon considering the action of $T_j^{(12)}$ on ``mixed'' Eisenstein-cusp form terms in $\Zd(\t_1,\t_2)$:
\e{}{T_j^{(12)}\(E^*_{s_n}(\t_1) \phi_n(\t_2)\) = \Bigg( \frac{\mathsf{a}^{(s)}_{j}}{2}  - \tilde \sfb_j^{(n)}\Bigg)E^*_{s_n}(\t_1) \phi_n(\t_2)}
This is nonzero on account of the inequality of Eisenstein and cusp form Fourier coefficients.  While such terms are present in $\Zd(\t_1,\t_2)$, they are eliminated by the Hecke projection \eqr{heckeproj} by design. Note that the object $\Zh(\t_1,\t_2)$ is defined via projection for {\it all} spins $j$.\foot{We are being rather conservative in defining Hecke projection for all spins $j$, in order to guard against the possibility that $ {\mathsf{a}^{(s)}_{j}}  = 2\tilde \sfb_j^{(n)}$ for very special choices of $n$ and $j$ -- though even this seems extremely unlikely (and perhaps is provably false). On the other hand, it is highly inconceivable that the Fourier coefficients could be equal for {\it all} $j$, given the numerical and statistical properties of cusp forms -- for example, the available numerical data at finite $n$ \cite{lmfdb}, and the Sato-Tate conjectures on the statistical distribution of $\tilde \sfb_j^{(n)}$ at $j\rar\i$ or $n\rar\i$.} The spectral decomposition of $\Zh(\t_1,\t_2)$ is thus
\es{ZHecke}{\boxed{\Zh(\t_1,\t_2) =  \int_{\mathcal{C}_{\rm crit}} |(\Zs,E_s)|^2 E_s(\t_1)E_{1-s}(\t_2) + \sn (\Zs,\phi_n)^2 \phi_n(\t_1)\phi_n(\t_2)}}
We may rephrase the special property of $\Zh(\t_1,\t_2)$ as a statement of enhanced symmetry: employing the terminology of \cite{Benjamin:2023nts}, we say that a modular-invariant function $f(\t_1,\t_2)$ enjoys \textit{Hecke symmetry} if 
\e{}{ \qquad \qquad T_j^{(12)}f(\t_1,\t_2)=0~~\forall~~j\in\Z_+\qquad (\textit{Hecke symmetry})}
In this sense, $\Zh(\t_1,\t_2)$ carries an enhanced symmetry with respect to both $\Zd(\t_1,\t_2)$ and $\Zs(\t_1)\Zs(\t_2)$: it is annihilated by an infinite set of commuting operators, one for each positive integer. 

Let us recapitulate the idea behind $\Zh(\t_1,\t_2)$. It is a fact of life on the $\sl$ fundamental domain that there is spectral degeneracy for the infinite discretuum of Laplace eigenvalues $\w=\w_n$ where Maass cusp forms exist. From the point of view of the trace formula and analogy to periodic orbit theory, the Eisensteins and Maass cusp forms with equal eigenvalues appear as different orbits: their actions are equal, but their one-loop determinants differ. The Hecke projection $\cP_\Hecke$ is designed to account for this innate quirk of 2d CFT. We are proposing Hecke projection as an enhanced form of coarse-graining. It would be nice to understand this more fundamentally, say, from a microcanonical perspective.

For these reasons, we view $\Zh(\t_1,\t_2)$ as the ``right'' two-copy partition function which diagonalizes the chaotic spectral correlations. This view will find further support in Section \ref{s4}, as the  Hecke projection really comes to life when we consider its dual gravitational interpretation. In particular, we will show that Hecke symmetry is an emergent feature of Euclidean torus wormholes in semiclassical AdS$_3$ gravity.  

\sec{Random Matrix Universality in Chaotic 2d CFTs}\label{sec:zdiag}

Having identified a 2d CFT trace formula using the $SL(2,\Z)$ spectral decomposition of partition functions, we presented a framework for studying coarse-grained CFT correlations with modular invariance and Virasoro symmetry baked in, focusing on the diagonal approximation in particular. We now analyze general properties of the diagonal partition function and the implications of chaos for the CFT. This will lead to a necessary and sufficient condition for the presence of a linear ramp in the spectral form factor (SFF).

The partition functions $\Zd(\t_1,\t_2)$ and $\Zh(\t_1,\t_2)$ admit double Fourier decompositions, e.g.
\e{}{\Zd(\t_1,\t_2) = \sum_{j_1=0}^\i (2-\delta_{j_1,0})\cos(2\pi j_1 x_1)\sum_{j_2=0}^\i (2-\delta_{j_2,0})\cos(2\pi j_2 x_2) Z_{\diag}^{(j_1,j_2)}(y_1,y_2)}
and likewise for $\Zh(\t_1,\t_2)$. In this section we specialize to the scalar sector, $(j_1,j_2)=(0,0)$; note that $\Zh^{(0,0)}(y_1,y_2) = \Zd^{(0,0)}(y_1,y_2)$ because the cusp forms have vanishing scalar component.\foot{We use the ``diag'' subscript in this section to emphasize the physical setting.} Besides reasons of clarity and simplicity, we do so because we will later consider a special class of amplitudes -- namely, wormhole amplitudes -- which are {\it fully determined} by the scalar sector. 

In the spectral basis,
\es{zdiag}{\Zd^{(0,0)}(y_1,y_2) =  {1\o 4\pi i}\ds \{\Zs,E_s\}\{\Zs,E_{1-s}\} E^*_{1-s,0}(y_1)E^*_{s,0}(y_2) }
where we have written the integral over the critical line explicitly. Inserting the zero mode
\e{vphidef}{E^*_{s,0}(y) = \L(s) \(y^s + \varphi(s) y^{1-s}\)\,,\quad \varphi(s) := {\L(1-s)\o\L(s)}}
%
and using symmetry under reflections $s \rar 1-s$ yields 
\e{Zdmain}{{\Zd^{(0,0)}(r,z) = {1\o 2\pi i} \ds|(\Zs,E_s)|^2 \big(r \,z^{\half-s}+ r^{2s}\varphi(1-s)\big)}}
where we have defined 
\e{}{r := \sqrt{y_1y_2}\,,\quad z:= {y_1\o y_2}}
Recall that $|f(s)|^2 := f(s) f(1-s)$. We can write this as 
\e{Zdmain2}{\Zd^{(0,0)}(r,z) =  r \sqrt{z}\mathcal{R}(z) + \mathcal{S}(r)}
where 
\es{Rdef}{\mathcal{R}(z) &:=\M^{-1}\big[|(\Zs,E_s)|^2;z\big]\\
\mathcal{S}(r) &:=\M^{-1}\big[|(\Zs,E_s)|^2 \varphi(1-s); r^{-2}\big]}
are inverse Mellin transforms of the squared overlap, defined in general as
\e{}{\cM^{-1}[f(s);x] = {1\o 2\pi i} \int_{\mathcal{C}}ds\, f(s) x^{-s}}
The contour $\mathcal{C}$ is a vertical contour within the critical strip of the inverse Mellin transform \cite{FLAJOLET19953}, which for our purposes may be taken to be $\text{Re}\, s= \half$.\foot{If there are poles on the contour, they must be regulated for the spectral decomposition to formally exist. This can be done, e.g. by shifting the poles to $\text{Re}\, s=\half\pm\eps$. We neglect this possibility in what follows.} 

Lest it appear that there are two independent functions in \eqr{Zdmain2}, we emphasize that $\mathcal{S}(r)$ is completely determined by $\mathcal{R}(z)$, since the latter may be used to reconstruct $|(\Zs,E_s)|^2$. This determinism can be formalized using techniques of Mellin convolution (see Appendix \ref{secconvS}), giving
\e{S00final}{{\mathcal{S}(r) = \sum_{n=1}^\i {\phi(n) \o n^2}  \int_0^1 {du\o \sqrt{u(1-u)}} \, \mathcal{R}\({u\o n^2r^2}\)}}
where $\phi(n)$ is the Euler totient function. An equivalent representation in terms of residues of $\mathcal{R}(z)$, given in \eqr{S00final3app}, may be obtained by blowing up the contour. So in total,
\e{Zdmain3}{\Zd^{(0,0)}(r,z) =  r \sqrt{z}\mathcal{R}(z) + \sum_{n=1}^\i {\phi(n) \o n^2}  \int_0^1 {du\o \sqrt{u(1-u)}} \, \mathcal{R}\({u \o n^2r^2}\)}
Note the $z \rar 1/z$ inversion symmetry of $\sqrt{z} \cR(z)$, inherited from the inverse Mellin transform.

The above expressions are valid for all temperatures. We now take the low-temperature limit:
\e{largey}{r\gg 1\,,\quad \text{$z$ fixed}\,.}
This limit includes two physically distinct regimes of interest (see Figure \ref{fig:contourz}), depending on how we take $y_i$ to infinity in the complex plane:
\begin{itemize}
\item In ``Euclidean'' signature where $y_i\in\R$, this is a low-temperature limit, and $z\in\R_+$.

\item In Lorentzian signature, analytically continuing as one does to construct the SFF, 
\e{sffkin}{y_1 \rar \b+iT\,,\quad y_2 \rar \b-iT\,.}
The limit \eqr{largey} is a simultaneous late-time/low-temperature limit of $T\gg1$ and fixed $T/\b$, with $z\in\HH$ on the unit circle:
\e{zangle}{z := e^{i\theta}\,,\quad \theta = 2\tan^{-1}\({T\o \b}\) \in\[0,{\pi}\]}
\end{itemize}
In either case, applying the limit \eqr{largey} to \eqr{Zdmain}, we must deform the contour of the second term to the left. Consequently the $\mathcal{S}(r)$ term is subleading:\foot{This is the same mechanism that operates in the spectral decomposition of $\cN=4$ super Yang-Mills observables in the large $N$ `t Hooft limit \cite{Collier:2022emf}. There, the spectral integrand contains a sum of terms with powers $N^{1-2s}$ and $N^0$. The former is suppressed, so the latter gives the full planar result.}
\e{Zdlinear}{\Zd^{(0,0)}(r,z) \approx r \sqrt{z} \mathcal{R}(z)  \qquad (r\gg1\,,~ z~\text{fixed})}
This result is interesting. First, the behavior is universally linear in $r$. More importantly, since $\mathcal{R}(z)$ fixes the entire function $\Zd^{(0,0)}(r,z)$ via \eqr{Zdmain3}, {\it the leading result at low temperatures fixes the result at all temperatures}. Indeed, it fixes the entire Eisenstein part of $\Zd(\t_1,\t_2)$. As we will see later, for two-copy partition functions with enhanced symmetries, $\mathcal{R}(z)$ fixes the function completely.

\ssec{A condition for a linear ramp}

The spectral form factor, $K_\b(T)$, is defined for quantum systems as
\e{}{K_\b(T) := Z(\b+i T)Z(\b-iT)}
where $Z(\b+iT) = \Tr \qty(e^{-(\b+iT)H})$  is the analytically-continued sum over states. In 2d CFT, it is cleaner to account for Virasoro descendants by defining the SFF with respect to the Virasoro primary partition function, 
\e{}{K_\b(T) := {1\o \sqrt{\b^2+T^2}} \<Z_p(\b+i T)Z_p(\b-iT)\>}
where the brackets denote a coarse-graining of some kind. The prefactor, which amounts to stripping off the $\sqrt{y_1y_2}$ coming from using the {\it primary} partition function (before analytic continuation), is a useful convention for the late-time limit and contact with previous literature (e.g. \cite{Cotler:2016fpe,Saad:2018bqo,Cotler:2020ugk}). 

We can also define a SFF using $\Zs(\t)$ instead,
\e{}{K_\spec(\b;T) := {1\o \sqrt{\b^2+T^2}} \<\Zs(\b+i T)\Zs(\b-iT)\>}
Now, since $\Zs(\t)$ and $Z_p(\t)$ differ only by terms which contribute to the slope/dip of the SFF \cite{Dyer:2016pou,Benjamin:2018kre},\foot{This is because the difference, $\widehat Z_L(\t)$, is a sum over self-averaging quantities. (If $Z_L(\t)$ sums over an infinite set of light operators, we are assuming that this sum can be regulated.) Supporting arguments from other perspectives can be found in the literature. In 2D gravity, the ramp is generated by a {\it new} bulk saddle \cite{Saad:2018bqo}, with similar arguments for one-point wormholes \cite{Saad:2019pqd} and in higher dimensions \cite{Cotler:2016fpe,Saad:2018bqo}. It was argued in \cite{Dyer:2016pou,Benjamin:2018kre} in the AdS$_3$ context that Poincar\'e completions of individual operators (i.e. sums over smooth saddles) contribute only to the dip. In the Poincar\'e completion of individual light states, the modular images are self-averaging, with a density that is a sum of smooth functions growing exponentially in energy (see e.g. eq. (2.3) of \cite{Benjamin:2020mfz}). Similar arguments can be made for enigmatic AdS$_3$ black holes, which are also smooth saddles (when they exist).} the difference between $K_\spec(\b;T)$ and $K_\b(T)$ vanishes at late times: 
\e{}{K_\b(T) \approx K_\spec(\b;T)\qquad (T \gg \b)}
Our spectral formalism is tailor-made to study the SFF in the late-time regime: the quantity $\Zd(\t_1,\t_2)$, constructed from $\Zs(\t)$, is designed precisely as a coarse-grained product of partition functions in the diagonal approximation, whose point is to produce the ramp. So our claim is that 
\e{KbT}{K_\b(T) \approx {1\o T}\,\Zd(\b;T)\qquad (T \gg \b)}
where $\Zd(\b;T)$ is the analytic continuation of $\Zd(\t_1,\t_2)$ to SFF kinematics. (Implicitly, here and below, $T$ is bounded above by the plateau time.) 

\begin{figure}[t]
\centering
{
\subfloat{\includegraphics[scale=0.15]{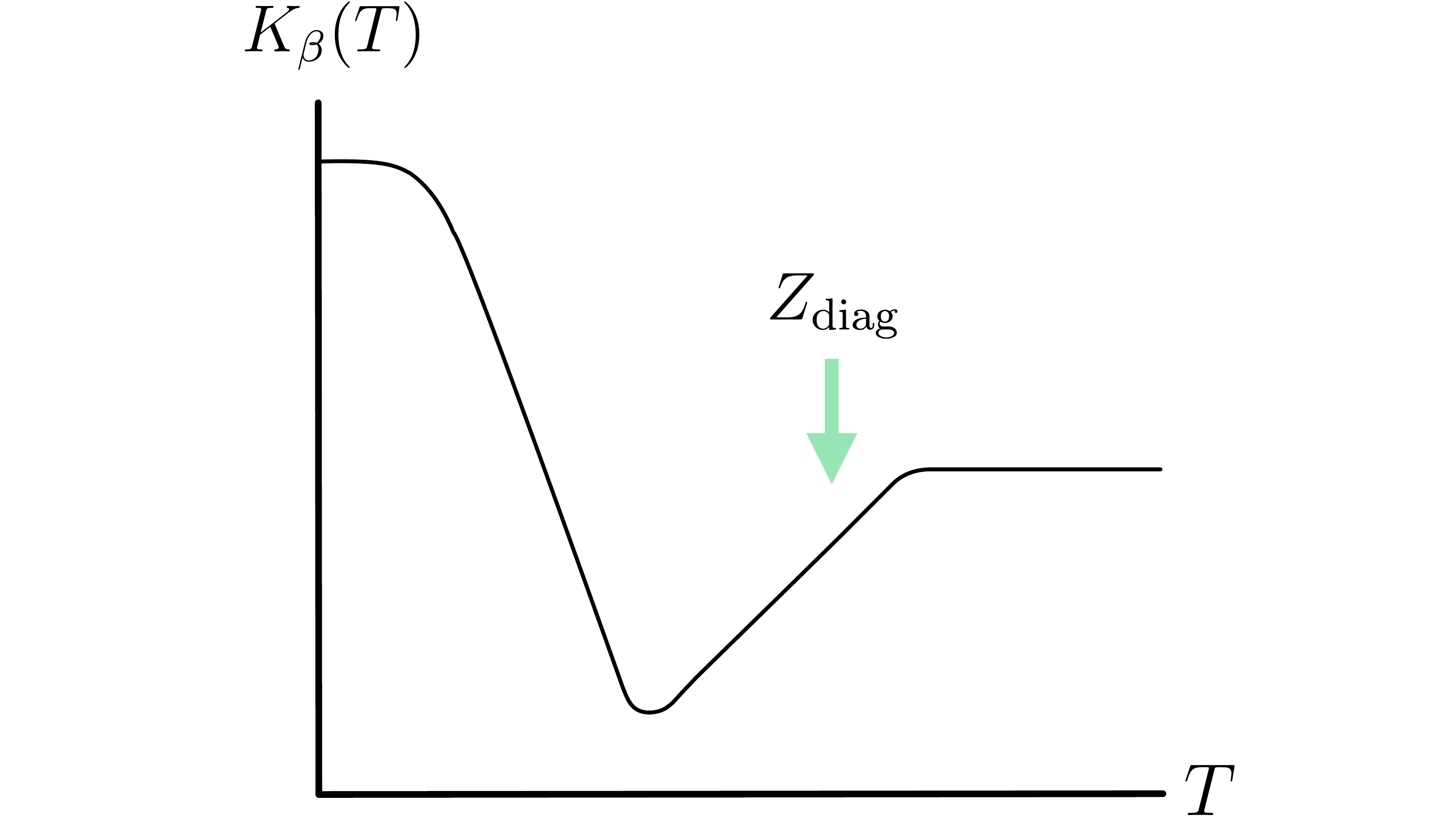}}
}
\caption{$\Zd$ captures the ramp in the spectral form factor $K_\beta(T)$ of a 2d CFT.}
\label{DRPfig}
\end{figure}

Starting from $\Zd(\t_1,\t_2)$, we construct the SFF by analytically continuing as in \eqr{sffkin}. Denote the spin-graded SFF as $K^{(j_1,j_2)}_\b(T)$. We specialize to the case $(j_1,j_2)=(0,0)$, whereupon \eqr{Zdlinear} and \eqr{KbT} imply that
\e{KhighT}{K^{(0,0)}_\b(T) = \sqrt{{\b+i T\o \b-iT}} \mathcal{R}\({\b+i T\o \b-iT}\) + \text{(subleading at $T\gg\b$)}}
We now impose the ramp. From \eqr{zangle} and \eqr{KhighT}, the presence of a {\it linear} ramp is equivalent to the presence of a simple pole in the analytic continuation of $\mathcal{R}(z)$ to $z=-1$:
\es{ramppole}{\boxed{K^{(0,0)}_\b(T\rar\i) \sim {T\o 4\pi\b}\mathsf{C}_{\rm RMT} \quad \Leftrightarrow \quad \mathcal{R}(z \rar -1) \sim {\mathsf{C}_{\rm RMT}\o 2\pi}{1\o 1+z}}}
$\mathsf{C}_{\rm RMT}$ is a constant that fixes the choice of RMT ensemble governing the ramp.\foot{The GOE ensemble is the relevant one for parity-invariant 2d CFTs \cite{Yan:2023rjh}.}  Since $\mathcal{R}(z)$ is simply the inverse Mellin transform \eqr{Rdef} of the spectral overlap, the pole condition \eqr{ramppole} is equivalent to an asymptotic property of the spectral overlap $(\Zs,E_s)$ on the critical line $s=\half+i\w$.  In the parameterization \eqr{zangle},  
\e{Rramp}{2\pi(z+1)\mathcal{R}(z)\Big|_{z=e^{i\theta}} = 2 \cos{\theta\o2} \int_{-\infty}^\i d\w\, |(\Zs,E_\crit)|^2 \cosh \w\theta}
This only satisfies \eqr{ramppole} if the integral diverges linearly. This implies the following condition: 
\es{lineardivfinal}{\boxed{{\text{(Linear ramp)} \quad \Leftrightarrow \quad|(\Zs,E_{\crit})|^2 \sim e^{-{\pi}\w}f(\w)\qquad (\w\rar\i)}}}
where $f(\w)$ is allowed to fluctuate around $\mathsf{C}_{\rm RMT}$, but is flat at infinity ``on average'',
\e{fcond}{\lim_{W\rar\i}{1\o W} \int_{\w_0}^{\w_0+W}d\w\,f(\w) = \mathsf{C}_{\rm RMT}}
for any finite $\w_0$. 

Equations \eqr{lineardivfinal} and \eqr{fcond}, equivalently \eqr{ramppole}, comprise a necessary and sufficient condition for the presence of a linear ramp in the (scalar) SFF of a 2d CFT, with the constant in \eqr{fcond} set by the chosen RMT ensemble governing the ramp. The linear ramp, a property of the coarse-grained theory, is transmuted into a quantitative property of the microscopic spectrum. The algorithm to detect it from a torus partition function is straightforward: form $\Zs(\t)$, then look for the asymptotic \eqr{lineardivfinal} in the spectral basis. 

Note for context that the convergence of the spectral decomposition only requires the much weaker falloff $|(\Zs,E_{\crit})| \lesssim \O(1)$ as $\w\rar\i$, up to power-law and logarithmic corrections (e.g.  \cite{Collier:2022emf}). Systems for which this upper bound is saturated would have an exponential rather than linear ramp.\foot{This was seen in e.g. the quadratic SYK model \cite{Winer:2020mdc}. Exponential ramps are also characteristic of arithmetic systems \cite{Bogomolny1997,Bogomolny1996} } So we see that in the $\sl$ spectral basis, random matrix universality becomes a condition of rapid decay at infinity.

\ssec{Universal corrections}
The symmetries of 2d CFT prescribe corrections to this result. Having extracted the condition \eqr{lineardivfinal} on the overlap, modular invariance implies an infinite set of other terms in $\Zd^{(0,0)}(r,z)$, coming from the second term in \eqr{Zdmain3} (the $\cS(r)$  term in \eqr{Zdmain2}). In SFF kinematics these terms give subleading corrections to the RMT ramp at $T\gg\b$, but are otherwise present and unsuppressed for generic temperatures.

Specifically, the falloff \eqr{lineardivfinal} implies an infinite set of square-root singularities at $z\in\R_-$. The simplest way to see this is to insert the polar behavior \eqr{ramppole} into the second term in \eqr{Zdmain3}. This yields 
\e{Sresult}{\mathcal{S}(r) \supset {\mathsf{C}_{\rm RMT}\o 2} \sum_{n=1}^\i {\phi(n)\o n} {{r\o\sqrt{1+n^2r^2}}}}
The branch point singularities \eqr{Sresult} are universal in any chaotic 2d CFT. Their existence is required by modular invariance (and Virasoro symmetry) upon imposing the linear ramp in the SFF.

Note that in detecting these singularities, we are probing analytic structure of the partition function in the complex-temperature plane: the $r^2 = -1/n^2$ locus is not a ``physical'' locus, neither in real time $T$ in SFF kinematics (where $r^2 = \b^2 + T^2$) nor in Euclidean kinematics (where $r^2 = y_1y_2$). In the limit of $r\rar\i$ with fixed $z$, these singularities indicate an accumulation of branch points at $z \rar -n^2 y_1^2$ and their images under inversion. At leading order in $r\rar\i$, the sum is constant in $r$ but linearly divergent (see \cite{Cotler:2020ugk} and Appendix \ref{secconvS}).

We note in passing the generalization to non-linear ramps. At leading order,
\es{nonlineardiv}{\text{($T^{1-\a}$ ramp)} \quad \Leftrightarrow \quad|(\Zs,E_{\crit})|^2 \sim \w^{-{\a}}e^{-{\pi}\w}f(\w)\qquad (\w\rar\i)}
where $f(\w)$ can again fluctuate around a constant at infinity. The constraints on $\mathcal{S}(r)$ may be likewise derived from \eqref{S00final}, yielding
\e{}{{\mathcal{S}(r) \sim \sum_{n=1}^\i  {\phi(n) \o n} {r\o (1+n^2 r^2)^{\half-\a}}\,,\qquad \(r^2 \rar -{1\o n^2}~~\forall~n\in\Z_+\)}}
We note that if $\a> \half$, the corrections are term-wise finite in $r$. This singles out $\sqrt{T}$ behavior of the late-time SFF as a somewhat notable threshold from the analyticity point of view.

\sec{Wormholes and Hecke Symmetry}\label{s4}

In Section \ref{diagsec} we introduced a coarse-grained two-copy partition function, $\Zh(\t_1,\t_2)$, which projects the factorized product $\Zs(\t_1)\Zs(\t_2)$ onto its diagonal subspace. The centrality of $\Zh(\t_1,\t_2)$ for analyzing chaotic spectral correlations in 2d CFT really comes to life upon considering its dual gravitational interpretation, as we now address a driving question of this work: where are the wormholes? As we show in this section, Hecke symmetry is an emergent feature of Euclidean torus wormholes in semiclassical AdS$_3$ gravity.  

\ssec{Interlude: A Wormhole Farey Tail}\label{farey}

The next subsection establishes some technical results about $SL(2,\Z)$ Poincar\'e sums over appropriate ``seed'' functions $f_0(\t_1,\t_2)$ of two complex moduli $\t_1$ and $\t_2$. Sums of this form may be taken as one definition of ``wormhole amplitudes'' in semiclassical theories of gravity. 

To motivate this, and to explain the title of this subsection, let us recall the (non-supersymmetric) black hole Farey tail \cite{Dijkgraaf:2000fq,Maloney:2007ud,Keller:2014xba}. The partition sum over all smooth saddles $\cM$ of semiclassical Einstein gravity with $\p\M = T^2$ can be written as a (regularized) $SL(2,\Z)$ Poincar\'e sum over the partition function of thermal AdS$_3$: writing the primary partition function, 
\e{zmwk}{Z_{\rm MWK}(\t) = \sum_{\g\in SL(2,\Z)/\G_\i} \sqrt{\Im(\g\t)} \big|q_\g^{-\xi} (1-q_\g)\big|^2}
The sum runs over the full ``$SL(2,\Z)$ family'' of BTZ black holes, one for each independent $SL(2,\Z)$ element. The essential point is that the one-to-one correspondence between smooth saddles and $SL(2,\Z)$ images of thermal AdS$_3$ is not merely a technical observation: these saddles are generated by large bulk diffeomorphisms acting on the vacuum contribution to the path integral, inducing a boundary $SL(2,\Z)$ action.

The idea that Poincar\'e sums are how semiclassical gravity implements boundary modular invariance in AdS$_3$/CFT$_2$ is independent of the number of asymptotic boundaries. As long as the bulk theory respects diffeomorphism invariance, large diffeomorphisms will again relate different contributions to the path integral, implementing $SL(2,\Z)$ transformations (or the appropriate higher-genus generalization) of boundary moduli. The obvious question is what the analog of the vacuum solution is. Consider now bulk topologies $\cM$ with two asymptotic torus boundaries, $\p\cM = T^2 \cup T^2$. In the example of semiclassical Einstein gravity, as we will recall in Section \ref{sec:CJ}, there is no smooth on-shell solution with this topology. Nevertheless, diffeomorphism invariance implies that if there exists such an amplitude -- perhaps off-shell, justifiable one way or another -- it should take the form of a Poincar\'e sum over a suitable seed function $f_0(\t_1,\t_2)$, with the image sum geometrizing the boundary modular invariance.\foot{This view receives indirect formal support from \cite{Collier:2023fwi}, where it is argued that the canonical quantization of AdS$_3$ gravity on hyperbolic manifolds $M$ produces sums of the form $Z(M)= \sum_{\g} Z_0(M_{\gamma})$, where $Z_0$ is the bulk gravity partition function on a fixed manifold and $\gamma \in \text{Map}(\p M)/\text{Map}(M,\p M)$, where $\text{Map}(\p M)$ and $\text{Map}(M,\p M)$ are the boundary and bulk relative mapping class groups, respectively. Torus wormholes are not hyperbolic. Nevertheless, the sum prescribed above would, if naively applied, lead to the picture in the text. Namely, for $M=T^2\times I$, the resulting sum would be over the coset $\sl \times \sl/\sl$, which can be parametrized as a sum over relative modular transformations of a seed which is invariant under simultaneous modular transformations.} 

This slight generalization of the familiar single-boundary ideology to the case $\p\cM = T^2 \cup T^2$ guides what we will define as a ``torus wormhole'' in semiclassical gravity, and accordingly begets the {\it wormhole Farey tail}: that is, {\it the identification of torus wormholes, i.e. $SL(2,\Z)$ Poincar\'e sums over suitable seed functions, as gravitational duals of $Z_\Hecke(\t_1,\t_2)$ in large $c$ CFTs.}

\ssec{Properties of wormholes}\label{sec:whprop}
We will study $SL(2,\Z) \x SL(2,\Z)$-invariant functions $f(\t_1,\t_2)\in L^2(\cF\times \cF)$ which admit representations as Poincar\'e sums over relative modular transformations\foot{Depending on context, the sum may instead be taken over $PSL(2,\Z)$. This distinction is not relevant for what follows, and indeed, in Section \ref{sec:CJ} we will study an $SL(2,\Z)$ sum.}
\e{poinc}{ f(\tau_1,\tau_2)=\sum_{\gamma \in SL(2,\Z)}  f_0(\tau_1,\gamma \tau_2).}
As a consequence,  the seed function $f_0(\tau_1,\tau_2)$ is invariant under simultaneous modular transformations,
\e{simul}{f_0(\t_1,\t_2) = f_0(\g\t_1,\g\t_2)\,,\quad \g\in SL(2,\Z).}
\noindent The results to follow may be tweaked to accommodate seeds that instead obey 
\e{simul'}{f_0(\t_1,\t_2) = f_0(\g\t_1,M\g M\t_2)\,,\quad \g\in SL(2,\Z)\,,\quad M = \begin{pmatrix}
-1 & 0 \\
0 & 1 
\end{pmatrix}}
where $M\t = -\t$ implements an orientation reversal. 

In Subsection \ref{sec:heckewh} we will assume slightly more of the seed, namely, an invariance under simultaneous $SL(2,\R)$ transformations.\foot{Actually, we will not use the full $SL(2,\R)$, but only the subset of matrices whose entries are given by $\sqrt{j}$ times integers, where $j\in\Z_+$.} One may justify why this is sensible from different points of view. As we will see, it will play nicely with the result of Subsection \ref{sec:distwh}, where we will prove that Hecke symmetry of $f(\t_1,\t_2)$ implies that the seed $f_0(\t_1,\t_2)$ is not only fully $SL(2,\R)$-invariant, but is solely a function of the $\HH$-invariant distance between $\t_1$ and $\t_2$. This being a very natural criterion from a physical standpoint, one may reasonably demand that the seed amplitudes depend only on this distance in the first place, justifying the assumption of simultaneous $SL(2,\R)$-invariance. At any rate, we expect that the logical connections among these ideas will be clear in the proofs to follow. 

We take $f(\t_1,\t_2)$ as an operational definition of torus wormhole amplitudes. What properties do these possess?

\sssec{Hecke symmetry}\label{sec:heckewh}

\begin{result}{I} {\it If $f(\t_1,\t_2)$ is a Poincar\'e sum \eqr{poinc} with a seed invariant under simultaneous $SL(2,\R)$ transformations,
\e{}{f_0(\t_1,\t_2) = f_0(\g\t_1,\g\t_2)\,,\quad \g\in SL(2,\R)}
then $f(\t_1,\t_2)$ is Hecke symmetric.}
\end{result}

A modular-invariant function $f(\tau_1,\tau_2)\in L^2(\cF\times \cF)$ is entirely specified by its spectral overlaps with the basis elements: if two such functions have the same overlaps, they are equal. Using this logic, we prove Hecke symmetry by showing that
\e{}{ (T_j^{(\tau_1)}f(\tau_1,\tau_2),\psi_\w(\tau_2)) = (T_j^{(\tau_2)}f(\tau_1,\tau_2),\psi_\w(\tau_2))}
where $\psi_\w(\t_2) := \{E_\crit(\t_2), \phi_n(\t_2)\}$ is an $\sl$ eigenfunction. We start from the expression
\es{}{ 
(T_j^{(\tau_1)}f(\tau_1,\tau_2),\psi_\w(\t_2))= \frac{1}{\sqrt{j}} \sum_{A\in SL(2,\Z)\backslash M_j} \int_{\HH}\frac{dx_2dy_2}{y_2^2} f_0(A\tau_1,\tau_2)\psi_\w(\t_2)
}
where we have used the unfolding trick for $f(\tau_1,\tau_2)$, since it is given by a Poincar\'e sum. We now use the invariance of the seed,\foot{Elements of $M_j$ may be mapped to elements of $SL(2,\R)$ by rescaling: given a matrix $B\in M_j$ with $\det B = j$, there exists a matrix $B' := B/\sqrt{j}$ with $\det B'=1$. Since $j\in\Z_+$, and $B$ is integer-valued, $B'\in SL(2,\R)$. Note that both $B$ and $B'$ act on $\t$ via the same fractional linear transformation.}
\e{}{f_0(A\tau_1,\tau_2) = f_0(\tau_1,A^{-1}\tau_2)\,,\quad A \in SL(2,\Z)\backslash M_j}
Next, we perform a change variables $\tau_2'=A^{-1}\tau_2$ in the integral, which leaves the Poincar\'e measure on $\HH$  invariant.\footnote{Similar manipulations can be found in Theorem 3.6.4, Point 4 of \cite{Terras_2013}.} This results in
\es{hecke1}{
(T_j^{(\tau_1)}f(\tau_1,\tau_2),\psi_\w(\t_2))&=\frac{1}{\sqrt{j}} \sum_{A\in SL(2,\Z)\backslash M_j} \int_{\HH}\frac{dx_2'dy_2'}{y_2'^2} f_0(\tau_1,\tau_2')\psi_\w(A\tau_2')\\&=\int_{\HH}\frac{dx_2'dy_2'}{y_2'^2} f_0(\tau_1,\tau_2')T^{(\tau_2')}_j\psi_\w(\tau_2')\\&=
(f(\tau_1,\tau'_2),T^{(\tau_2')}_j\psi_\w(\tau'_2))}
In the second line we brought the sum inside the integral. Using self-adjointness of the Hecke operators,
\e{hecke2}{(T_j^{(\tau_2)}f(\tau_1,\tau_2),\psi_\w(\tau_2))= (f(\tau_1,\tau_2),T_j^{(\tau_2)}\psi_\w(\tau_2))}
Noting equality of \eqr{hecke1} and \eqr{hecke2}, we have
\e{heckeres}{T_j^{(12)} f(\t_1,\t_2) = 0}
which is the desired result. The proof for a seed invariant under \eqr{simul'} is essentially identical. Following the same steps, the final result follows from the invariance of the right-hand side of \eqr{heckedef} upon taking $b \rar -b$ in the summand.

\sssec{Strong Spectral Determinacy}

\begin{result}{II}
{\it If $f(\t_1,\t_2)$ is a Hecke-symmetric Poincar\'e sum \eqr{poinc} with seed obeying \eqr{simul}, then the Eisenstein and cusp form overlaps are equal.}
\end{result}

A Hecke-symmetric function has spectral overlaps proportional to the respective basis elements, as seen in \eqr{ZHecke}. We parameterize these as
\es{}{(f(\t_1,\t_2),E_s(\tau_2))&= f_E(s) \,E_{1-s}(\tau_1)\\
      (f(\tau_1,\tau_2),\phi_n(\tau_2))&=f_\phi(s_n) \,\phi_n(\tau_1)}
In general, $f_E(s)$ and $f_\phi(s_n)$ are different functions of their arguments. Using the unfolding trick for $f(\t_1,\t_2)$, we write these overlaps explicitly as
\es{}{ 
f_E(s)\, E_{1-s}(\tau_1)&=\int_{\HH}\frac{dx_2dy_2}{y_2^2} f_0(\tau_1,\tau_2) E_{1-s}(\tau_2)\\
f_\phi(s_n) \,\phi_n(\tau_1)&= \int_{\HH}\frac{dx_2dy_2}{y_2^2} f_0(\tau_1,\tau_2) \phi_n(\tau_2)}
We now expand the functions on the right-hand side in a Fourier series, and then project onto a spin $k\neq0$ with respect to $\tau_1$. In doing so we employ the Fourier decomposition of a seed $f_0(\t_1,\t_2)$ invariant under simultaneous $SL(2,\Z)$ transformations \eqr{simul},
\e{f0fourier}{f_0(\t_1,\t_2) = \sum_{\ell\in\Z} e^{4\pi i \ell x_+} f_0^{(\ell)}(x_-,y_1,y_2)~~~\text{where}~~~ x_\pm := {x_1 \pm x_2\o 2}}
Changing variables to $x_{\pm}$, we can perform the integral over $x_+$ which sets $j=|k-2\ell |$ and yields (recall that $\mathsf{a}_{j}^{(s)}=\mathsf{a}_{-j}^{(s)}$ and likewise for $\mathsf{b}_{j}^{(n)}$)
%
\es{eq:ssdproof1}{ \mathsf{a}_k^{(1-s)} f_E(s) \(\sqrt{y_1}K_{s-\half}(2\pi k y_1)\)&=\sum_{\ell\in\Z} (2-\delta_{k-2\ell,0})\mathsf{a}_{k-2\ell}^{(1-s)}\,\cI^{(s)}_{k,\ell}(y_1)\\
\mathsf{b}_k^{(n)}f_\phi(s_n)\(\sqrt{y_1}K_{s_n-\half}(2\pi k y_1)\)&=2\sum_{\ell\in\Z} \mathsf{b}_{k-2\ell}^{(n)}\,\cI^{(s_n)}_{k,\ell}(y_1)}
where 
\es{}{
\cI^{(s)}_{k,\ell}(y_1):=\int\frac{dy_2}{y_2^2} \int dx_- e^{2\pi ix_-(-k+|k-2\ell|)}f_0^{(\ell)}(x_-,y_1,y_2) \sqrt{y_2}K_{s-\half}(2\pi|k-2\ell|y_2)
}
The factor in parentheses in \eqr{eq:ssdproof1} is the spin-$k$ Fourier mode with the coefficient stripped off. Evaluating the first line of  \eqr{eq:ssdproof1} at $s=s_n$ and taking the ratio of the two equations, 
the left-hand side is independent of $y_1$. This is incompatible with the right-hand side unless 
\e{eq:seedspinzero}{f^{(\ell)}_0(x_-,y_1,y_2) \propto \delta_{\ell,\ell^*}}
for some fixed $\ell^*$. In fact, the only possible choice is $\ell^*=0$, such that the Fourier coefficients cancel completely, because the equation must be satisfied for {\it any} $k\in\Z_+$. Therefore,
\e{cuspeis}{f_\phi(s_n) = f_E(s_n)}
That is, the cusp form overlap is equal to the Eisenstein overlap evaluated at $s=s_n$. 

As a statement about Fourier decomposition, this means that the $(j_1,j_2)=(0,0)$ mode of $f(\t_1,\t_2)$, which determines the Eisenstein part of the spectral decomposition, actually determines the entire function. Building on the language of \cite{Kaidi:2020ecu,Benjamin:2021ygh}, we call this {\it strong spectral determinacy}. This is compatible with the general result of \cite{Haehl:2023tkr} because we are imposing extra structure on $f(\t_1,\t_2)$.

\sssec{Dependence on hyperbolic distance}\label{sec:distwh}

\begin{result}{III}
{\it A function $f_0(\t_1,\t_2)$ that obeys \eqr{simul} and has spin $\ell=0$ under the Fourier decomposition in $x_1+x_2$ is solely a function of the invariant hyperbolic distance,
\e{seedgeod}{f_0(\t_1,\t_2) = f_0(\sigma(\t_1,\t_2))\,,\quad  \text{\rm where} \quad \sigma(\t_1,\t_2) := {|\t_1-\t_2|^2\o y_1y_2}\,.}
This result applies to wormhole amplitudes because the seed for \eqr{poinc} is such a function (cf. below \eqr{eq:seedspinzero}).}
\end{result}

The spin $\ell=0$ property implies that $f_0(\t_1,\t_2) = f_0(x_-,y_1,y_2)$. Without loss of generality we can write this as $f_0(\t_-,y_1,y_2)$, where $\t_- := (\t_1-\t_2)/2$. This is invariant under simultaneous $T$-transformations.  Under simultaneous $S$-transformation, 
\e{}{f_0(\t_-,y_1,y_2) \mapsto f_0\({\t_-\o \t_1\t_2},{y_1\o |\t_1|^2},{y_2\o |\t_2|^2}\)}
Demanding simultaneous $S$-invariance then fixes the functional form to be as in \eqr{seedgeod},\foot{This is a two-variable generalization of the fact that a modular-invariant function $f(y)$ must be a constant. The proof in that case is similar: an $S$-transformation introduces $x$-dependence.} where $\sigma(\t_1,\t_2)$ is the $\HH$-invariant distance between $\t_1$ and $\t_2$, related to the geodesic distance $d(\t_1,\t_2)$ as
\e{}{{\sigma(\t_1,\t_2)} = 4\sinh^2\( {d(\t_1,\t_2)\o2}\) \,.}
This concludes the proof. Essentially the same result holds for a seed obeying the condition \eqr{simul'}. In this case, $f_0(\t_1,\t_2)$ has a Fourier decomposition \eqr{f0fourier} with the swap $x_+ \leftrightarrow x_-$. The same argument then leads to
\e{}{f_0(\sigma_-(\t_1,\t_2))\,,\quad \text{where}\quad \sigma_-(\t_1,\t_2) := \sigma(\t_1,-\t_2)  = {|\t_1+\t_2|^2\o y_1y_2}}
That is, the seed depends on geodesic distance with the orientation-reversal $\t_2 \rar -\t_2$. 

That a Hecke-symmetric Poincar\'e sum of the form  \eqr{poinc} has a seed which depends only on the $\HH$-invariant distance ties nicely into the earlier proof of Hecke symmetry: $\sigma(\t_1,\t_2)$ is fully $SL(2,\R)$-invariant.

\ssec{Summary and comments}\label{sec:heckesum}

The results of the previous subsection suggest the following general nomenclature: we define ``wormhole amplitudes'' $Z_\WH(\t_1,\t_2)$ as those functions which have the following spectral decomposition:
\e{zwh}{\boxed{Z_\WH(\t_1,\t_2) =  \int_{\ccrit} f_\WH(s) E_{s}(\tau_1)E_{1-s}(\tau_2) +\sum_{n=1}^\i f_\WH(s_n)\phi_n(\tau_1)\phi_n(\tau_2)}}
In our spectral framework, $Z_\WH(\t_1,\t_2)$ are to be regarded as instances of $\Zh(\t_1,\t_2)$ with enhanced structure. Given a $\Zs(\t)$ of a CFT,
\e{fareyid}{f_\WH(s) = |(\Zs,E_s)|^2\,.}
The property $\Zh(\t_1,\t_2) = Z_\WH(\t_1,\t_2)$ thus holds if and only if 
\e{cuspeisequal}{(\Zs,\phi_{n})^2 = |(\Zs,E_{s_n})|^2}
Poincar\'e sums of the form \eqr{poinc} are but one example of such amplitudes. We may also view this as a prediction for large $c$ CFTs: if we take the Farey tail sum \eqr{poinc} as a wormhole ansatz in semiclassical gravity, \eqr{cuspeisequal} should hold in the dual large $c$ CFT.

The functions $Z_\WH(\t_1,\t_2)$ furnish a highly-symmetric class of square-integrable $\sl \x \sl$-invariant functions: not only are they Hecke-symmetric, but they are fixed by a single overlap function, $f_\WH(s)$. Let us put this last point in sharper focus. In the language of Section \ref{sec:zdiag}, recall that the leading low-temperature term of the $(j_1,j_2)=(0,0)$ Fourier mode was controlled by the function $\mathcal{R}(z)$. This term fixes other Fourier modes of our various amplitudes, {\it away} from the low-temperature limit, according to their degree of symmetry:

\vs
\begin{table}[h]
    \makegapedcells
\centering
\begin{tabular}{c|c}

   Amplitude         & Modes $(j_1,j_2)$ fixed by $\mathcal{R}(z)$ \\ \midrule
$\Zd(\t_1,\t_2)$        & $(0,0)$          \\
$\Zh(\t_1,\t_2)$        &  $(0,j)$             \\
$Z_\WH(\t_1,\t_2)$ & All\\

\end{tabular}
\label{tab:R0modes}
\end{table} 
\vs

\noindent For $Z_\WH(\t_1,\t_2)$ in particular, $\mathcal{R}_\WH(z)  = \cM^{-1}[f_\WH(s);z]$ fixes {\it all} Fourier modes, and hence the entire function (``strong spectral determinacy'').

While the notation $Z_\WH(\t_1,\t_2)$ invokes bulk language to reflect the properties of Poincar\'e sums derived in the previous subsection, other functions could in principle assume the functional form \eqr{zwh}. However, there is clearly a gravitational nature to $Z_\WH(\t_1,\t_2)$, and to $\Zh(\t_1,\t_2)$ more generally. The projection $\cP_\Hecke$ is in some sense capturing the ``wormhole part'' of the factorized product $\Zs(\t_1)\Zs(\t_2)$: it isolates the correlations between the two copies which, when there is an AdS$_3$ bulk dual, are geometrized by a smooth, two-boundary, connected Euclidean spacetime. This interpretation is on firmest footing at large $c$, where it becomes a semiclassical geometric statement that is supported by our proofs above for wormholes-qua-$SL(2,\Z)$ Poincar\'e sums. 

These results suggest the following conceptual interpretation. In gravitational language, the Hecke operators $T_j^{(\t)}$, which act on a single boundary component, are AdS$_3$ ``half-wormhole detectors.'' That $T_j^{(\t)} \Zs(\t) \neq 0$ reflects the fact that $\Zs(\t)$ by definition receives no contribution from smooth geometric saddles in gravity: it only counts off-shell, or non-geometric (e.g. matter), degrees of freedom. Then forming the difference operator $T_j^{(12)}= T_j^{(\t_1)} - T_j^{(\t_2)}$, the projection onto its kernel via $\cP_\Hecke$ implements the pairing of these degrees of freedom to form a smooth two-boundary spacetime. To say this slightly differently, $T_j^{(12)}$ has the flavor of a non-local bulk operator, which can be freely moved to the ``left'' or ``right'' boundary tori: Hecke symmetry is then the statement that these operators are, in a sense, topological. Conversely, a nonzero action of $T_j^{(12)}$ detects non-geometric contributions to the wormhole amplitude. It would of course be nice to find an explicit {\it bulk} construction of Hecke operators, or of $\cP_\Hecke$ as a topology-changing operator. 

We will next make these ideas explicit in semiclassical AdS$_3$ pure gravity. 


\sec{Pure Gravity as MaxRMT}\label{sec:CJ}

We now apply our framework to semiclassical AdS$_3$ pure gravity. We begin our treatment with the two-boundary torus wormhole of Cotler and Jensen \cite{Cotler:2020ugk}. 

\ssec{Torus wormhole}
The Cotler-Jensen (CJ) wormhole was originally presented in \cite{Cotler:2020ugk} as a Poincar\'e sum of the form considered in Section \ref{s4},\foot{\cite{Cotler:2020ugk} wrote this as a $PSL(2,\Z)$ image sum, but considerations of discrete symmetries suggest that the result should be doubled \cite{Yan:2023rjh}. For this reason, and to streamline some of the expressions to follow, we work with the GOE version of the CJ wormhole, which amounts to overall multiplication of the expressions in \cite{Cotler:2020ugk} by $\mathsf{C}_{\rm GOE}=2$.}
\e{CJpoinc}{ Z_\CJ(\tau_1,\tau_2)=\sum_{\gamma \in SL(2,\Z)} f_{0,\CJ}(\t_1,\g\t_2)}
with seed given by the inverse of the (orientation-reversed) hyperbolic distance:
\e{CJseed}{f_{0,\CJ}(\t_1,\t_2) = {1\o 2\pi^2} \,\s^{-1}_-(\t_1,\t_2) = {1\o 2\pi^2} {y_1y_2\o |\t_1+\t_2|^2}}
The CJ wormhole is an off-shell contribution to the path integral of AdS$_3$ pure gravity with the topology of a torus times an interval, $T^2\times I$. It is a constrained instanton, namely it becomes on-shell after adding a constraint \cite{Cotler:2020lxj}. The CJ computation uses a dynamical theory of only boundary gravitons \cite{Cotler:2018zff}. This differs from the corresponding quantity in the Chern-Simons formulation of AdS$_3$ gravity\cite{Cotler:2020ugk,Collier:2023fwi,Eberhardt:2022wlc}.
 
The point is now to understand {\it $Z_\CJ(\t_1,\t_2)$ as an instance of $Z_{\Hecke}(\t_1,\t_2)$}: that is, as a coarse-grained two-copy partition function of an underlying chaotic CFT.

Since $Z_\CJ(\t_1,\t_2)$ is a Hecke-symmetric wormhole amplitude, the entire amplitude is fixed by the $(0,0)$ Fourier mode. From \cite{Cotler:2020ugk} eq. (4.12) (and doubling the result), we read off the functions in \eqr{Zdmain2} as
\es{cjrs}{\mathcal{R}_{\CJ}(z) &= {1\o\pi}{1\o 1+z}\\
\mathcal{S}_{\CJ}(r) &= \sum_{n=1}^\i {\phi(n)\o n}{r\o \sqrt{1+n^2r^2}}}
with $r = \sqrt{y_1y_2}$ and $z=y_1/y_2$. Recall that $\mathcal{R}_{\CJ}(z)$ is the leading low-temperature term, which fixes the entire zero mode by \eqr{Zdmain3}. 

By the general proofs of Section \ref{sec:whprop} for Poincar\'e sums of the form \eqr{CJpoinc}, $Z_\CJ(\tau_1,\tau_2)$ admits a spectral decomposition\foot{We thank Scott Collier for sharing a note on the spectral decomposition of the CJ wormhole, which helped inform our general perspective.} of wormhole form \eqr{zwh}: we need only determine the spectral overlap, call it $f_\CJ(s)$, by Mellin transform of $\mathcal{R}_\CJ(z)$. A simple integral yields
\es{CJoverlap}{f_\CJ(s) &= \cM\[\mathcal{R}_\CJ(z);s\]\\
&= {1\o\pi}\,\G(s)\G(1-s)}
This determines the full $Z_\CJ(\tau_1,\tau_2)$. Let us write it explicitly for the sake of clarity:
\e{cjspec}{\pi Z_\CJ(\t_1,\t_2) = \int_{\ccrit} \G(s)\G(1-s) E_{s}(\tau_1)E_{1-s}(\tau_2) +\sum_{n=1}^\i \G(s_n)\G(1-s_n)\phi_n(\tau_1)\phi_n(\tau_2)}
 It will prove useful to rewrite \eqr{cjspec} in terms of the completed Eisenstein series:
\e{cjspec*}{\boxed{\pi Z_\CJ(\t_1,\t_2) = \int_{\ccrit} {\pi \o \z(2s)\z(2-2s)}E^*_{s}(\tau_1)E^*_s(\tau_2) +\sum_{n=1}^\i \G(s_n)\G(1-s_n)\phi_n(\tau_1)\phi_n(\tau_2)}}
The Eisenstein overlap takes an intriguing form in terms of the Riemann zeta function. 

The spectral decomposition \eqr{cjspec} is a very clean way to package the Fourier modes for arbitrary spins, determined as they are by the low-temperature scalar term $\mathcal{R}_{\CJ}(z)$. (For comparison, the modes $Z_\CJ^{(j_1,j_2)}(y_1,y_2)$ are given in eq. (4.20) of \cite{Cotler:2020ugk}.)  Note that there is no constant term in the decomposition. This follows our general discussion earlier in the paper, but is additionally bound up with the fact that the (0,0) mode is actually divergent \cite{Cotler:2020ugk} (cf. the $r\rar\i$ limit of $\cS_\CJ(r)$).\foot{The divergence, and subsequent regularization, of \cite{Cotler:2020ugk} may be recovered in the spectral formalism for functions $f(\t_1,\t_2) \in L^2(\cF \x \cF)$ by taking the appropriate residues of the Eisenstein overlap; see Appendix \ref{app:l2space}. Note also that this divergence is not special to the CJ wormhole, as shown in \eqr{S00div}.} Given this, the constant term is subject to the choice of regularization scheme, and the finite constant may be set to zero; this is the meaning of \eqr{cjspec}.

\ssec{MaxRMT}\label{maxRMT}
The CJ wormhole has some very special features. Notice that \eqr{CJoverlap} realizes the RMT ramp falloff condition \eqr{lineardivfinal}, with $f(\w)=\mathsf{C}_{\rm GOE} = 2$ exactly -- in particular, with no fluctuations. Likewise, not only does \eqr{cjrs} manifestly realize the universal singularities \eqr{ramppole} and \eqr{Sresult} required by the presence of a linear ramp in the scalar SFF, but the CJ wormhole is given {\it exactly} as the sum over those singularities! 

This last property is remarkable. Having shown that general wormhole amplitudes are fully determined by the single function $\cR(z)$ after accounting for Virasoro and modular symmetries, $\cR_\CJ(z)$ is exactly equal to the double-scaled RMT result. Stating this from the Lorentzian point of view, the scalar SFF of the CJ wormhole is, for {\it all} fixed $T/\b$, given by ``just the ramp,'' plus corrections fully fixed by the symmetries; in turn, the entire amplitude for {\it all} moduli $\t_1$ and $\t_2$, including Euclidean and Lorentzian temperatures, is its minimal completion. 

These properties of the torus wormhole reveal that the spectrum of semiclassical AdS$_3$ pure gravity and its dual 2d CFT exhibit random matrix statistics to the maximal extent possible. As stated in the Introduction, this is a signature of what we call 

\begin{itemize}
[leftmargin=+1.3in] 
\item [{\sf {MaxRMT}:}] {\it the maximal realization of random matrix universality consistent with \\Virasoro symmetry and modular invariance.}
\end{itemize}

\noindent This extends the sense in which AdS$_3$ pure gravity may be understood as an extremal theory: within the space of consistent theories of AdS$_3$ gravity, the statistical correlations among black hole microstates of pure gravity are as random as possible (in the sense of RMT). 

At early times, well before random matrix behavior sets in, theories of semiclassical Einstein gravity (in any spacetime dimension) saturate the chaos bound on the Lyapunov exponent of out-of-time-order correlators \cite{MSS}. What our analysis shows is that pure gravity is also maximally chaotic at late times: viewing $Z_\CJ(\t_1,\t_2)$ as computing coarse-grained spectral correlations of a dual microscopic CFT at large $c$, its level statistics are ``maximally random'' for a 2d CFT.\foot{To give some indicative timescales, the early-time Lyapunov chaos happens well before the scrambling time, logarithmic in entropy, whereas the RMT ramp occurs after the dip time, exponential in entropy.} In other words, random matrix universality enjoys a maximally extended regime of validity. This is a refinement, incorporating chaos, of the statement \cite{Hartman:2014oaa} that the Cardy density of states enjoys an extended regime of validity beyond the asymptotic regime $\D\rar\i$, all the way down to $\D\sim c/6$ (or possibly to $c/12$, for certain exceptional theories), in a sparse CFT at leading order in large $c$. 

Rephrasing this slightly, we have recovered the CJ wormhole amplitude as an extremal two-copy partition function. In particular, the CJ wormhole amplitude is the unique solution to the following large $c$ CFT bootstrap problem: find a $\Zh(\t_1,\t_2)$ that {\it i)} admits a Poincar\'e sum representation of wormhole form \eqr{poinc}, and {\it ii)} reproduces the double-scaled RMT SFF at late times $T$ for any fixed $T/\beta$. The first criterion specializes to large $c$, though not to pure gravity specifically, by imposing the geometric structure of semiclassical wormholes (see Subsection \ref{farey}). Upon imposing the SFF of RMT, Virasoro symmetry and modular invariance take care of the rest, leading to $Z_\CJ(\t_1,\t_2)$.\foot{The CJ wormhole was also ``bootstrapped'' in \cite{Cotler:2020hgz} from a different set of conditions, the main one being a specific prescription for the zero mode volume in the two-boundary gravity path integral. The origin of that important factor is rather subtle \cite{Collier:2023fwi, Yan:2023rjh, Eberhardt:2022wlc}.} This is similar in spirit to the bootstrap approach to four-point functions in AdS$_5 \x S^5$ supergravity in \cite{Rastelli:2017udc}.

\sssec{Comments}

\sssec*{Ensembles, RMT wormholes and ETH wormholes }

Is semiclassical pure gravity dual to an ensemble of CFTs? A more grounded statement,  supported by explicit computations in the literature, is that saddle point partition functions of semiclassical pure gravity are dual to ensembles of CFT data. This appears to hold for saddles of arbitrary boundary topology. On the other hand, we are able to interpret $Z_\CJ(\t_1,\t_2)$, a bulk {\it off-shell} wormhole amplitude encoding spectral data alone, in terms of a {\it microscopic} CFT. This provides a realization of the large $c$ notions articulated in \cite{Schlenker:2022dyo}: indeed, our formalism amounts to a dynamical mechanism of the ``apparent averaging'' of \cite{Schlenker:2022dyo} in AdS$_3$/CFT$_2$, by showing how wormholes can emerge from the high-energy spectral statistics of a large $c$ limit of a family of chaotic CFTs. It is satisfying that the torus wormhole admits a clean microscopic interpretation in bona fide CFT$_2$, without requiring an ensemble interpretation (though it does not prohibit one, if desired). 

Note that, in contrast, saddle point wormholes in AdS$_3$ gravity capture different physics. Unlike what one might call ``RMT wormholes'' that encode spectral statistics, such as the CJ wormhole, on-shell wormholes \cite{Chandra:2022bqq,Chandra:2022fwi,Chandra:2023dgq,Collier:2023fwi} are ``ETH wormholes'': they are instead computing averaged matrix elements that are in turn fixed by leading-order spectral (Cardy) and OPE (DOZZ) asymptotics of Virasoro primaries \cite{Cardy:1986ie,Collier:2019weq,Belin:2020hea}, independent of level statistics.  All on-shell wormholes in AdS$_3$ gravity are of this type \c{Maldacena:2004rf}.\foot{As an illustrative example, for a two-boundary torus wormhole to be on-shell in gravity, one must instead consider a one-point wormhole with boundary operators $\O$ inserted, where $\D_\O< 2\xi$ is dual to a bulk matter field \c{Maldacena:2004rf}. The trace over Hilbert space is an OPE sum, and the result is fixed by Virasoro ETH asymptotics. Consistent with this, if one studies the late-time behavior of the ``one-point SFF'', the on-shell one-point wormhole does not produce a ramp \cite{Yan:2023rjh}.} The partition function of any such wormhole saddle can be reduced to a moment problem for a probability distribution of CFT data that encodes universal asymptotics \cite{Chandra:2022bqq},\foot{This is not fully crossing symmetric on average, but is (almost) $S$-invariant on average. Small violations of average $S$-invariance may be fixed using a matrix-tensor model with a potential controlled by the variance under $S$ \cite{jafferisetall}. For other proposals on AdS$_3$ gravity with a similar treatment of partial crossing symmetry, see \cite{Cotler:2018zff,Mertens:2022ujr}.} insensitive to chaos and the structure of black hole microstates. For this reason, on-shell AdS$_3$ wormholes and their possible ensemble interpretations are of a rather different nature than wormholes in JT gravity.

\sssec*{Quantum chaos from arithmetic chaos}
The CJ amplitude exhibits a linear ramp in all spin sectors \cite{Cotler:2020ugk}. Given the amplitude \eqr{cjspec}, we can compute the Eisenstein contribution to any Fourier mode $Z_\CJ^{(j_1,j_2)}(y_1,y_2)$. The necessary integral was performed in \cite{Haehl:2023tkr} for $j_1>0$, yielding 
\es{}{ 
Z_\CJ^{(j_1,j_2)}(y_1,y_2)\big|_{\rm Eis} \approx \lambda_{j_1} \delta_{j_1,j_2} e^{-2\pi(j_1y_1+j_2y_2)}\sqrt{\frac{y_1y_2}{y_1+y_2}}+\ldots, \qquad (y_1,y_2\rightarrow \infty)
}
where $\lambda_{j_1}$ are $\cO(1)$ numerical factors. After analytic continuation and stripping off a factor $\sqrt{y_1y_2}$ (see \eqr{KbT}), this gives an $\cO(1)$ contribution to the SFF. Therefore, the late-time linear ramp at nonzero spin comes from the cusp form contribution alone. This observation establishes a relation between the \textit{arithmetic} chaos of Maass cusp forms and \textit{quantum} chaos of AdS$_3$ pure gravity as defined by RMT. \footnote{This connection has been developed further in \cite{Haehl:2023wmr,Haehl:2023mhf}. }

\sssec*{On $c$-independence of wormhole amplitudes}
The CJ wormhole amplitude is independent of Newton's constant $G_N = 3L_{\rm AdS}/2c$. This is consistent with the fact that it is not a semiclassical saddle in the bulk, but otherwise calls for a fundamental explanation (in contrast to the JT double-trumpet, whose $\O(1)$ scaling follows from the topological expansion of 2D gravity). 

Our formalism gives an indirect explanation of this fact: it is a consequence of the linear ramp in the SFF and the ``wormhole'' form of the amplitude. In short, if there were a $c$-dependent factor in the wormhole amplitude, it would violate random matrix universality.  Demanding a linear ramp in the scalar SFF with RMT coefficient implies \eqr{ramppole}, a $c$-independent condition. In general, given a family of chaotic CFTs $\{\mathcal{T}_c\}$ which admits a $c\rar\i$ limit, any dependence on $c$, or other parameters (e.g. exactly marginal couplings) which we collectively denote as $\l$, must sit in an additive correction term,
\e{}{\mathcal{R}^{(c)}(z|\l) = {\mathsf{C}_{\rm RMT}\o 2\pi}{1\o 1+z} + \delta \mathcal{R}^{(c)}(z|\l)}
At large $c$, this is finite. Since any wormhole amplitude $Z_\WH(\t_1,\t_2)$ of the form \eqr{zwh} is fixed completely by this quantity, as shown in Section \ref{s4}, the entire amplitude is independent of $c$. What sets the CJ wormhole apart is that the correction exactly vanishes -- a property of MaxRMT.

\sssec*{Wormholes in AdS$_3\x \cM$ gravity}\label{sec:otherworm}

In a theory of AdS$_3$ pure gravity, corrections to MaxRMT behavior should be non-perturbative in $G_N$. We will discuss these in Subsection \ref{sec:rmtboot}. On the other hand, in theories of gravity coupled to matter, such as AdS$_3 \x \cM$ string or M-theory compactifications with large extra dimensions $\cM$, the corrections will generically be $\O(1)$. These theories should not have MaxRMT statistics. The RMT wormholes with $T^2 \x I$ topology will receive large matter contributions to the quantity $\delta \mathcal{R}^{(c)}(z|\l)$ written above, thus modifying the geometry on macroscopic scales. By virtue of diffeomorphism invariance their partition functions should take the form \eqr{zwh} and, being chaotic, preserve the linear ramp of the SFF, but we expect their overlaps $f_\WH(s)$ to differ substantially from $f_\CJ(s)$.

We may parameterize these overlaps as
\e{whoverlap}{f_{\WH}(s) = f_{\CJ}(s) g(s)}
with $\mathcal{R}_\WH(z)$ and $G(z) = \delta(z-1) + \delta G(z)$ the corresponding inverse Mellin transforms, respectively. Mellin convolution implies
\e{}{\mathcal{R}_\WH(z) = {1\o\pi}{1\o 1+z} + \delta \mathcal{R}_\WH(z)}
where
\e{deltaFWH}{\delta \mathcal{R}_\WH(z) := \int_0^\i dy {\delta G(y)\o 1+z y}}
This correction must be finite at $z=-1$. Singularities of $\delta \mathcal{R}_\WH(z) $ for $z\in\mathbb{C}\backslash \{-1\}$ can arise either from delta functions (and possibly their derivatives) in $\delta G(y)$ on the contour, or poles off the contour. These singularities must not violate the physical requirement that $\mathcal{R}_\WH(z) $ be regular at $|z|= 1$ (no spurious poles in the SFF) and $z\in \R_{\geq 0}$ (finite Euclidean low-temperature limit).\foot{In SFF kinematics, a pole $1/(z+z_*) \supset \delta\cR_\WH(z)$ gives a decaying contribution $\sim 1/T$ as $T\rar\i$. Note that these are not the same as the non-ergodic modes/massive modes of \cite{Altland:2017eao,Altland:2020ccq}, which decay as $T e^{-\# T}$ in the SFF. Those are exponential corrections to the dip, whereas ours are power-law corrections to the ramp.} 

In Appendix \ref{app:toywh}, we provide a toy family of RMT wormholes in AdS$_3 \x \cM$ gravity, for which both $f_\WH(s)$ and $\mathcal{R}_\WH(z) $ can be computed explicitly. The spectral overlaps are of the form \eqr{whoverlap} with 
\e{}{g(s) = {\mathsf{C}_\RMT\o 4}\big(\z(s+p) + \z(p+1-s)\big)\,,\quad \text{where}~~ p\in\Z_+\,,}
while their dual functions $\mathcal{R}_\WH(z) $ in \eqr{whharmfinal} are harmonic sums over an infinite set of simple poles, with the correctly normalized RMT ramp \eqr{ramppole} at $z\rar-1$. We note here that $g(s)$ non-trivially satisfies the falloff condition \eqr{lineardivfinal}, which follows from the asymptotic $|\z(\s+i\w)| \rar 1$ as $\w\rar\i$ for any $\s>1$. 

\ssec{Black hole microstates of AdS$_3$ pure gravity}\label{sec:BHmicro}

The above framework actually lets us extract {\it non}-coarse-grained microstructure of the black  hole spectrum of AdS$_3$ pure gravity by factorizing the CJ wormhole; see Figure \ref{fig:fact}. In particular, we can deduce what $\Zs(\t)$ gives rise to the correlations encoded in $Z_\CJ(\t_1,\t_2)$ upon coarse-graining by using the powerful fact, highlighted earlier, that the 2d CFT trace formula expands $\Zs(\t)$ over a complete eigenbasis. Combined with the high degree of symmetry of wormhole amplitudes, the result is nearly unique.

We wish to solve
\e{}{{Z_{\rm CJ}(\t_1,\t_2) = \cP_{\text{Hecke}}\[Z_{\rm RMT}(\t_1)Z_{\rm RMT}(\t_2)\]}}
That is, $Z_\RMT(\t)$ is the $Z_\spec(\t)$ that generates the CJ wormhole upon coarse-graining. We deduce the spectral overlaps $(Z_\RMT,E_s)$ and $(Z_\RMT,\phi_n)$, and hence $Z_\RMT(\t)$, from \eqr{cjspec}. The cusp form overlaps are fixed by \eqr{cuspeisequal} in terms of the Eisenstein overlap, 
because of the wormhole form \eqr{zwh} of the CJ amplitude. The Eisenstein overlap, in turn, is easily determined using \eqr{cjspec*} in which the reflection symmetry is manifest:
\e{zrmteis}{\boxed{\{Z_{\rm RMT},E_s\}^2 = {1\o \z(2s)\z(2-2s)}}}
On the critical line, $\{Z_{\rm RMT},E_{\half+i\w}\}^2 = {|\z(1+2i\w)|^{-2}}$. So a putative CFT dual to AdS$_3$ pure gravity has a spectral partition function $\Zs(\t) \approx Z_{\rm RMT}(\t)$ where 
\es{zrmt}{Z_{\rm RMT}(\t) &= \int_{\ccrit} (Z_{\rm RMT},E_s) E_s(\t) + \sum_{n=1}^\i |(Z_{\rm RMT},E_{s_n})| \phi_n(\t)\,,}
with overlaps given above. In other words, by factorizing the wormhole we have extracted a contribution to the torus partition function of a CFT dual to pure gravity,
\es{Zgravrmtapprox}{Z_{\rm grav}(\t) \approx Z_{\rm MWK}(\t) + Z_\RMT(\t) \,.}
This holds up to exponentially small corrections in $c$, discussed further in Subsection \ref{sec:rmtboot}. 

$Z_\RMT(\t)$ is almost, but not quite, uniquely determined: its overlaps are fixed up to overall signs. This is inherent in our derivation-by-factorization of the two-boundary wormhole, which is quadratic in $Z_\RMT(\t)$. One would thus like to identify a computation, or a physical criterion, which selects one branch of the square root. While we so far lack a proof, the trace formula \eqr{eq:2dtrace} suggests that the positive branch may be the correct one, both for the Eisenstein and cusp form overlaps, because the spectral overlaps play the role of orbit densities: just as semiclassical periodic orbit densities must be positive for physical systems, one may reasonably expect the same to hold for their 2d CFT analog. A full determination of these signs is important. Either way, it is quite intriguing that $\{Z_\RMT,E_s\}$ is {\it sign-definite} on the critical line, a feature which is surprising from the point of view of the individual functions themselves: neither $Z_\RMT(\t)$ nor $E_s(\t)$ obeys any manifest sign constraint for $\t\in \cF$ (for example, the integral of $E_s(\t)$ over $\cF$ vanishes). This appears to be a special feature of wormhole amplitudes under the CFT identification \eqr{fareyid}.\foot{One easily verifies that the positivity holds for Narain wormholes \c{Collier:2021rsn} as well. Conversely, one may ask whether imposing the positivity of $f_\WH(\half+i\w)$, as in the identification \eqr{fareyid}, may be used to constrain the allowed seeds of Poincar\'e sums \eqr{poinc} representing wormhole amplitudes in gravity (or, possibly, whether not all reasonable wormholes actually possess such positivity). We thank David Berenstein for this question, and Scott Collier for a subsequent discussion.} It suggests that the chaotic spectrum $Z_\spec(\t)$ of any large $c$ CFT must obey such a sign constraint.

\begin{figure}[t]
\centering
{
\subfloat{\includegraphics[scale=0.7]{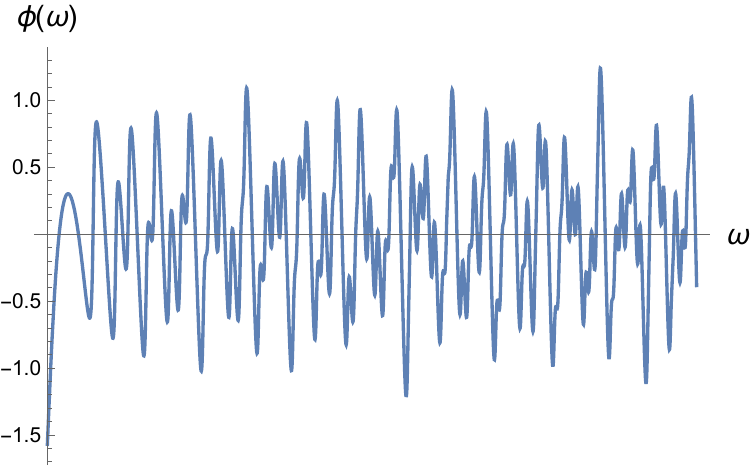}}
}
\caption{A plot of the Riemann zeta phase $\phi(\w) = \text{arg}\big({\z(1+2i\w)}\big)$ for $\w\in[0,100]$.}
\label{zetaphasefig}
\end{figure}

We view $Z_\RMT(\t)$ as a half-wormhole of AdS$_3$ pure gravity: it is a single-boundary, non-self-averaging quantity which, when squared and glued appropriately, generates the smooth spacetime wormhole. See Figure \ref{fig:halfWH}. As is by now hopefully clear, the right procedure is prescribed by the 2d CFT trace formula, gluing the half-wormholes in $\sl$ spectral space. The detailed form of $Z_\RMT(\t)$ is ``jiggly,'' with spectral overlaps characterized by erratic fluctuations. We can make this manifest by writing the Eisenstein overlap on the critical line as (choosing the positive branch for definiteness) 
\e{}{(Z_{\rm RMT},E_\crit) = {1\o \sqrt{\cosh(\pi \w)}}e^{i(\chi(\w) + \phi(\w))}}
where
\e{}{\chi(\w) = \text{arg}\(\pi^{-i\w}\G\(\half+i\w\)\)\,,\quad \phi(\w) = \text{arg}\big(\z(1+2i\w)\big)}
In addition to the regular ``background'' phase $\chi(\w)$, the overlap is dressed by the ``Riemann zeta phase'' $\phi(\w)$, a quantum-chaotic object: $\phi(\w)$ oscillates erratically along the critical line (see Figure \ref{zetaphasefig}), producing the non-self-averaging behavior characteristic of half-wormholes.\foot{The Riemann zeta phase also appears in the Gutzwiller model of chaotic scattering on the ``leaky torus'' as a phase in the S-matrix,  producing poles at the locations of non-trivial Riemann zeros \cite{leakytorus}.} Upon squaring and projecting $Z_\RMT(\t)$ to form the CJ wormhole, the phase cancels, leaving the smooth integrand of \eqr{cjspec}.

This dovetails nicely with expectations of how the wormholes and half-wormholes of 2D gravity should generalize to higher dimensions. The factorized product $Z_{\rm RMT}(\t_1)Z_{\rm RMT}(\t_2)$ explicitly contains the wormhole $Z_\CJ(\t_1,\t_2)$, isolated by the Hecke projection $\cP_\Hecke$; the terms that are projected out, presumably to be thought of as ``linked half-wormholes" of some kind, combine with the wormhole to restore factorization. We can say this in the closely related but some what more general ``broken cylinders'' picture of \cite{Saad:2021uzi} (see Sections 3.6 and 5.2). A broken cylinder in the ``$b$ basis'' is geometrically a trumpet with a geodesic boundary carrying a random boundary condition, $\Psi(b)$. Two broken cylinders are glued into a smooth wormhole by averaging over these boundary conditions, integrating over $b$ with measure 
\e{Psimeasure}{\<\Psi(b)\Psi(b')\>_\Psi = {1\o b}\delta(b-b')}
The random function $\Psi(b)$ is a bulk effective description of coarse-graining the dual theory. There is an intuitive analogy between AdS$_2$ broken cylinders and AdS$_3$ half-wormholes:
\begin{equation*}
\begin{split}
 b \qquad &\longleftrightarrow \qquad \w\\
\Psi(b) \qquad &\longleftrightarrow \qquad e^{i \phi(\w)} \\
\eqr{Psimeasure} \qquad &\longleftrightarrow \qquad \cP_\Hecke
\end{split}
\end{equation*}
We recall that $\cP_\Hecke$ projects onto the diagonal  the factorized double sum over eigenfunctions,
\e{}{\cP_\Hecke\[\sumint_\w \tilde f_\w\, \psi_\w(\t_1) \sumint_{\w'} \tilde f^*_{\w'}\, \psi^*_{\w'}(\t_2)\] = \sumint_\w |\tilde f_\w|^2 \psi_\w(\t_1) \psi^*_{\w}(\t_2)}
where we used the unified notation \eqr{specunified}. An interesting aspect here is that $\w$ is not manifestly a geometric quantity in AdS$_3$, unlike the geodesic length $b$ in AdS$_2$.

\begin{figure}[t]
\centering
\vspace{.4in}
{
\subfloat{\includegraphics[scale=0.21]{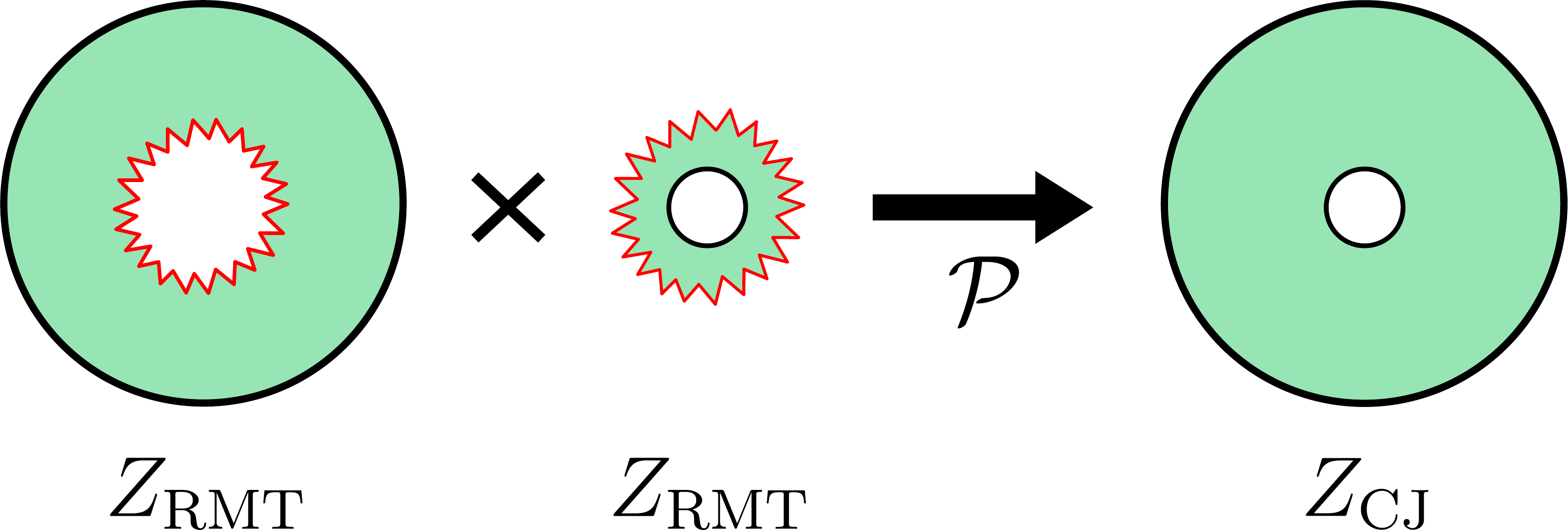}}
}
\vspace{.2in}
\caption{An alternate depiction of Figure \ref{fig:halfWH} emphasizing the analogy to 2D gravity, with the extra $S^1$ direction suppressed.  Each half-wormhole is viewed as a deformed annulus times a circle, in which one of the annular boundaries is replaced by an erratic/random boundary. In the spectral representation this boundary is precisely the Riemann zeta phase $e^{i\phi(\w)}$. The CJ wormhole is formed by gluing along the deformed boundaries. We have drawn the second deformed annulus ``inside out'' for visual aid.}
\label{fig:defannuli}
\end{figure}

This picture emphasizes that, while the CJ wormhole is topologically an annulus $\x S^1$, it should be viewed as a gluing of two constituent topologies, namely, of two ``deformed solid tori'' (i.e. deformed annuli times a circle). See Figures \ref{fig:halfWH} and \ref{fig:defannuli}. This is somewhat schematic because we do not yet have a bulk geometric derivation of $Z_\RMT(\t)$. What we do know is that $Z_\RMT(\t)$ probes the part of the black hole Hilbert space that is precisely {\it not} captured by semiclassical saddles: whatever off-shell configurations should be included in the path integral with a single torus boundary, they must generate this term.  

We note that the Riemann zeta phase $\phi(\w)$ is the phase for any $(\Zs,E_\crit)$, as one can show from its functional equation in $\w$. This supports our earlier half-wormhole characterization of $\Zs(\t)$ in the discussion of Hecke projection (Subsection \ref{sec:heckesum}), even absent a large $c$ limit. For this reason, $\Zs(\t)$ is closely analogous to the ``brane one-point function'' $Z_{\rm brane}(\b)$ in the 2D gravity model of \cite{Blommaert:2021fob}, an additive correction to the JT disk partition function that is non-perturbatively exact in the entropy $S_0$. What we are finding is that this same mechanism operates in the bona fide AdS$_3$/CFT$_2$ setting in broad generality, with $\Zs(\t)$ playing the role of the brane. 

\sssec{Pure gravity path integral}\label{sec:rmtboot}

The identification of $Z_\RMT(\t)$ augments our knowledge of $Z_{\rm grav}(\t)$, the primary partition function of semiclassical AdS$_3$ pure gravity with a single torus boundary, as shown in \eqr{Zgravrmtapprox}. Any primary partition function $Z_p(\t)$ for which the vacuum is the only Virasoro primary with\foot{Recall our definitions $t = \text{min}(h,\hb) - \xi$ and $\xi=(c-1)/24$.}  $t<0$ can, in general, be written 
\e{zpmwk}{Z_p(\t) = Z_{\rm MWK}(\t) + \Zs(\t) }
By leveraging $Z_\CJ(\t_1,\t_2)$, we have shown that $\Zs(\t)$ is not a free function in semiclassical pure gravity, leading to \eqr{Zgravrmtapprox}. Including further corrections, 
\e{Zgravrmt}{{Z_{\rm grav}(\t) = Z_{\rm MWK}(\t) + Z_\RMT(\t) + \delta Z_\RMT(\t)}}
There are three contributions:
\begin{itemize}
\item $Z_{\rm MWK}(\t)$ sums over smooth bulk saddles $\cM$ with $\p\cM = T^2$.
\item $Z_\RMT(\t)$ encodes the leading RMT fluctuations of the black hole spectrum.
\item $\delta Z_\RMT(\t)$ gives possible corrections to $Z_\RMT(\t)$.
\end{itemize}
This refines the usual point of view taken in the modular bootstrap approach to $Z_{\rm grav}(\t)$, by adding the (off-shell) fine-grained spectral fluctuations of the black hole/heavy spectrum to the sum over (on-shell) BTZ black hole geometries. The ``brane'' terms $Z_\RMT(\t) + \d Z_\RMT(\t)$ are exponentially suppressed relative to the sum over saddles.

The corrections $\delta Z_\RMT(\t)$ are important, even though they are expected to be exponentially small in $c\sim 1/G_N$. Were they perturbative instead, they would lead, upon squaring and projecting, to $1/c$ corrections to the CJ wormhole: this would violate reasonable expectations inferred from the genus expansion of 2D gravity for the structure of AdS$_3$ wormholes, as well as the method of computation of \cite{Cotler:2020ugk}.\foot{It should again be emphasized, given the unusual nature of the computation \cite{Cotler:2020ugk}, that it is not known how (or whether) higher topologies beyond $T^2 \x I$ organize themselves in a small $G_N$ expansion, and what their moduli dependence is. However, they should be present, and we are giving some expectations for their scaling.} Instead, $\delta Z_\RMT(\t)$ should represent ``higher genus'' terms that encode UV data of a (say) strongly-coupled string theory, or M-theory, compactification. Indeed, it could well be the case that the string landscape hosts many compactifications to AdS$_3$ pure gravity which share a parametrically identical spectral gap, differing only in the detailed correlations among black hole microstates: the quantity $\d Z_\RMT(\t)$ encodes exactly these contributions. When glued to form a wormhole, these would give $\O(e^{-\a c^{\b}})$ corrections to $Z_\CJ(\t_1,\t_2)$ for some $\a,\b>0$, perhaps corresponding to wormholes with torus boundary but higher topology in the interior.

Actually, not only should such corrections be present on general grounds, they {\it must} be present in a consistent CFT: if $\d Z_\RMT(\t)=0$, then the partition function \eqr{Zgravrmt} is non-unitary. To see this, recall that the MWK partition function $Z_{\rm MWK}(\t)$ contains an exponentially large negative density of states in a near-extremal limit of $t \rar 0_+$ and $j\rar\i$ \cite{Benjamin:2019stq} for odd $j$: suppressing power-law prefactors,
\e{mwkneg}{\int_0^{e^{-\half S_0(j)}} dt\, \rho_{{\rm MWK},j}(t) \approx (-1)^j \,e^{{1\o4}S_0(j)}\,,\quad \text{where} \quad S_0(j) := 4\pi\sqrt{\xi |j|}}
is the extremal spin-$j$ Cardy entropy. But this cannot be cured by $Z_\RMT(\t)$, which is independent of the central charge. 

The above results altogether lead us to a rather interesting conclusion. If a unitary $Z_{\rm grav}(\t)$ exists, the non-perturbative corrections $\delta Z_\RMT(\t)$ must not only be present, but must resum in some manner to give a contribution to the density of states below the semiclassical black hole threshold. Were this not the case, then for all values of the modulus  $\t$ we could safely drop $\delta Z_\RMT(\t)$ in the large $c$ limit; but as reviewed above, the resulting partition function would be non-unitary. To avoid this order-of-limits issue between large $c$ and near-extremality, the corrections $\delta Z_\RMT(\t)$ should resum to shift the black hole threshold slightly below $t=0$, giving {\it non}-square-integrable contributions to $Z_{\rm grav}(\t)$.\foot{We thank Daniel Jafferis for a discussion.} In other words, the BTZ black hole threshold is strictly below $t=0$.

On the CFT side, the large $c$ modular bootstrap quest is thus transformed to the following: {\it find $\delta Z_\RMT(\t)$ such that \eqr{Zgravrmt} is unitary.} In the bulk, $\delta Z_\RMT(\t)$ is an off-shell contribution to the AdS$_3$ path integral that presumably captures non-perturbative degrees of freedom. 

\sssec{On the Maxfield-Turiaci proposal}\label{sec:MT}

There is one proposal in the literature for at least some such off-shell contributions, due to Maxfield and Turiaci \cite{Maxfield:2020ale}, which realizes precisely the resummation mechanism described above. They proposed that $Z_{\rm grav}(\t)$ should include a sum over a class of Seifert manifolds. These smooth 3-manifolds are $S^1$ fibrations over a disc $D_2$ with orbifold points.\foot{An excellent review of Seifert manifolds may be found in \cite{Closset:2018ghr}.} Their inclusion in $Z_{\rm grav}(\t)$ was motivated by studying the near-extremal limit, where the AdS$_3$ theory admits an effective dimensionally-reduced description as JT gravity coupled to conical defects. Maxfield and Turiaci pointed out that, from the dimensionally-reduced point of view, the MWK negativity \eqr{mwkneg} is cured upon summing over multiple-defect configurations, not just the saddle points. This shifts the density near extremality to\foot{The result of \cite{Maxfield:2020ale} formally holds for $j\rar\i$, but the functions appearing in the calculation depend on the combination $\xi j$, which partially justifies an extrapolation to all nonzero spins at large $\xi$.}
\e{rhoMT}{\rho_{{\rm MT},\,j}(t) \approx e^{S_0(j)} \sqrt{2(t-t_0(j))}\,,\quad \text{where}~~t_0(j) \approx -{1\o 2(2\pi)^2}(-1)^j e^{-\half S_0(j)}}
plus exponentially small corrections to the shift. The corrections, which are required for modular invariance of the AdS$_3$ result, would come from including the full set of defects. More explicitly, each species of JT defect of \cite{Maxfield:2020ale} is labeled by an integer $q$ (corresponding to a $\Z_q$ orbifold point on the base $D_2$), and an associated coupling 
\e{}{\l_{1/q} \sim e^{-S_0(j)\({1-{1\o q}}\)}}
In \cite{Maxfield:2020ale}, the approximation $t_0(j) \approx \l_{1/2}$ is made: that is, the sum is only over the gas of $q=2$ defects. The full sum over arbitrary numbers of arbitrary species of defects was proposed to have an $S^1$ uplift to a sum over Seifert manifolds,
\e{Zgrav}{Z_{\rm MT}(\t) = Z_{\rm MWK}(\t) + Z_{\rm Seifert}(\t) + \ldots}
The $\ldots$ represents other, non-Seifert contributions which near-extremal physics does not constrain. $Z_{\rm Seifert}(\t)$ has not been computed.

There has been ample reason to believe that the Maxfield-Turiaci proposal is correct. Our argument above supports this further. If it is correct, the Seifert manifolds of \cite{Maxfield:2020ale} should contribute to $Z_\RMT(\t) + \delta Z_{\rm RMT}(\t)$. Understanding how this works is a very worthwhile direction for future work. Let us take some initial steps toward understanding this connection. First, recall that $Z_\RMT(\t)$ is independent of $c$, and hence of the entropy. A generic defect configuration term has coefficient
\e{Sscaling}{e^{S_0(j)}\prod_{q=2}^\i (\l_{1/q})^{n_q} \sim e^{S_0(j)\Big(1-\sum_q n_q\(1-{1\o q}\)\Big)}\,,\quad n_q \in \Z_{\geq 0}} 
It is rather tidy that the only term which is independent of entropy is the $q=n_2=2$ term. So we see that $Z_\RMT(\t)$ should include only the Seifert manifold with $n_2=2$ exceptional fibers, each over a $\Z_2$ orbifold point on the base. On the other hand, $Z_\RMT(\t)$ is $SL(2,\Z)$-invariant, whereas the $q=n_2=2$ Seifert partition function is not. This confirms that $Z_{\rm grav}(\t)$ should include other, non-Seifert, off-shell topologies which give an $\O(1)$ contribution, as anticipated in \cite{Maxfield:2020ale}. (More explicitly, in the Maxfield-Turiaci realization of \eqr{Zgravrmt}, $\delta Z_\RMT(\t)$ would include the other Seifert manifolds that are further entropically suppressed, while $Z_\RMT(\t)$ would include non-Seifert manifolds.)

\sec{Future Directions}\label{sec:s7}

We have presented a framework for quantifying random matrix statistics of microscopic 2d CFT spectra, manifestly compatible with the requisite symmetries. The identification of a 2d CFT trace formula suggests deep and robust connections with random matrix theory and chaotic quantum systems. While we focused mostly on the diagonal approximation to the product of partition functions, a clear path forward is to understand how the full non-perturbative structure of RMT embeds itself in the dynamics of high-energy states in 2d CFTs and their dual black holes of AdS$_3$ gravity, and how to excavate it within our spectral framework. 

Let us highlight a few specific directions of interest.

One target therein is to understand how encounter theory \cite{Sieber2001,PerOrb1,PerOrb2,haake}, which initiates the transition from ramp to plateau in the SFF of quantum chaotic systems, should be phrased in the language of the $\sl$ spectral decomposition. This subject was nicely reviewed and ported over to the JT gravity setting in \cite{SSSY}. The goal vis-\`a-vis 2d CFT is to go beyond the diagonal approximation to $\Zs(\t_1)\Zs(\t_2)$, organizing the off-diagonal terms in a way that mimics encounter theory.  Our framework gives a clear meaning to ``off-diagonal terms'': they are correlations among $\sl$ eigenfunctions with unequal spectral parameters, $\w_1 \neq \w_2$.  The question is how to develop a systematic approach to these correlations. and whether they give rise to different types of corrections to (say) the SFF. 

A related, more specific starting point is the following. The SFF of double-scaled random matrix ensembles admits a ``$\uptau$-scaling limit'' in which $T\rar\i$ with $\uptau := T e^{-S_0}$ held fixed. The expansion in $\uptau\ll1$ has been reproduced to all orders by encounter theory \cite{PerOrb1,PerOrb2}.   Is there a $\uptau$-scaling limit in 2d CFT? If so, how do we compute it in our formalism? Understanding this could help to extend the meaning of MaxRMT to the non-perturbative level \cite{Keating1993,BogomolnyKeat,PerOrb3,PerOrb4,MVBerry_1990,KeatingRS,BerryKeating1992,Keating2007}.

It is interesting to notice that there are two distinct classes of corrections to the SFF in the $SL(2,\Z)$ spectral framework: corrections to the spectral overlaps themselves, and the inclusion of terms beyond the diagonal approximation. The former give corrections to the SFF beyond the strict ramp term, even {\it without} incorporating off-diagonal terms in $\Zs(\t_1)\Zs(\t_2)$. The latter are obviously crucial to reproduce the constant late-time plateau; but in the interpolation from ramp to plateau, the two types of corrections should in principle combine to give the full SFF. It is  important to disentangle and stratify these various effects, and to connect them to non-perturbative corrections to the CJ wormhole of AdS$_3$ pure gravity. 

The condition \eqr{lineardivfinal} for the SFF to have an RMT linear ramp at $T\gg\b$ is an asymptotic condition on $Z_\spec(\t)$ in $\sl$ spectral space. Given the torus partition function of a CFT, this condition can be checked unambiguously. If we accept that generic irrational CFTs are chaotic, then to the extent that one believes (a CFT version of) the Bohigas-Giannoni-Schmit conjecture \cite{BGS} that ties chaos to random matrix statistics, the condition \eqr{lineardivfinal} should be obeyed by generic irrational CFTs \cite{Kudler-Flam:2019kxq}. It is worth asking whether this condition can be tied to other, ideally low-energy, properties of irrational CFT data. For example, it is not known whether a nonzero primary twist gap above the vacuum state is sufficient to imply that the CFT is chaotic in an appropriate sense.\foot{It is not clear to us that the twist gap is a relevant criterion. A twist gap can vanish due either to extra conserved currents, or an accumulation to zero twist at asymptotically large spin. One should always use the maximally extended chiral algebra to formulate the notion of twist gap \cite{Moore:1988ss}, so we ignore the first possibility. But we are not aware of any solid argument that a zero-twist accumulation point implies an absence of chaos/RMT statistics or vice-versa (even though it does invalidate standard Cardy-type arguments about spectral asymptotics).} Perhaps one can determine whether the twist gap is related to \eqr{lineardivfinal}. 

On the bootstrap front, it is of clear interest to further pursue the large $c$ modular bootstrap program \cite{Hellerman:2009bu,Friedan:2013cba,Collier:2016cls,Afkhami-Jeddi:2019zci,Hartman:2019pcd} for the partition function \eqr{Zgravrmt}. This raises the larger question of how to incorporate RMT spectral statistics into modular and conformal bootstrap algorithms. 

Our formalism is not restricted to studying partition functions. One immediate generalization is to study coarse-grained products of torus one-point functions of local primary operators, $\<\O\>_\t$. One-point functions are modular-covariant and, like partition functions, can be dressed and massaged to be modular-invariant and square-integrable on $\cF$, hence admitting a $\sl$ spectral decomposition and thus a trace formula. The spectral overlaps encode the OPE statistics. In the AdS$_3$ bulk, the projections $\cP_\Hecke[\<\O\>_{\t_1}\<\O\>_{\t_2}]$ would be holographically dual to ``one-point RMT wormholes'': that is, off-shell connected configurations with a profile for the field dual to $\O$ threading the wormhole, with local sources at each torus boundary. Essentially this quantity was studied in the 2D gravity context in \cite{Saad:2019pqd,Jafferis:2022wez,Jafferis:2022uhu}. These generalizations of the CJ wormhole are not the Maldacena-Maoz one-point ETH wormholes \cite{Maldacena:2004rf}, instead incorporating the level statistics of the states in the OPE. One can formulate a straightforward one-point generalization of the CJ wormhole \eqr{cjspec} that combines the RMT level statistics with Virasoro ETH asymptotics; this will be presented elsewhere.

We have introduced a notion of MaxRMT in 2d CFTs. How does it generalize to higher dimensions? A natural conjecture that we would like to make is that {\it the black hole spectrum of semiclassical Einstein gravity in AdS$_{D\geq 3}$ exhibits MaxRMT statistics.} The meaning of this in $D>3$ is somewhat vague. In the AdS$_3$/CFT$_2$ context, the MaxRMT nature of pure gravity is a {\it quantitative} statement, built on the infrastructure of the $\sl$ spectral decomposition as a trace formula. That allowed us to fully process the symmetries of the spectrum, leaving only those spectral correlations that are unrelated by symmetry; MaxRMT then states that these correlations are exactly those of RMT. This latter characterization in principle generalizes MaxRMT to higher dimensions. What, then, is the CFT$_d$ trace formula for the density of states? Given such a formula, one could construct a diagonal approximation to the two-point correlator and proceed as we did here. It would be very interesting to make this idea concrete, and to ask how MaxRMT relates to the familiar holographic CFT conditions of large $N$ and large higher-spin gap. 

Our work may also have something to say about lower dimensions. In JT gravity an analogous description of wormholes as emergent from non-disordered systems via a trace formula has been anticipated since  \cite{Saad:2019lba}. Given the existence of an emergent Schwarzian sector in the near-extremal limit of 2d CFTs \cite{Ghosh:2019rcj,Mertens:2017mtv,Maxfield:2020ale}, it seems reasonable to seek a dimensional reduction of the CJ wormhole that yields the JT double-trumpet. One could then hope to translate the diagonality of the CJ wormhole \eqr{cjspec} to a diagonal approximation interpretation of the JT double-trumpet. In that case, identifying the reduction of $Z_{\rm RMT}(\t)$ would yield new microscopic information about the microstates of JT gravity in the form of a trace formula for $\rho(E)$. It would be interesting to compare it to other proposals for restoring factorization in JT gravity \cite{Harlow:2018tqv,Saad:2021rcu,Saad:2021uzi,Blommaert:2021fob,Marolf:2020xie, Johnson:2022pou, Altland:2022xqx}. 

\sec*{Acknowledgments}
We thank David Berenstein, Veronica Collazuol, Luca Iliesiu, Daniel Jafferis, Dalimil Mazac, Julian Sonner, Douglas Stanford, Yiannis Tsiares, Pierfrancesco Urbani and especially Scott Collier for helpful discussions. EP and GD thank the Kavli Institute for Theoretical Physics, Santa Barbara for support during the course of this work. EP thanks the ICTP Trieste, where this work was first presented, for hospitality. This research was supported by ERC Starting Grant 853507, and in part by the National Science Foundation under Grant No. NSF PHY-1748958.

\appendix

\section{Glossary}

The following is the short exact sequence of partition functions appearing in this work. All objects below are modular-invariant with respect to each independent argument. 

\begin{itemize}
[leftmargin=+1in]
\item [{$Z_p(\t)$}:~~] Virasoro primary-counting torus partition function.

\item [{$Z_\spec(\t)$}:~~] Spectral partition function, obtained by subtracting light states and their modular completions from $Z_p(\t)$. Admits an $\sl$ spectral decomposition.

\item [{$Z_\diag(\t_1,\t_2)$}:~~] Diagonal projection of the factorized product $\Zs(\t_1)\Zs(\t_2)$, defined by projection onto the kernel of $\D_{12} := \D_{\t_1}-\D_{\t_2}$, where $\D_\t$ is the hyperbolic Laplacian. Pairs eigenvalues of the Laplacian. 

\item [{$Z_\Hecke(\t_1,\t_2)$}:~~] Hecke projection of the factorized product $\Zs(\t_1)\Zs(\t_2)$, defined by projection onto the kernel of $T_j^{(12)} := T_j^{(\t_1)} - T_j^{(\t_2)}$, where $T_j^{(\t)}$ is a spin-$j$ Hecke operator for $SL(2,\Z)$, for all spins $j\in\Z_+$. Pairs eigenfunctions of the Laplacian. 

\item [{$Z_\WH(\t_1,\t_2)$}:~~] A name for $Z_\Hecke(\t_1,\t_2)$ when Eisenstein and cusp form overlaps are equal. Includes off-shell torus wormhole amplitudes in semiclassical AdS$_3$ gravity.

\item [{$Z_\CJ(\t_1,\t_2)$}:~~] Cotler-Jensen torus wormhole amplitude of AdS$_3$ pure gravity. An instance of $Z_\WH(\t_1,\t_2)$.

\item [{$Z_\RMT(\t)$}:~~] The $\Zs(\t)$ whose coarse-grained product generates $Z_\CJ(\t_1,\t_2)$. That is, a solution of $Z_\CJ(\t_1,\t_2) = \cP_\Hecke[Z_\RMT(\t_1)Z_\RMT(\t_2)]$. 

\end{itemize}

\section{Spectral decomposition on $L^2(\mathcal{F}\times \mathcal{F})$}\label{app:l2space}

We present  a few details regarding the functional spaces appearing in the discussion and their different spectral decompositions.  Consider the manifold $\cF=\HH/\sl$ and the associated space of square-integrable functions $f(\t) \in L^2(\cF)$ with respect to the Petersson inner product defined by the  measure $dx dy/y^2$. A  complete basis of functions for $L^2(\cF)$ is given by the eigenfunctions of the Laplacian: 
\es{}{
\Delta_{\tau}\psi_{\w}(\tau)=\l_\w \psi_{\w}(\tau), \qquad f(\t) = \sumint_\w \tilde f_\w \,\psi_\w(\t)}
where
\e{doublebasis}{
\psi_{\w}(\t)=\{\phi_0, E_\crit(\t),\phi_n(\t)\}
}
with  $\phi_0=\text{vol}(\cF)^{-\half} = \sqrt{3/\pi}$ and $n\in\Z_+$. The overlaps are denoted here by $\tilde f_{\w} = (f(\t),\psi_\w(\t))$, with implicit degeneracy for every $\w=\w_n$ for which a Maass cusp form exists. The tensor product  space $L^2(\cF)\otimes L^2(\cF)$  of factorized functions $f(\tau_1)g(\tau_2)$ admits a simple decomposition which is just given by  the product of the individual decompositions of $f(\tau_1)$ and $g(\tau_2)$. 

Consider now the space of square integrable functions of two moduli $\tau_1,\tau_2$, i.e. $L^2(\cF_1\times\cF_2)$. A function $f(\tau_1,\tau_2) \in L^2(\cF_1\times\cF_2)$ is in general not factorizable. The functions appearing in the body of the text ($Z_\diag, Z_\Hecke, Z_\WH, Z_\CJ$) are elements of this space. The Stone-Weierstrass theorem \cite{Canzani} states that the subspace of factorized functions $f(\tau_1)g(\tau_2)\in L^2(\cF)\otimes L^2(\cF)$ is dense in $L^2(\cF_1\times\cF_2)$. We then have a complete basis of functions for $L^2(\cF_1\times\cF_2)$ given by:
\es{ft1t2decomp}{ 
f(\tau_1,\tau_2)=\sumint_{\w_1,\w_2} \tilde f_{\w_1,\w_2} \,\psi_{\w_1}(\t_1)\psi_{\w_2}(\t_2)
}
The overlaps $\tilde f_{\w_1,\w_2}$ are computed by taking the inner product  with the basis elements, with the ``double inner product'' defined by the product metric
\es{}{ 
(f,g):= \int_{\cF_1} \frac{dx_1dy_1}{y_1^2}\int_{\cF_2} \frac{dx_2dy_2}{y_2^2} f(\tau_1,\tau_2)\bar{g}(\tau_1,\tau_2),
}

As an example,  the constant term in the spectral decomposition can be computed by taking residues of a double Rankin-Selberg transform:
\es{}{ 
\la f(\tau_1,\tau_2)\ra= \frac{1}{\text{vol}(\cF)^2} \int_{\cF_1} \frac{dx_1dy_1}{y_1^2}\int_{\cF_2} \frac{dx_2dy_2}{y_2^2} f(\tau_1,\tau_2)= \,\underset{s_1=1}{\Res}\,\underset{s_2=1}{\Res} R_{s_1,s_2}[f],
}
where the double Rankin-Selberg transform is defined as:
\es{}{ 
R_{s_1,s_2}[f]:= \int_{\cF_1} \frac{dx_1dy_1}{y_1^2}\int_{\cF_2} \frac{dx_2dy_2}{y_2^2} f(\tau_1,\tau_2) E_{s_1}(\tau_1)E_{s_2}(\tau_2).
}
The overlaps with a single Eisenstein can also be computed in terms of the Rankin-Selberg transform on the critical line $\Re{s}=\half$:
\es{}{ 
(f,\phi_0 E_s(\tau_1))&= \frac{1}{\sqrt{\text{vol}(\cF)}} \int_{\cF_1} \frac{dx_1dy_1}{y_1^2}\int_{\cF_2} \frac{dx_2dy_2}{y_2^2} f(\tau_1,\tau_2) E_{s}(\tau_1)\\&= \sqrt{\text{vol}(\cF)} \underset{r=1}{\Res}R_{1-s,r}[f],
}
and similarly for $(f,\phi_0 E_s(\tau_2))$.

\ssec{Regularization of the Cotler-Jensen wormhole}
As an application, we show how the divergence of the Cotler-Jensen wormhole can be obtained and regularized from the constant term in the spectral decomposition.
The constant term is given by taking residues of the double Rankin-Selberg transform: 
\es{}{ 
\la Z_\CJ\ra &= ~\underset{s_1=1}{\Res}\,\underset{s_2=1}{\Res} \int_{\cF_1} \frac{dx_1dy_1}{y_1^2}\int_{\cF_2} \frac{dx_2dy_2}{y_2^2} Z_\CJ(\tau_1,\tau_2) E_{s_1}(\tau_1)E_{s_2}(\tau_2)\\&=
\frac{1}{\text{vol}(\cF)} \underset{s_2=1}{\Res} \int_{\cF_1} \frac{dx_1dy_1}{y_1^2} \frac{\Gamma(s_2)\Gamma(1-s_2)}{\pi}E_{s_2}(\tau_1).
}
The Rankin-Selberg transform has a double pole at $s_2=1$, indicating the need for regularization. Simply taking the residue one obtains
\es{cjres}{ 
\la Z_\CJ\ra&=- \frac{1}{\pi\text{vol}(\cF)} \int_{\cF_1} \frac{dx_1dy_1}{y_1^2}\widehat{E}_1(\tau_1).
}
where
\e{}{\widehat E_1(\t) := \lim_{s\rar 1} \(E_s(\t) -  {3\o\pi}{1\o s-1}\)}
is the ``finite part'' of $E_s(\t)$ as $s\rar 1$. The integral \eqr{cjres} is divergent, as $\widehat{E}_1(\tau_1)$ diverges linearly as $y_1\rar\i$.  We can regularize the overlaps by splitting the pole with a new parameter $r$ in the following way, preserving the $s \rar 1-s$ reflection symmetry: 
\es{}{ 
\la Z_\CJ^{(\rm reg)}\ra&= \frac{1}{\pi\text{vol}(\cF)} \underset{s=1}{\Res} \int_{\cF_1} \frac{dx_1dy_1}{y_1^2} \Gamma(s+r-1)\Gamma(r-s)E_s(\tau_1)\\&=  \frac{1}{\pi\text{vol}(\cF)^2} \int_{\cF_1} \frac{dx_1dy_1}{y_1^2} \Gamma(r)\Gamma(r-1)\\&=  \frac{1}{\pi\text{vol}(\cF)}  \Gamma(r)\Gamma(r-1)}
As $r \rar 1$, this has a divergence that is constant in $\tau_1,\tau_2$, so we can regularize by subtracting this term: 
\es{}{ 
Z_\CJ^{\rm (reg)}=\underset{r\rightarrow 1}{\lim}\qty(Z_\CJ(\tau_1,\tau_2)- \frac{3}{\pi^2} \frac{1}{r-1} )
}
This is the same regularization originally proposed by \cite{Cotler:2020ugk}, now obtained from regularizing the constant term in $L^2(\cF\times \cF)$. The remaining finite constant is a free parameter of the regularization scheme.

\ssec{Diagonal projection}

Let us make a comment on the diagonal projection of Section \ref{sec:zdiag}, defined as projection of $\Zs(\t_1)\Zs(\t_2)$ onto ker$(\D_{12})$. Note that functions for which $\D_\t f(\t)$ is constant are annihilated by $\D_{12}$. When $\D_\t f(\t)=0$, $f(\t)$ is a constant. When $\D_\t f(\t)\neq 0$, the unique solution up to rescaling is given by $\widehat E_1(\t)$, which obeys 
\e{}{\D_\t \widehat E_1(\t) = -{3\o \pi}}
Thus, given a $\Zd(\t_1,\t_2)$, we may define
\e{Zdshift}{\Zd'(\t_1,\t_2) = \Zd(\t_1,\t_2) + c_1(\widehat E_1(\t_1) + \widehat E_1(\t_2)) + c_2}
which also obeys $\Zd'(\t_1,\t_2) \in \text{ker}(\D_{12})$, where $c_1$ and $c_2$ are constants. The choice of these constants is thus a choice of scheme for the diagonal projection. In the main text, we choose $c_1=c_2=0$. One way to motivate this choice is to note that $\Zd'(\t_1,\t_2) \notin L^2(\cF\x\cF)$ when $c_1\neq 0$: in particular, $\widehat E_1(\t)$ is not part of the basis \eqr{doublebasis}. So the choice $c_1=c_2=0$ may be rephrased as the condition that the diagonal projection preserve the square integrability, present in the original product $\Zs(\t_1)\Zs(\t_2)$, and moreover sets the constant term to zero.  

\section{Continuity of the $j\rightarrow 0$ limit }\label{app:jzero}
In this appendix we show that the scalar Fourier mode of the Eisenstein series $E_{s,0}(y)$ can be obtained from a smooth $j\rightarrow 0$ limit of the spin $j$ mode $E_{s,j}(y)$,  and similarly for the Eisenstein density $\rho_{s,j}(t)$.

We want to evaluate the following limit
\es{}{ 
\lim_{j\rightarrow 0} E^*_{s,j}(y)=\lim_{j\rightarrow 0} \frac{2\sigma_{2s-1}(j) }{j^{s-\half} } \sqrt{y}K_{s-\half}(2\pi jy)
}
and show its equality with
\e{}{E^*_{s,0}(y)= \Lambda(s)y^s +\Lambda(1-s)y^{1-s}}
The limit $j\rightarrow 0$ for fixed $y$ corresponds to the small argument expansion for the Bessel function which is,  at leading order
\es{}{ 
K_{s-\half}(2\pi jy)\approx \frac{(2\pi jy)^{s-\half}}{2^{s+\half}} \Gamma\qty(\half-s) +( s\rightarrow 1-s) 
}
Using $\Lambda(s)=\Lambda\qty(\half-s)=\pi^{s-\half}\Gamma\qty(\half-s)\zeta(1-2s)$ we can simplify various factors to obtain
\es{}{ 
\lim_{j\rightarrow 0} \frac{2\sigma_{2s-1}(j) }{j^{s-\half}} \sqrt{y}K_{s-\half}(2\pi jy)= \lim_{j\rightarrow 0} \qty(\Lambda(s)y^s  \frac{\sigma_{2s-1}(j) }{\zeta(1-2s)} + ( s\rightarrow 1-s) )
}
Remembering that the divisors of zero are all the natural numbers we have
\es{siglimit}{ 
\lim_{j\rightarrow 0}\sigma_{2s-1}(j)=\zeta(1-2s)
}
which correctly gives the scalar result.

The same manipulations may be applied to the corresponding density.  The spin-$j$ Eisenstein density is
\e{}{
\rho_{\crit,j}^*(t)=
2 \frac{\sigma_{2i\omega}(j)}{j^{i\omega}}\theta(t)\frac{\cos(\omega \cosh^{-1}\qty(\frac{2t}{j}+1))}{\sqrt{t(t+j)}}
}
To arrive at the scalar density we need to compute the following limit:
\e{}{ 
\lim_{j\rightarrow 0} \frac{\sigma_{2i\omega}(j)}{j^{i\omega}} e^{i\omega \cosh^{-1}\qty(\frac{2t}{j}+1) }= (4t)^{i\omega }\lim_{j\rightarrow 0} \frac{\sigma_{2i\omega}(j)}{j^{2i\omega}}.
}
We use the identity
\e{}{ 
 \frac{\sigma_{2i\omega}(j)}{j^{2i\omega}}= \sigma_{-2i\w}(j)
}
which, together with the previous limit \eqr{siglimit}, recovers the $j=0$ density from the $j\rightarrow 0$ limit, 
\es{}{ 
\lim_{j\rightarrow 0} \rho^*_{\half +i\omega,j}(t)&=\rho^*_{\half +i\omega,0}(t)\\&=\frac{\zeta(2i\omega)}{t}(4t)^{i\omega}+(\omega\rightarrow -\omega).
}

\section{Correlations among spins}\label{app:pindelta}
In this appendix we prove that modular invariance correlates unequal spins in the diagonal approximation: 
\e{}{ 
Z_{\diag}^{(j_1,j_2)}(y_1,y_2)\propto \delta_{j_1,j_2} \quad \Longleftrightarrow \quad Z_{\diag}(\t_1,\t_2)=0
}
We proceed by contradiction, assuming that $Z_{\diag}^{(j_1,j_2)}(y_1,y_2)\propto \delta_{j_1,j_2}$. 

First suppose the scalar sector is non-vanishing, $Z_{\diag}^{(0,0)}(y_1,y_2)\neq0$. Then the Eisenstein overlaps are non-vanishing:
\e{}{ 
Z_{\diag}^{(0,0)}(y_1,y_2)\neq0 \quad \Rightarrow \quad (Z_{\spec},E_s)\neq 0.
}
The spin-$(0,j)$ mode, which by assumption is equal to zero, is given by 
\es{}{ 
0=Z_{\diag}^{(0,j)}(y_1,y_2)&= \int_{\cC_{\rm crit}}  \{Z_{\spec},E_s\}^2 E^*_{s,0}(y_1)\mathsf{a}_j^{(s)} \sqrt{y_2}K_{s-\half}(2\pi jy_2)\\+&
\sum_{n=1}^\i \{Z_{\spec},E_{s_n}\}(Z_{\spec},\phi_n)\qty(E^*_{s_n,0}(y_1) \mathsf{b}_j^{(n)} \sqrt{y_2}K_{s_n-\half}(2\pi jy_2)+(y_1\leftrightarrow y_2))
}
We can single out the first line by using the Bessel orthogonality relation
\es{}{ 
\int_0^\infty dx \frac{K_{it}(x)K_{i\omega}(x)}{x}= \frac{\pi^2}{2t\sinh(\pi t)}\delta(t-\omega),
}
which allows us to project onto a chosen, non-degenerate eigenvalue $\omega\neq \omega_n$ on the critical line. This implies $\{Z_{\spec},E_s\}=0$, in contradiction with the starting assumption that $Z_{\diag}^{(0,0)}(y_1,y_2)\neq 0$. 

If instead $Z_{\diag}^{(0,0)}(y_1,y_2)=0$, then the  Eisenstein overlaps vanish. But nonzero cusp form overlaps imply $Z_{\diag}^{(j_1,j_2)}(y_1,y_2)\neq0$ for $j_1\neq j_2$, which is in contradiction with the $\delta_{j_1,j_2}$ behavior.

Thus the only modular-invariant $\Zd(\t_1,\t_2)$ proportional to $\delta_{j_1,j_2}$ is zero. 

\sec{Derivation of \eqref{S00final}}\label{secconvS}
Here we derive \eqr{S00final}, the ``corrections'' to the RMT ramp that are mandated by $SL(2,\Z)$ invariance, starting from the definition of $\mathcal{S}(r)$ in \eqr{Rdef}. 

To prove \eqr{S00final} we first write 
\e{}{\mathcal{S}(r) = r^2 \cM^{-1}[\cM[\mathcal{R}(z);1-s]\varphi(s);r^2]}
We now use the Mellin convolution theorem: given two functions $f(s)$ and $g(s)$ with inverse Mellin transforms\foot{Defined within the relevant strips.} $F(z)$ and $G(z)$, respectively,
\e{mellinconv1}{\cM^{-1}[f(s)g(s);x] = \int_0^\i {dy\o y} F\({x\o y}\) G(y)}
If $g(s) = g(1-s)$, then $G(x) = x^{-1} G(x^{-1})$, and we can instead write
\e{mellinconv2}{\cM^{-1}[f(s)g(s);x] = \int_0^\i dy\, F(x y) G(y)}
Since $\cM[\mathcal{R}(z);s] = \cM[\mathcal{R}(z);1-s]$ we can apply \eqr{mellinconv2} to write
\e{SD4}{\mathcal{S}(r) = r^{2} \int_0^\i dy \, \cM^{-1}[\varphi(s);r^2y] \mathcal{R}(y)}
The function $\varphi(s)$, defined in \eqr{vphidef}, may be written as
\e{vphiexp}{\varphi(s) = \frac{\sqrt{\pi } \Gamma \left(s-\frac{1}{2}\right)}{\Gamma (s)}{\zeta (2 s-1) \o \z(2s)}}
We next use
\e{zexp}{{\zeta (2 s-1) \o \z(2s)} = \sum_{n=1}^\i {\phi(n)\o n^{2s}}}
where $\phi(n)$ is the Euler totient function. Using \eqr{vphiexp} and \eqr{zexp},
\es{}{\cM^{-1}[\varphi(s);r^2y] &= \sum_{n=1}^\i \phi(n) \cM^{-1}\[n^{-2s}\frac{\sqrt{\pi } \Gamma \left(s-\frac{1}{2}\right)}{\Gamma (s)};r^2y\]\\
&=  \sum_{n=1}^\i {\phi(n) \o n} \frac{\Theta \left(1-n^2 r^2 y\right)}{\sqrt{r^2 y \left(1-n^2 r^2y\right)}}}
Plugging into \eqr{SD4} and moving the sum over $n$ outside the integral, this can be massaged into the form 
\e{S00finalapp}{{\mathcal{S}(r) = \sum_{n=1}^\i {\phi(n) \o n^2}  \int_0^1 {du\o \sqrt{u(1-u)}} \, \mathcal{R}\({u \o n^2 r^2}\)}}
This is \eqr{S00final}. 

As a check, we plug in the CJ/MaxRMT result, $\cR_\CJ(z) =\pi^{-1}/(1+z)$, which gives 
\e{}{\cS_\CJ(r) =\sum_{n=1}^\i {\phi(n)\o n}{r\o \sqrt{1+n^2 r^2}}}
This reproduces the correct result \eqr{cjrs}. 

Note that if $\mathcal{R}(0)$ is finite and nonzero, then $\mathcal{S}(r)$ has a linear, $r$-independent, divergence as $r\rar\i$:
\e{S00div}{\mathcal{S}(r) \approx \pi \mathcal{R}(0) \sum_{n=1}^\i {\phi(n) \o n^2}  \qquad (r\rar\i)}
This is the case for the CJ/MaxRMT wormhole, where $\mathcal{R}_\CJ(0)=\pi^{-1}$. 

We can also rewrite \eqr{S00finalapp} as a sum over residues by contour deformation. Since the integral is along the branch cut from zero to one,\foot{$\mathcal{R}(z)$ is analytic for all $z>0$ because it is the low-temperature limit $y_1,y_2\rar\i$ for fixed $z=y_1/y_2$, which is non-singular on physical grounds.}  we can replace it with a contour surrounding the branch cut, which we then blow up to pick up residues elsewhere in $\mathbb{C}$. For example, simple poles of $\mathcal{R}(z)$ contribute as
\e{S00final3app}{\mathcal{S}(r) \supset \pi \sum_{n=1}^\i {\phi(n) \o n}   \sum_{z_*} \sqrt{r^2\o z_*(n^2 r^2 z_*-1)}\underset{z=z_*}{\Res} \mathcal{R}(z)}
This may of course be generalized to include higher-order poles or non-meromorphic behavior.

\sec{Toy wormholes in AdS$_3 \times \mathcal{M}$ gravity}\label{app:toywh}

In this appendix we construct an explicit family of torus wormhole amplitudes which are toy models (or perhaps candidates) for off-shell wormholes of AdS$_3$ gravity with light matter, such as AdS$_3 \x \cM$ compactifications with large $\cM$. These theories are not MaxRMT theories, and their wormholes should reflect this. Specifically, the function $\mathcal{R}_\WH(z)$, which determines the entire amplitude (see \eqr{Rdef} and \eqr{zwh}), should have poles in the complex $z$ plane away from the ramp pole at $z=-1$ with coefficients of $\O(c^0)$, thus augmenting the CJ pure gravity wormhole with large corrections. The examples given below have the added property that both the spectral overlap $f_\WH(s)$ and their Mellin transforms $\mathcal{R}_\WH(z)$ are computable, with a linear ramp. 

We first register a useful Mellin transform,
\e{}{\cM\[{1\o 1+nz};s\] = n^{-s} \G(s)\G(1-s)}
Note that reflection symmetry $f_\WH(s) = f_\WH(1-s)$ requires the poles to appear in pairs, at $z=z_*$ and $z=z_*^{-1}$. 
The reflection-symmetric combination is
\es{ccomb}{\cM\[{1\o 1+n z} + {1\o n+z};s\] &= (n^{-s} +n^{s-1})\G(s)\G(1-s)}
A well-known \cite{FLAJOLET19953} set of Mellin dual pairs involves harmonic sums in $z$, i.e. taking $F_\WH(z)$ to be an infinite sum over simple poles with no regular term. So long as we include the pole at $z=-1$ with the correct normalization, the wormhole reproduces the linear ramp. In accordance with the comments in Section \ref{sec:otherworm}, the remaining poles should avoid $z\in\R_+$ and $|z|=1$. For this reason, harmonic sums over poles along $ z\in\R_-$ is not only mathematically, but also physically, well-motivated here. 

Summing over $n$ in \eqr{ccomb} with an insertion of $n^{-p}$, we of course have
\e{whzeta}{\sum_{n=1}^\i n^{-p-s}= \z(s+p)}
and its reflection. For integer $p\in\Z_+$, the inverse Mellin transform of \eqr{whzeta} defines a convergent harmonic sum, plus a simple $p$-dependent shift:
\e{whharm}{\sum_{n=1}^\i {n^{-p}\o 1+n z} = (-z)^{p-1} H(z^{-1}) + \sum_{q=2}^p \z(q) (-z)^{p-q} }
So the previous two equations are Mellin partners. Upon adding the reflected term, we have an explicit pair for every $p\in\Z_+$. That is, written in terms of $f_\CJ(s) = \pi^{-1} \G(s)\G(1-s)$,
\e{}{f_\WH(s) = f_\CJ(s) \x {\mathsf{C}_\RMT\o 4}\big(\z(s+p) + \z(p+1-s)\big)}
and its inverse Mellin transform
\e{whharmfinal}{\mathcal{R}_\WH(z) = {\mathsf{C}_\RMT\o 4\pi} \Bigg((-z)^{p-1} H(z^{-1}) + \sum_{q=2}^p \z(q) (-z)^{p-q} +(-z)^{-p} H(z) - \sum_{q=2}^p \z(q) (-z^{-1})^{p+1-q}\Bigg)}
Inserting this overlap into the wormhole ansatz \eqr{zwh} defines a family of toy model RMT wormholes for a theory of gravity with light matter, one for every $p\in\Z_+$.\foot{It would be nice to ask whether one can construct an amplitude of this form from a Poincar\'e sum over relative modular transformations. The structure of such Poincar\'e sums might also constrain the allowed singularity structures in the complex $z$ plane.}

\begin{figure}[t]
\centering
{
\subfloat{\includegraphics[scale=0.75]{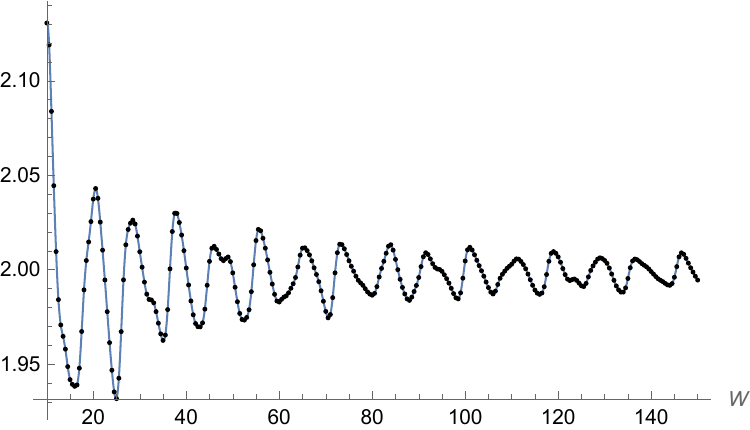}}
}
\caption{A plot of the integral in \eqr{F8int} for $p=1, \w_0=100$ and $W\in[10,150]$, exhibiting oscillations  around the asymptotic value of two. Black points are numerical data points, and the blue line is an interpolation.}
\label{fig:toywh}
\end{figure}

By construction, this has the RMT ramp:
\e{}{\mathcal{R}_\WH(z \rar -1) \sim {\mathsf{C}_\RMT\o 2\pi}{1\o 1+z}}
On the Mellin side, one readily verifies that
\e{F8int}{{1\o W}\int_{\w_0}^{\w_0+W} d\w \(\z\(p+\half+i\w\) + \z\(p+\half-i\w\) \) \approx 2\qquad (W\rar\i)}
for all $p\geq 1$ and any finite $\w_0$, because $|\z(\s+i\w)| \rar 1$ asymptotically as $\w\rar\i$ for any $\s>1$. This can be seen numerically in Figure \ref{fig:toywh}. Together with the exactly constant falloff of $f_\CJ(s)$, the total $f_\WH(s)$ satisfies the linear ramp condition \eqr{lineardivfinal}. 

\end{spacing}

\bibliographystyle{JHEP}
\bibliography{rmtbib_sample}

\end{document}